\documentclass[a4paper,11pt]{article}
\pdfoutput=1 % if your are submitting a pdflatex (i.e. if you have
\usepackage{jheppub} % for details on the use of the package, please
\usepackage[T1]{fontenc} % if needed

\usepackage{etoolbox}% http://ctan.org/pkg/etoolbox
\makeatletter
\patchcmd{\maketitle}{\@fpheader}{}{}{}
\makeatother

\newcommand{\pip}{\ensuremath{\pi^{+}}}
\newcommand{\pim}{\ensuremath{\pi^{-}}}

\newcommand{\pimp}{\ensuremath{\pi^{\mp}}}
\newcommand{\piz}{\ensuremath{\pi^{0}}}
\newcommand{\rhop}{\ensuremath{\rho^{+}}}
\newcommand{\rhom}{\ensuremath{\rho^{-}}}
\newcommand{\rhoz}{\ensuremath{\rho^{0}}}
\newcommand{\fz}{\ensuremath{f_{0}}}
\newcommand{\aone}{\ensuremath{a_{1}^{\pm}}}
\newcommand{\aonep}{\ensuremath{a_{1}^{+}}}
\newcommand{\aonem}{\ensuremath{a_{1}^{-}}}

\newcommand{\Kp}{\ensuremath{K^{+}}}
\newcommand{\Km}{\ensuremath{K^{-}}}
\newcommand{\Ks}{\ensuremath{K^{0}_{S}}}
\newcommand{\Kz}{\ensuremath{K^{0}}}
\newcommand{\Kstp}{\ensuremath{K^{*+}}}

\newcommand{\Kone}{\ensuremath{K_{1}}}
\newcommand{\Koneap}{\ensuremath{K_{1A}^{+}}}

\newcommand{\Koneaz}{\ensuremath{K_{1A}^{0}}}
\newcommand{\Konea}{\ensuremath{K_{1A}}}
\newcommand{\Konebz}{\ensuremath{K_{1B}^{0}}}
\newcommand{\Koneb}{\ensuremath{K_{1B}}}

\newcommand{\Bp}{\ensuremath{B^{+}}}
\newcommand{\Bm}{\ensuremath{B^{-}}}
\newcommand{\Bz}{\ensuremath{B^{0}}}
\newcommand{\Bzb}{\ensuremath{\bar B^{0}}}

\newcommand{\lcp}{\ensuremath{\lambda_{CP}}}
\newcommand{\Acp}{\ensuremath{{\cal A}_{CP}}}

\newcommand{\phione}{\ensuremath{\phi_{1}}}
\newcommand{\phitwo}{\ensuremath{\phi_{2}}}
\newcommand{\phithree}{\ensuremath{\phi_{3}}}

\title{\boldmath Resolving the \phitwo\ ($\alpha$) ambiguity in $B^0 \to a_1^\pm \pi^\mp$}

\author{J. Dalseno}
\affiliation{Instituto Galego de F\'{i}sica de Altas Enerx\'{i}as (IGFAE), Universidade de Santiago de Compostela,\\R\'{u}a de Xoaqu\'{i}n D\'{i}az de R\'{a}bago, Santiago de Compostela, Spain}

\collaborationImg{\hspace{-29pt}\includegraphics[height=100pt,width=!]{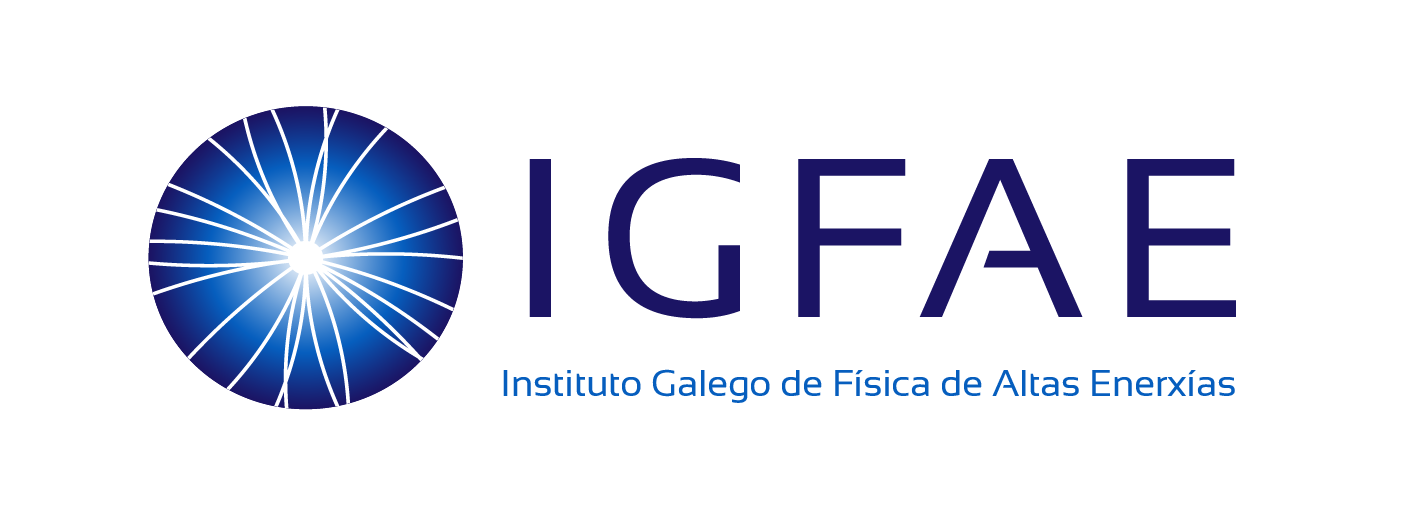}\vspace{15pt}}

\emailAdd{jeremy.peter.dalseno@cern.ch}

\abstract{
  I propose an alternative method for measuring the $CP$ violating phase $\phi_2$ ($\alpha$) without ambiguity in an extended SU(3) flavour symmetry analysis, which can ultimately be achieved by exploiting interference effects between $B \to AP$ and $B \to VV$ decay channels, where $A, V, P$ indicates an axial-vector, vector and pseudo-scalar meson, respectively. Under certain assumptions on the relevant decays based on current experimental results and minimal theoretical input, I demonstrate with an idealised amplitude model that a programme to extract a single solution for $\phi_2$ in the range [0,$\pi$], with the added possibility to simultaneously constrain non-factorisable SU(3)-breaking effects, could be executed to similar precision using Run~3 data at LHCb and the final Belle~II sample.
}

\begin{document}

%\linenumbers

\maketitle
\flushbottom

\section{Introduction}

Violation of the combined charge-parity symmetry ($CP$ violation) in the Standard Model (SM) arises from a single irreducible phase in the Cabibbo-Kobayashi-Maskawa~(CKM) quark-mixing matrix~\cite{Cabibbo,KM}. Various processes offer different yet complementary insight into this phase, which manifests in a number of experimental observables over-constraining the Unitarity Triangle. The measurement of such parameters and their subsequent combination is important as New Physics (NP) contributions can present themselves as an inconsistency within the triangle paradigm.

\begin{figure}[tbp]
  \centering
  \includegraphics[height=115pt,width=!]{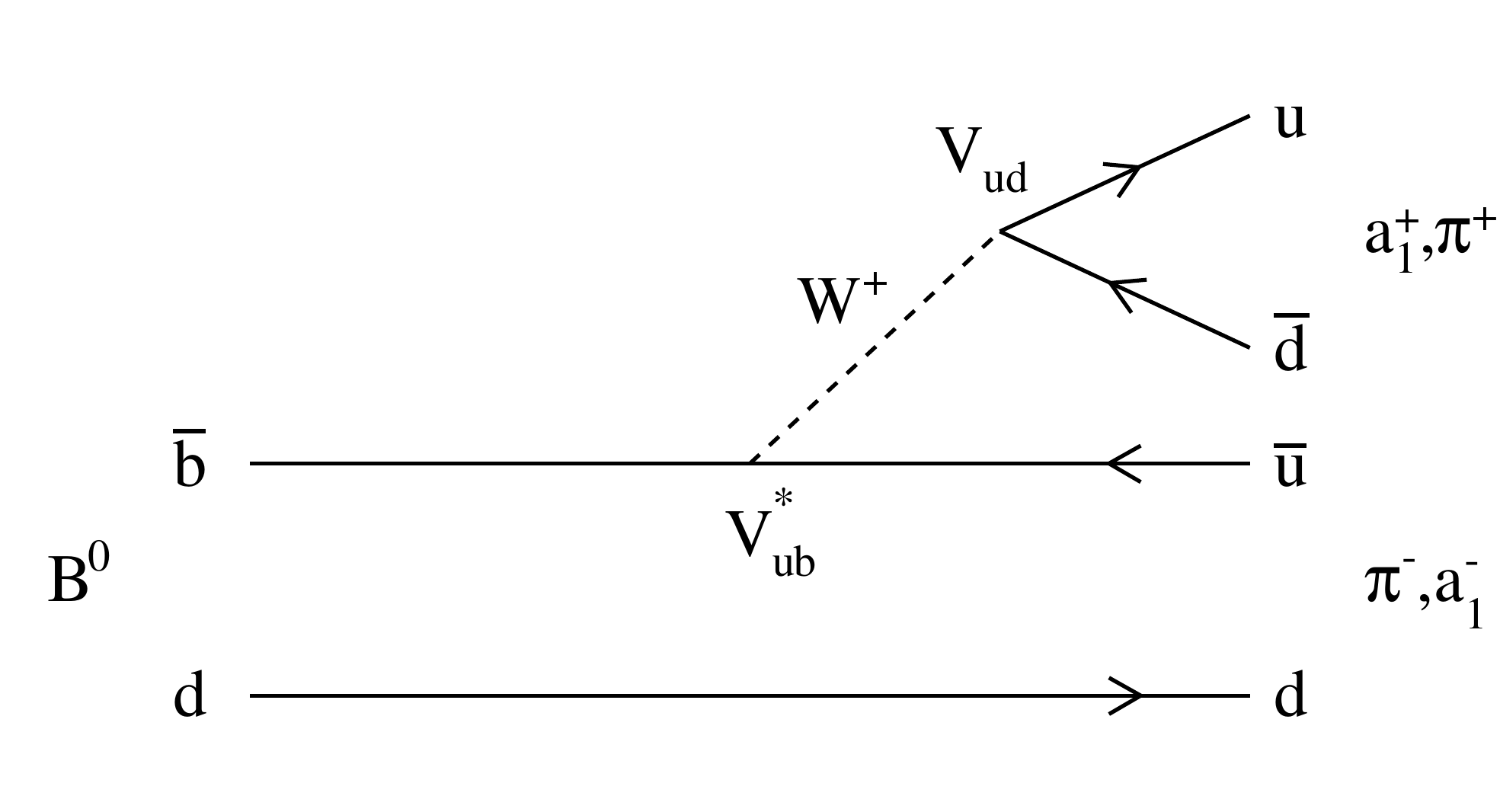}
  \includegraphics[height=115pt,width=!]{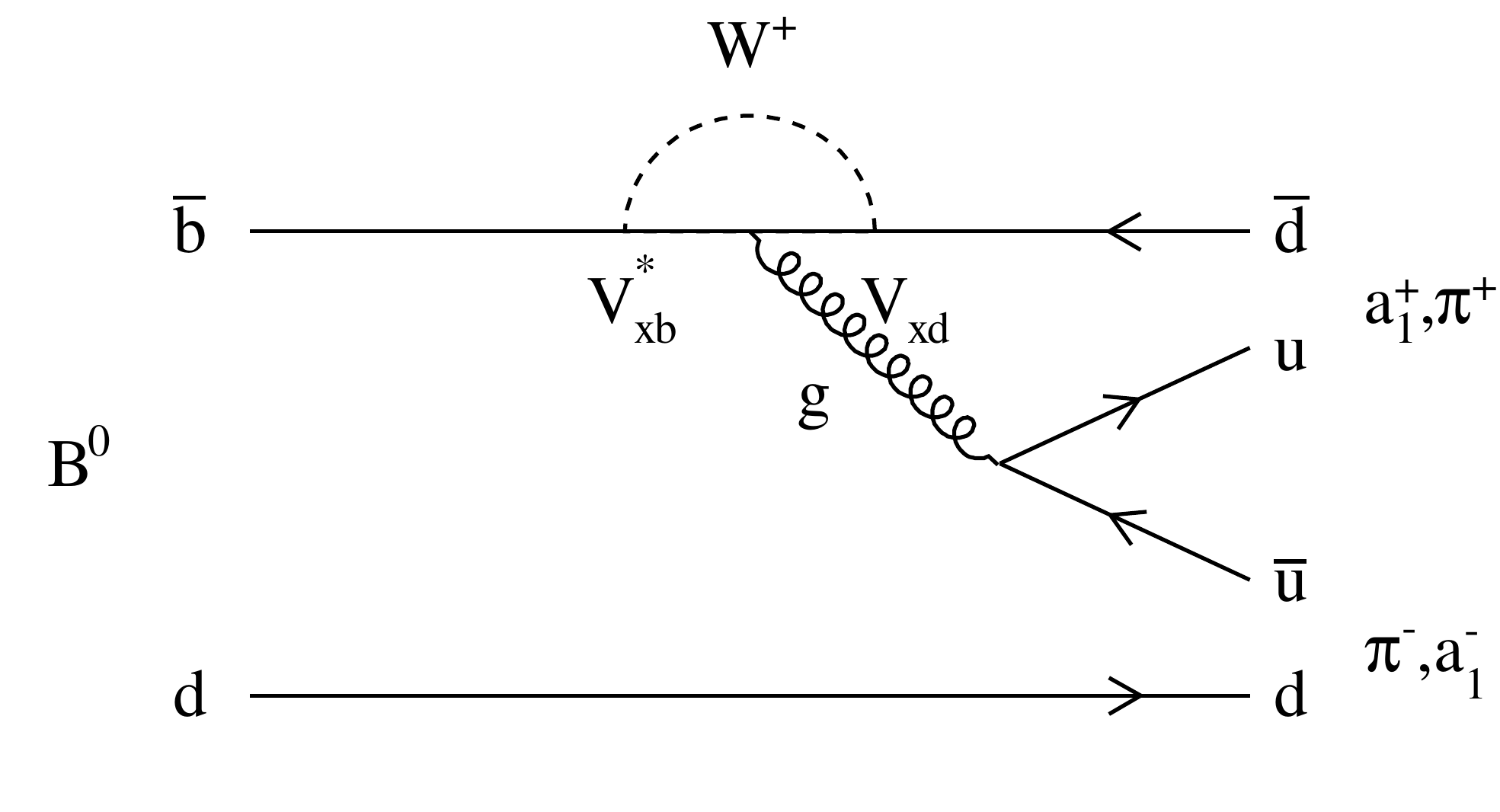}
  \put(-425,110){(a)}
  \put(-210,110){(b)}
  \caption{\label{fig:pipi} Leading-order Feynman diagrams shown producing $\Bz \to \aone \pimp$ decays, though the same quark transition can also produce $\Bz \to \pip \pim$, $\rho^{\pm} \pimp$ and $\rho^+ \rho^-$. (a) depicts the dominant (tree) diagram while (b) shows the competing loop (penguin) diagram. In the penguin diagram, the subscript $x$ in $V_{xb}$ refers to the flavour of the intermediate-state quark $(x=u,c,t)$.}
\end{figure}

Decays that proceed predominantly through the $\bar b \rightarrow \bar u u \bar d$ tree transition (figure~\ref{fig:pipi}a) in the presence of \Bz--\Bzb\ mixing are sensitive to the interior angle of the Unitarity Triangle $\phitwo = \alpha \equiv \arg(-V_{td}V^{*}_{tb})/(V_{ud}V^{*}_{ub})$, which can be accessed through %then comprises
mixing-induced $CP$ violation observables measured from time-dependent, flavour-tagged analyses.
This quark process manifests itself in multiple systems, including $B \to \pi \pi$~\cite{phi2_pipi1,phi2_pipi2,phi2_pipi3,phi2_pipi4,phi2_pipi5,phi2_pipi6}, $\rho \pi$~\cite{phi2_rhopi1,phi2_rhopi2}, $\rho \rho$~\cite{phi2_rhorho1,phi2_rhorho2,phi2_rhorho3,phi2_rhorho4,phi2_rhorho5,phi2_rhorho6,phi2_rhorho7} and $\aone \pimp$~\cite{phi2_a1pi1,phi2_a1pi2,phi2_a1pi3}, where the angle \phitwo\ has so far been constrained with an overall uncertainty of around $4^\circ$~\cite{phi2_gronau,CKMFitter1,UTfit}. However, one of the salient features of the overall \phitwo\ combination is the persistence of degenerate solutions within the range $[0, \pi]$: up to the $2\sigma$ level, two solutions currently remain, while beyond this further solutions emerge.

In a previous work, I showed how the ratio of time-dependent decay amplitudes, better known as the complex $CP$ violating parameter $\lcp \equiv \exp{(-2i\phitwo)}\bar A/A$, could be measured at amplitude level in $\Bz \to \rhoz \rhoz$ leading to a unique solution for \phitwo\ in the $B \to \rho\rho$ system~\cite{rhorho_dalseno}. What was not discussed at the time however, was that due to the procedure in which degenerate solutions are searched for in time-dependent amplitude analyses, $\lcp$ is also resolved for both $\Bz \to \aonep \pim$ and $\aonem \pip$ individually with the same significance as it would be in $\Bz \to \rhoz \rhoz$.

In this paper, I expound further on that idea with a proposal to additionally resolve the \phitwo\ solution degeneracy in $B^0 \to \aone \pimp$, once again achieved by harnessing interference effects unique to multibody decays. Essentially this involves relating the experimentally measured amplitudes of $B \to K3\pi$ final states to that of the formerly discussed $\Bz \to \pip \pim \pip \pim$. I open in section~\ref{sec:su3}, with a description of the SU(3)-based approach for controlling distortions in experimental \phitwo\ measurements arising from the ever-present strong-loop penguin processes. Following this, I outline an extension in section~\ref{sec:su3ext}, which permits a single solution for \phitwo\ to be obtained that is free of contamination from strong penguins and possibly distortion from non-factorisable SU(3) breaking effects. To demonstrate the capabilities of this proposed concept, section~\ref{sec:model} describes the rudimentary models used to generate pseudo-experiments containing the potential interference effects in the $\Bp \to \Kz \pip \pim \pip$ phase space allowed by current experimental limitations. The results of the pseudo-experiment study are discussed for various future experimental milestones in section~\ref{sec:results} and finally conclusions are drawn in section~\ref{sec:conc}.

\section{\boldmath Strong-penguin containment in \phitwo\ constraints}
\label{sec:su3}

In general, the extraction of \phitwo\ is complicated by the presence of interfering amplitudes that distort the experimentally determined value of \phitwo\ from its SM expectation and would mask any NP phase if not accounted for. These effects primarily include $\bar b \rightarrow \bar d u \bar u$ strong-loop decays (figure~\ref{fig:pipi}b), although isospin-violating processes such as electroweak penguins, $\piz$--$\eta$--$\eta^\prime$ mixing, $\rho^0$--$\omega$--$\phi$ mixing~\cite{iso_mixing} and the finite $\rho$ width in $B \to \rho\rho$~\cite{rhowidth1,rhowidth2} can also play a role.

The original method creating the possibility to remove the isospin-conserving component of this contamination invokes SU(2) arguments in a triangular analysis with input coming from the three $B \to \pi \pi$ or $\rho\rho$ charge configurations for an 8-fold degeneracy in \phitwo~\cite{pipi_th}. This can be reduced to two solutions if mixing-induced $CP$ violation can be measured in their respective colour-suppressed channels~\cite{phi2_gronau} and reduced further to a single solution in $B \to \rho\rho$ if the complex $CP$ violating parameter of $\Bz \to \rhoz \rhoz$ can be measured directly in a time-dependent, flavour-tagged amplitude analysis~\cite{rhorho_dalseno}. For flavour-non-specific channels such as $\Bz \to \rho^\pm \pimp$ and $\aone \pimp$, the original idea was subsequently extended to isospin pentagonal relations~\cite{rhopi_th1}. Then for $B \to \rho \pi$, a time-dependent flavour-tagged amplitude analysis of the $\Bz \to \pip \pim \piz$ final state was suggested to eliminate the problem of multiple solutions all without the need to involve the charged $B$ modes~\cite{rhopi_th2}, though they can still be combined with constructs from the former method to improve the constraint if desired.

In principle, it is possible to resolve the \phitwo\ ambiguity with a $\Bz \to (\rho\pi)^0$-style analysis in $\Bz \to (a_1\pi)^0 \to \pip \pim \piz \piz$ through interference with $\Bz \to \rhop \rhom$, although experimentally this is a highly unattractive prospect due to the presence of two \piz's in the final state. In order to obtain a meaningful constraint on \phitwo\ in this system, a more realistic approach is outlined in ref.~\cite{a1pi_th}. This time operating within the confines of SU(3) flavour symmetry, the rates of $\Bp \to \Koneaz \pip$, $\Kz \aonep$ and $\Bz \to \Koneap \pim$, $\Kp \aonem$ can be combined with the quasi-two-body time-dependent $CP$ violation parameters of $\Bz \to \aone \pimp$ for an 8-fold degeneracy in \phitwo. In spectroscopic notation, the $\Konea$ flavour eigenstate is the SU(3) ${}^3P_1$ partner of the $a_1$ and is an admixture of the $K_1(1270)$ and $K_1(1400)$ mass eigenstates. However, armed with knowledge of the $\Bz \to \pip \pim \pip \pim$ amplitude as suggested in ref.~\cite{rhorho_dalseno}, combined with measurements of the $\Bp \to \Kz \pip \pim \pip$ and $\Bz \to \Kp \pim \pip \pim$ amplitudes proposed here, a single solution for \phitwo\ is also attainable in $B \to a_1 \pi$.

\section{\boldmath Extension to the SU(3) flavour symmetry analysis}
\label{sec:su3ext}

In this paper, I employ the frequentist approach adopted by the CKMfitter Group~\cite{CKMFitter1} where a $\chi^2$ is constructed comparing theoretical forms for relations between parameters of interest and the physical observables with the experimentally measured values of those observables. The value of $\Delta \chi^2$ across the range of \phitwo\ can then be converted into a $p$-value scan, assuming it is distributed with one degree of freedom, from which confidence intervals can be derived.

Ultimately, my proposal relies on the ability to measure the strong phase difference between the axial-vector resonances contributing to the $B \to K 3\pi$ final states that contain the spectator quark and those that do not. If this can be achieved, the machinery to extract a unique solution for \phitwo\ can be inferred from ref.~\cite{a1pi_th}. Beginning with $\Bz \to \aone \pimp$, the amplitudes are given by
\begin{eqnarray}
  A(\Bz \to \aonep \pim) &=& T^+ e^{+i\phithree} + P^+, \nonumber \\
  A(\Bz \to \aonem \pip) &=& T^- e^{+i\phithree} + P^-,
\end{eqnarray}
where $T$ and $P$ represent complex amplitudes only involving strong dynamics and $\phithree = \gamma \equiv \arg(-V_{ud}V_{ub}^{*})/(V_{cd}V_{cb}^{*})$. Naturally, the $CP$-conjugate amplitudes are then given by
\begin{eqnarray}
  \bar A(\Bzb \to \aonem \pip) &=& T^+ e^{-i\phithree} + P^+, \nonumber \\
  \bar A(\Bzb \to \aonep \pim) &=& T^- e^{-i\phithree} + P^-.
\end{eqnarray}
For the purposes of a \phitwo\ constraint, \phithree\ should be parametrised as $\pi -\phione -\phitwo$, where $\phione=\beta \equiv \arg(-V_{cd}V^{*}_{cb})/(V_{td}V^{*}_{tb})$ is the phase of \Bz--\Bzb\ mixing. The $CP$ violating parameters of $\Bz \to \aone \pimp$ are thus,
\begin{eqnarray}
  \lambda^+_{CP} = \frac{\bar A(\Bzb \to \aonem \pip)}{A(\Bz \to \aonep \pim)} e^{i(2\pi-2\phione)}, \nonumber \\
  \lambda^-_{CP} = \frac{\bar A(\Bzb \to \aonep \pim)}{A(\Bz \to \aonem \pip)} e^{i(2\pi-2\phione)},
\end{eqnarray}
assuming no $CP$ violation in mixing, $|q/p|=1$. Note that the otherwise redundant appearance of $2\pi$ in the exponent is a technical necessity in the $\chi^2$ calculation due to the unitarity constraint applied when making the choice to express \phithree\ in terms of the other weak phases in the Unitarity Triangle. The overall effective weak phase is denoted by $\phitwo^\pm \equiv \arg(\lcp^\pm)/2$.

\subsection{Minimal SU(3) analysis}

The minimal SU(3)-related decay channel that would have to be studied is $\Bp \to \Kz \pip \pim \pip$, whose axial-vector contributions are pure penguin processes with amplitudes,
\begin{eqnarray}
  A(\Bp \to \Koneaz \pip) &=& -\frac{1}{\bar \lambda} \frac{f_{K_1}}{f_{a_1}} P^+, \nonumber \\
  A(\Bp \to \Kz \aonep) &=& -\frac{1}{\bar \lambda} \frac{f_{K}}{f_{\pi}} P^-,
\end{eqnarray}
where $\bar \lambda = |V_{us}|/|V_{ud}| = |V_{cd}|/|V_{cs}|$ and $f$ represents decay constants calculated with QCD. These CKM elements arising from the different form factors and decay constants that allow comparison of the $\Delta S = 0$ and $\Delta S = 1$ amplitudes of SU(3)-related channels are generally referred to as factorisable SU(3)-breaking corrections. Non-factorisable SU(3)-breaking effects will be discussed later on in section~\ref{sec:su3:nf} and the means by which to measure an amplitude relative to the $\Konea$ will be addressed in section~\ref{sec:model:k1}.

In principle, \phitwo\ could be determined directly from the amplitudes measured by experiment, though the amplitude analyses themselves would have to take care that the amount of phase space available to all analyses is comparable. This inconvenience can be mitigated through the use of branching fractions instead of the magnitudes coming directly from fits to data. These are related through
\begin{equation}
  \frac{{\cal B}}{\tau_B} = \frac{|\bar A|^2 + |A|^2}{2},
\end{equation}
where $\tau_B$ is the lifetime of the $B$ meson.

For the sake of argument, I now assume that the combined analysis of $\Bz \to \rhoz \rhoz$ and $\Bz \to \aone \pimp$ suggested in ref.~\cite{rhorho_dalseno} had since been performed at Belle and somehow they managed, albeit unrealistically given their data sample size, to resolve the weak phase solution degeneracy. In this test, the first solution is taken for the phase difference between $\Bz \to \aonem \pip$ and $\Bz \to \aonep \pim$ and the central values of $\phitwo^\pm$ are taken from QCD factorisation~\cite{fk1} as they cannot be inferred from the quasi-two-body Belle result. I then take input from BaBar on the $\Bp \to \Koneaz \pip$~\cite{phi2_a1pi3} and $\Bp \to \Kz \aonep$~\cite{a1k} branching fractions and leave their phase difference out of the $\chi^2$ calculation as it is currently unknown. The parameters for the minimal SU(3) analysis are listed in table~\ref{tab:su3_min}.

\begin{figure}[tbp]
  \centering
  \includegraphics[height=120pt,width=!]{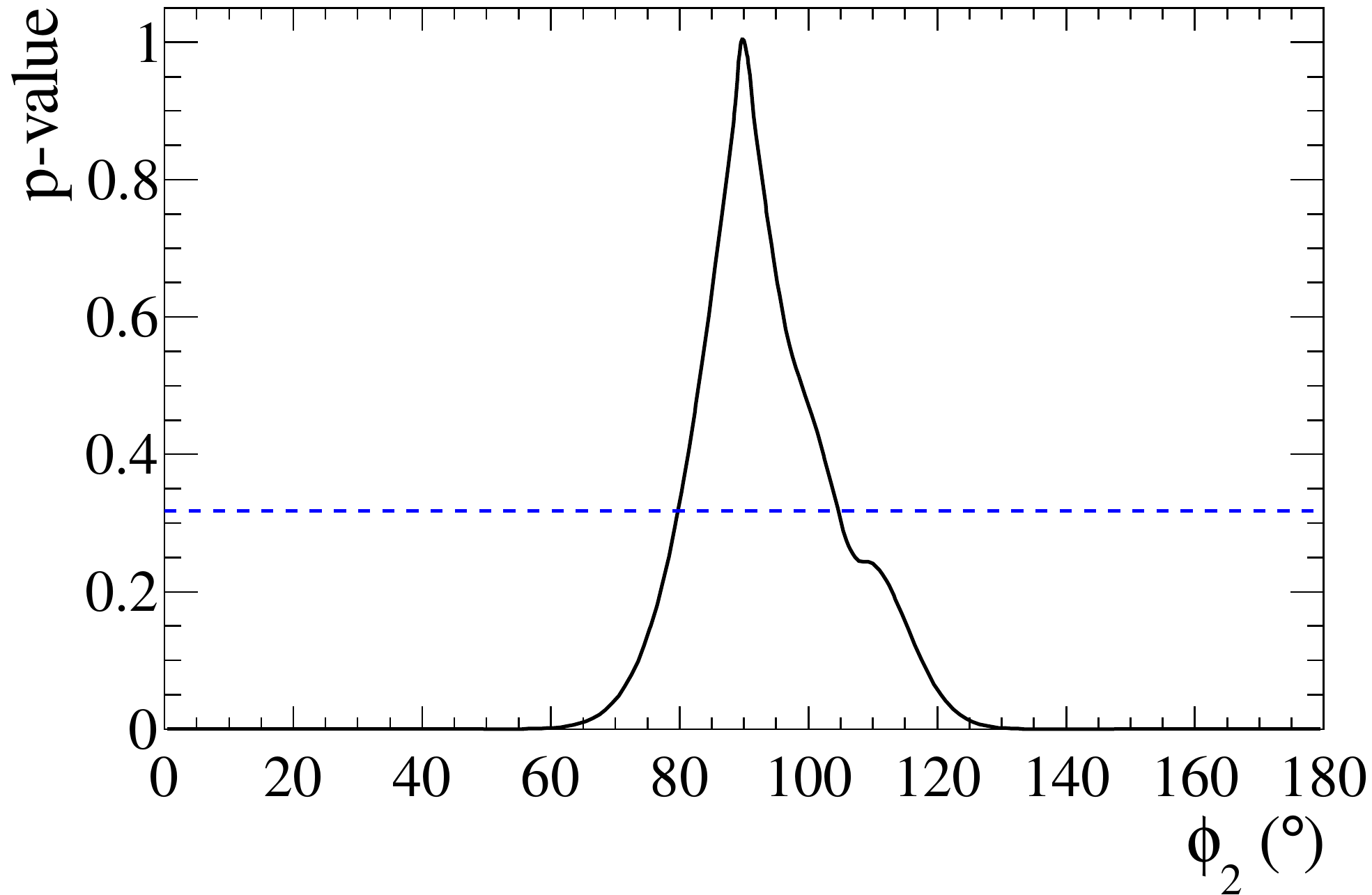}
  \includegraphics[height=120pt,width=!]{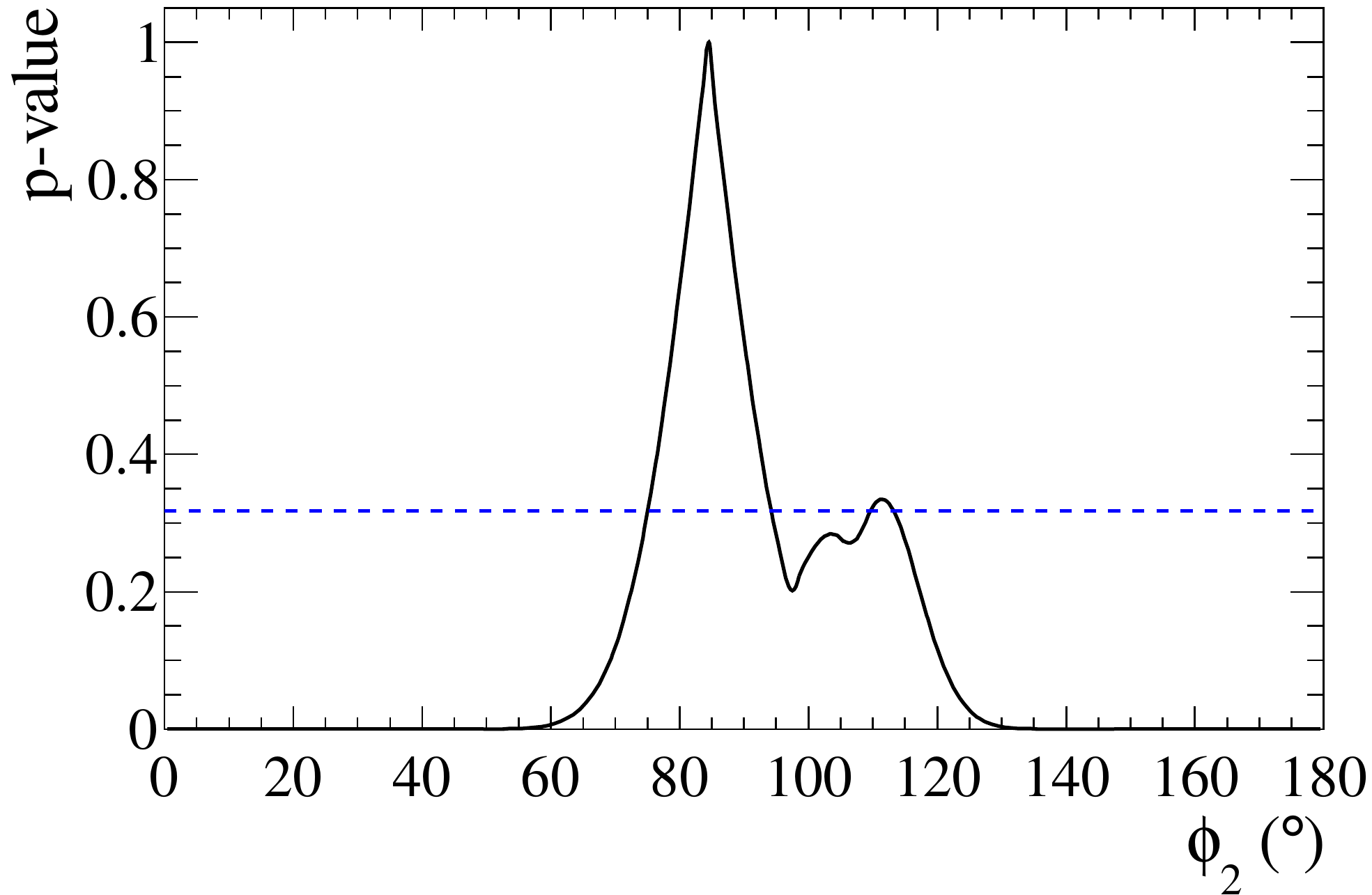}
  \put(-215,105){(a)}
  \put(-28,105){(b)}

  \caption{\label{fig:su3_min:mpv}
    $p$-value scans of \phitwo\ where the $\Bp \to \Koneaz \pip$ branching fraction is set to (a) the most probable value and (b) the mean value obtained by BaBar, with no constraint on the phase difference between $\Bp \to \Kz \aonep$ and $\Bp \to \Koneaz \pip$. The horizontal dashed line shows the $1\sigma$ bound.
  }
\end{figure}

For the $\Bp \to \Koneaz \pip$ branching fraction, the most probable value of $2.0 \times 10^{-6}$ and mean value of $16.0 \times 10^{-6}$ is tested with half their $1\sigma$ C.L. interval taken as the uncertainty. In the $\chi^2$ minimisation, the magnitude of $T^+$ is free in the fit as it has to scale to match the experimental rate, while its phase is fixed to zero as an absolute phase carries no physical meaning. This system is already over-constrained with 8 unknown parameters for 10 physical observables. The resulting $p$-value distributions for the \phitwo\ scans are shown in figure~\ref{fig:su3_min:mpv}.

A single solution for \phitwo\ is already preferred at the $1\sigma$ level even without knowledge on the phase difference between $\Bp \to \Kz \aonep$ and $\Bp \to \Koneaz \pip$. However, what these plots really indicate is that if the analysis from ref.~\cite{rhorho_dalseno} had been performed with as little as the full Belle data set, realistically there could already be an only two-fold \phitwo\ solution degeneracy in the $\Bz \to \aone \pimp$ system at $1\sigma$ instead of the 8 solutions it currently has today~\cite{phi2_a1pi3}.

\begin{table}[tbp]
  \centering
  \begin{tabular}{|c|c|c|}
    \hline
    Parameter & Value & Reference\\ \hline
    $\tau_{\Bz}$ & $1.520 \pm 0.004$ ps  & \cite{PDG}\\
    $\tau_{\Bp}$ & $1.638 \pm 0.004$ ps  & \cite{PDG}\\
    $|V_{cd}|$ & $0.224608^{+0.000254}_{-0.000060}$ & \cite{CKMFitter2}\\
    $|V_{cs}|$ & $0.973526^{+0.000050}_{-0.000061}$ & \cite{CKMFitter2}\\
    $\phione$ & $(22.2 \pm 0.7)^\circ$ & \cite{HFAG}\\
    $f_\pi$ & $130.2 \pm 1.7$ MeV & \cite{fpi1,fpi2,fpi3,fpi4,fpi5,fpi6,PDG}\\
    $f_K$ & $155.6 \pm 0.4$ MeV & \cite{fpi1,fpi2,fpi3,fpi4,fpi5,fpi6,fK1,fK2,fK3,PDG}\\
    $f_{a_1}$ & $203 \pm 18$ MeV & \cite{fa1}\\
    $f_{\Konea}$ & $207 \pm 20$ MeV & \cite{fk1,phi2_a1pi3}\\

    ${\cal B}(\aonep \pim)$ & $(16.0 \pm 2.9) \times 10^{-6}$ & \cite{phi2_a1pi2}\\
    ${\cal B}(\aonem \pip)$ & $(6.2 \pm 1.8) \times 10^{-6}$ & \cite{phi2_a1pi2}\\
    $\arg(A(\aonem \pip)/A(\aonep \pim))$ & $(0.6 \vee 179.4 \pm 8.8)^\circ$ & \cite{phi2_a1pi2}\\
    $|\lambda_{CP}^+|$ & $0.98 \pm 0.13$ & \cite{phi2_a1pi2}\\
    $|\lambda_{CP}^-|$ & $0.97 \pm 0.45$ & \cite{phi2_a1pi2}\\
    $\phitwo^+$ & $(97.2 \pm 9.3)^\circ$ & \cite{fk1,phi2_a1pi2}\\
    $\phitwo^-$ & $(107.0 \pm 16.9)^\circ$ & \cite{fk1,phi2_a1pi2}\\

    ${\cal B}(\Koneaz \pip)$ & $(2.0 \vee 16.0 \pm 10.5) \times 10^{-6}$ & \cite{phi2_a1pi3}\\
    ${\cal B}(\Kz \aonep)$ & $(34.9 \pm 6.7) \times 10^{-6}$ & \cite{a1k}\\
    $\arg(A(\Kz \aonep)/A(\Koneaz \pip))$ & $[-180, +180]^\circ$ & ---\\
    \hline
  \end{tabular}
  \caption{Parameters for the minimal SU(3) \phitwo\ constraint, where an unknown central value is indicated by a range. A double reference indicates that the central value comes from theory while the uncertainty derives from experiment.}
  \label{tab:su3_min}
\end{table}

\begin{figure}[tbp]
  \centering
  \includegraphics[height=120pt,width=!]{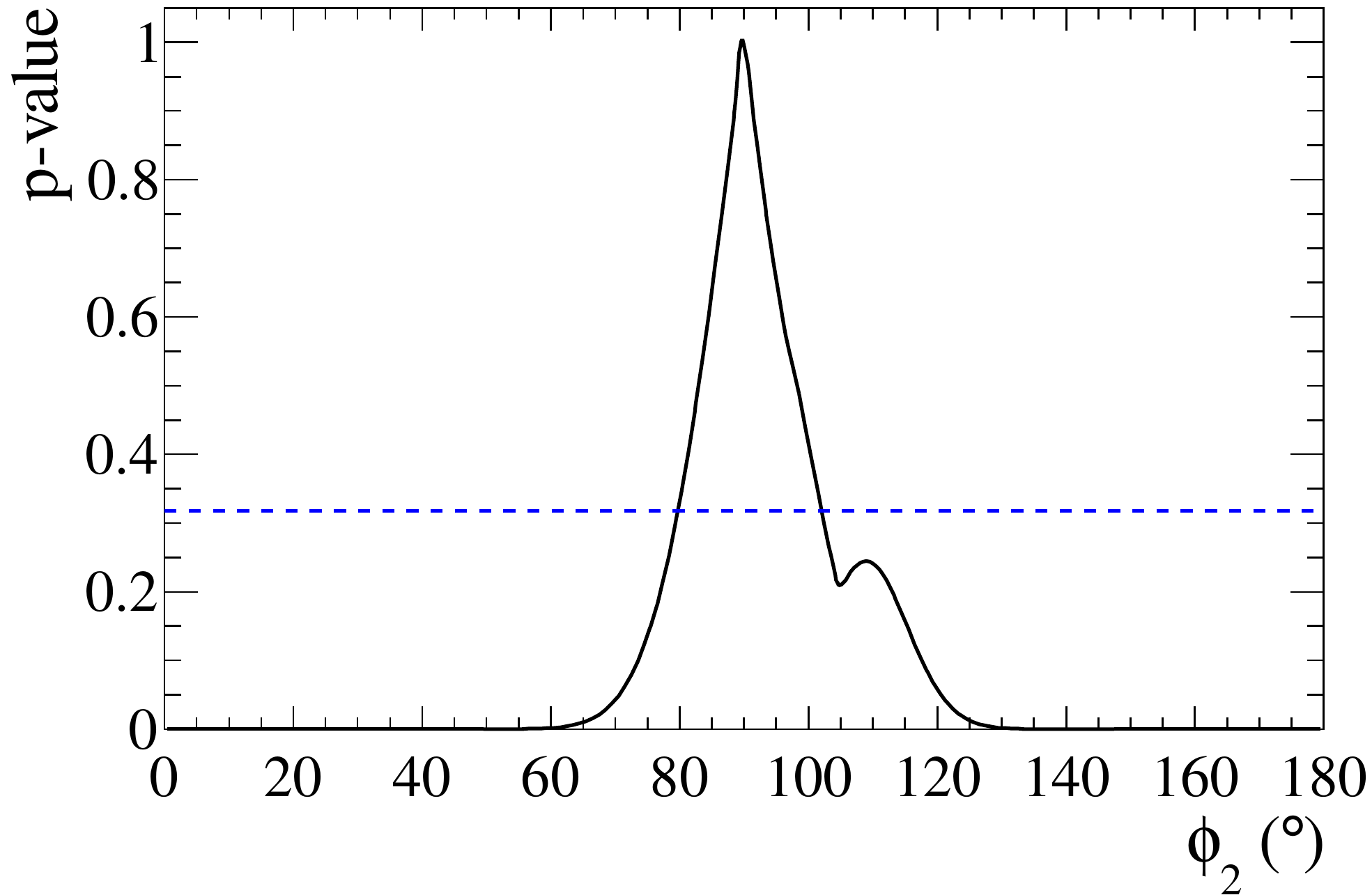}
  \includegraphics[height=120pt,width=!]{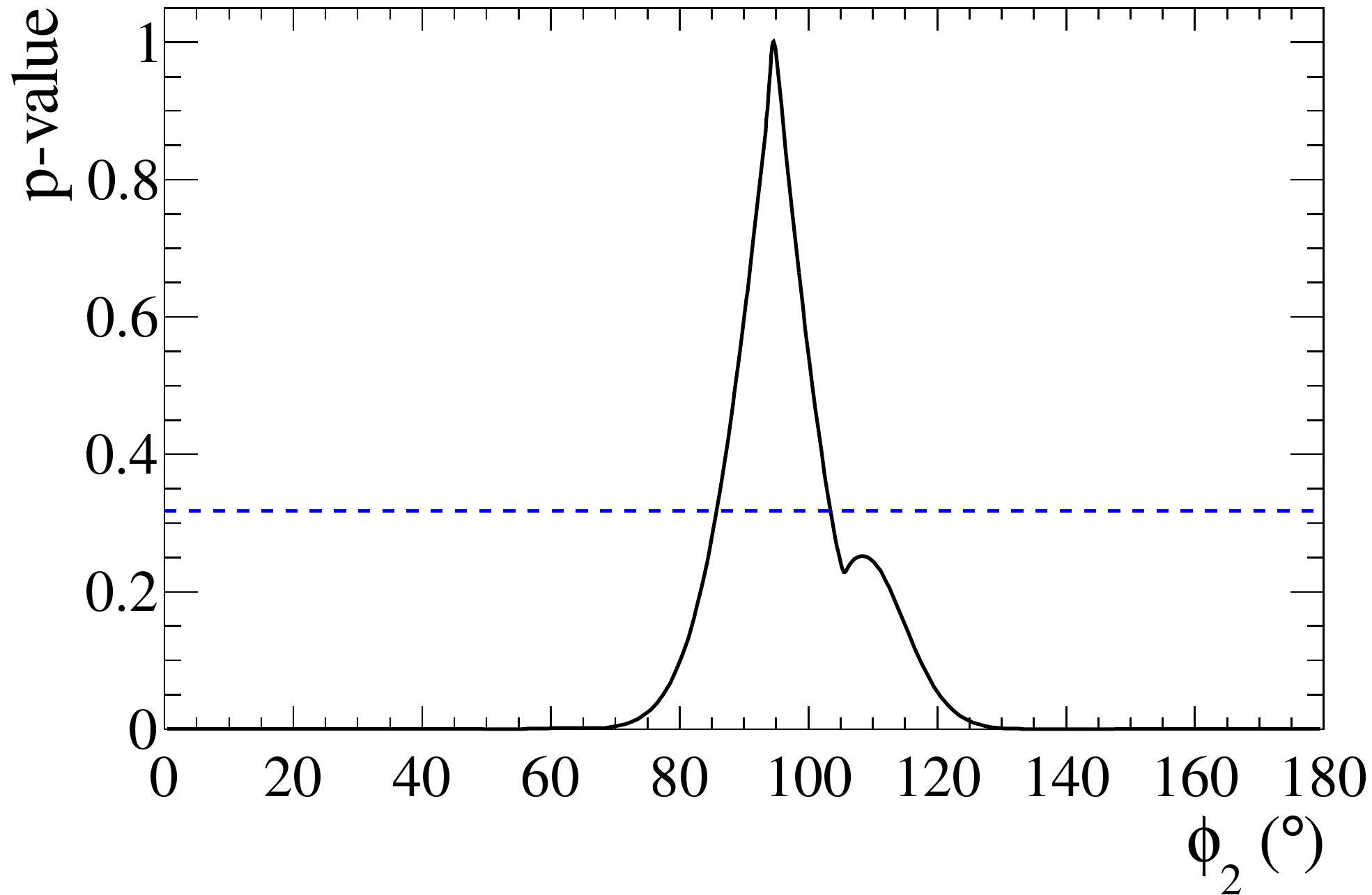}
  \put(-215,105){(a)}
  \put(-28,105){(b)}

  \includegraphics[height=120pt,width=!]{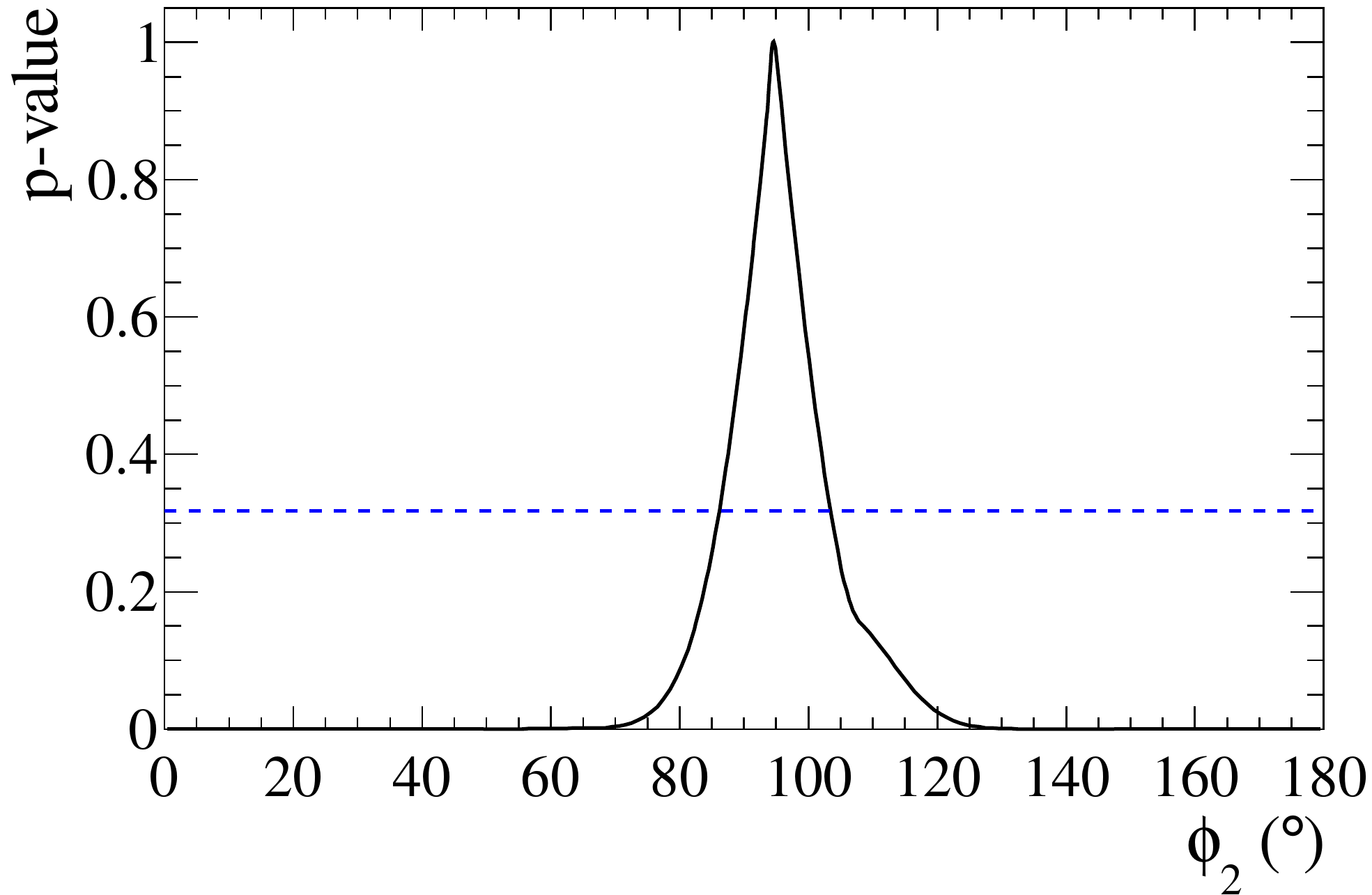}
  \includegraphics[height=120pt,width=!]{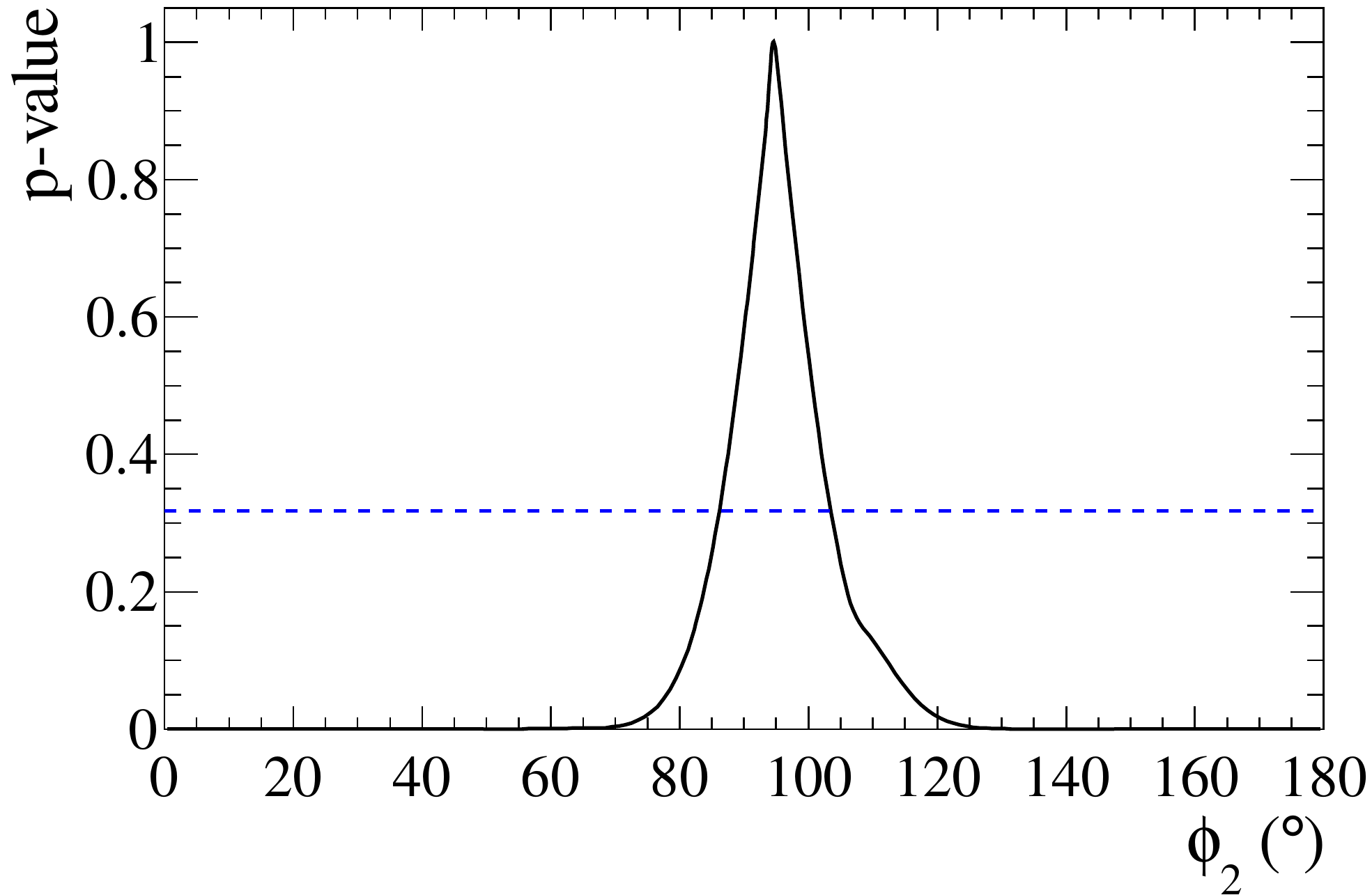}
  \put(-215,105){(c)}
  \put(-28,105){(d)}

  \includegraphics[height=120pt,width=!]{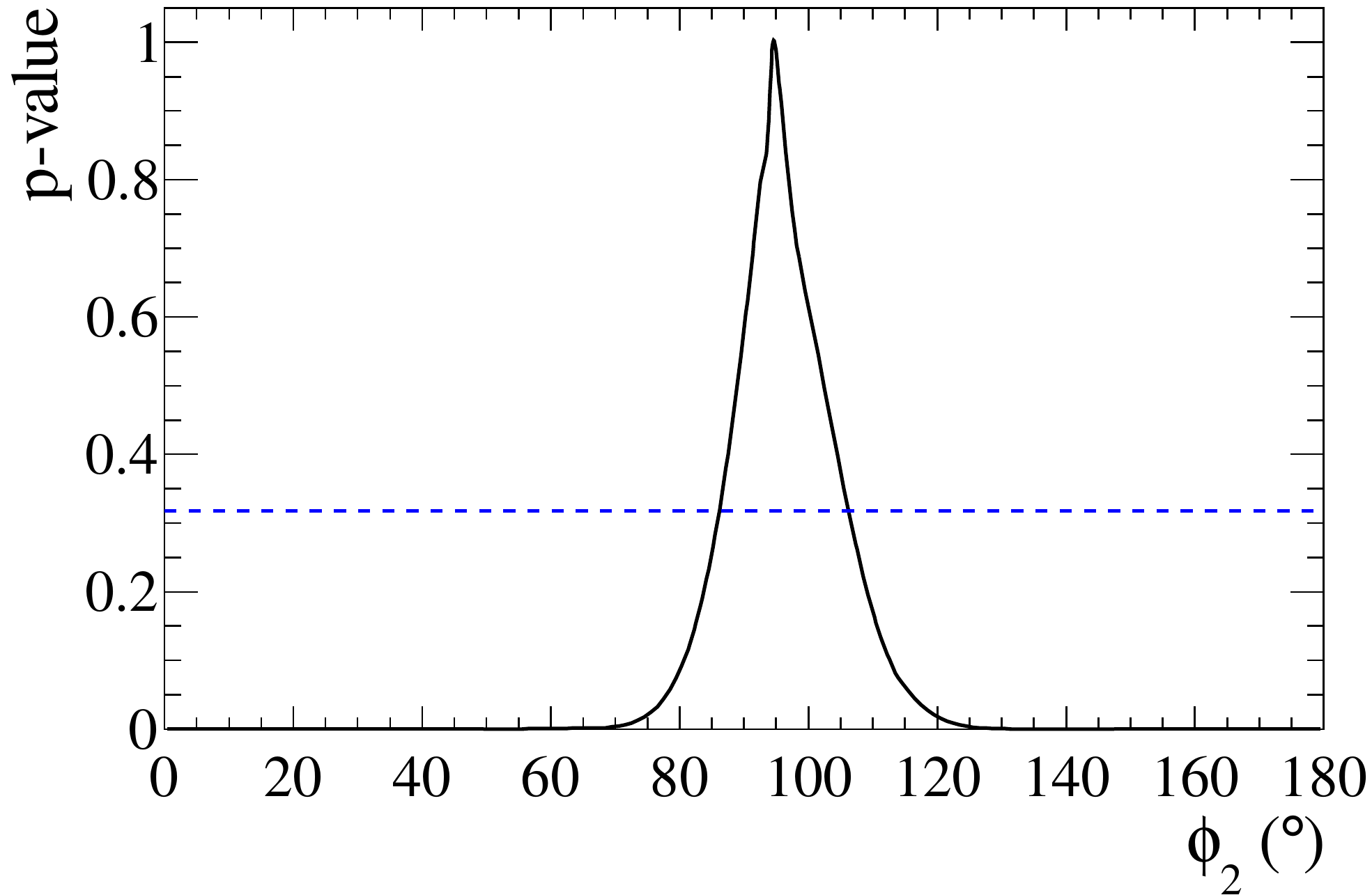}
  \includegraphics[height=120pt,width=!]{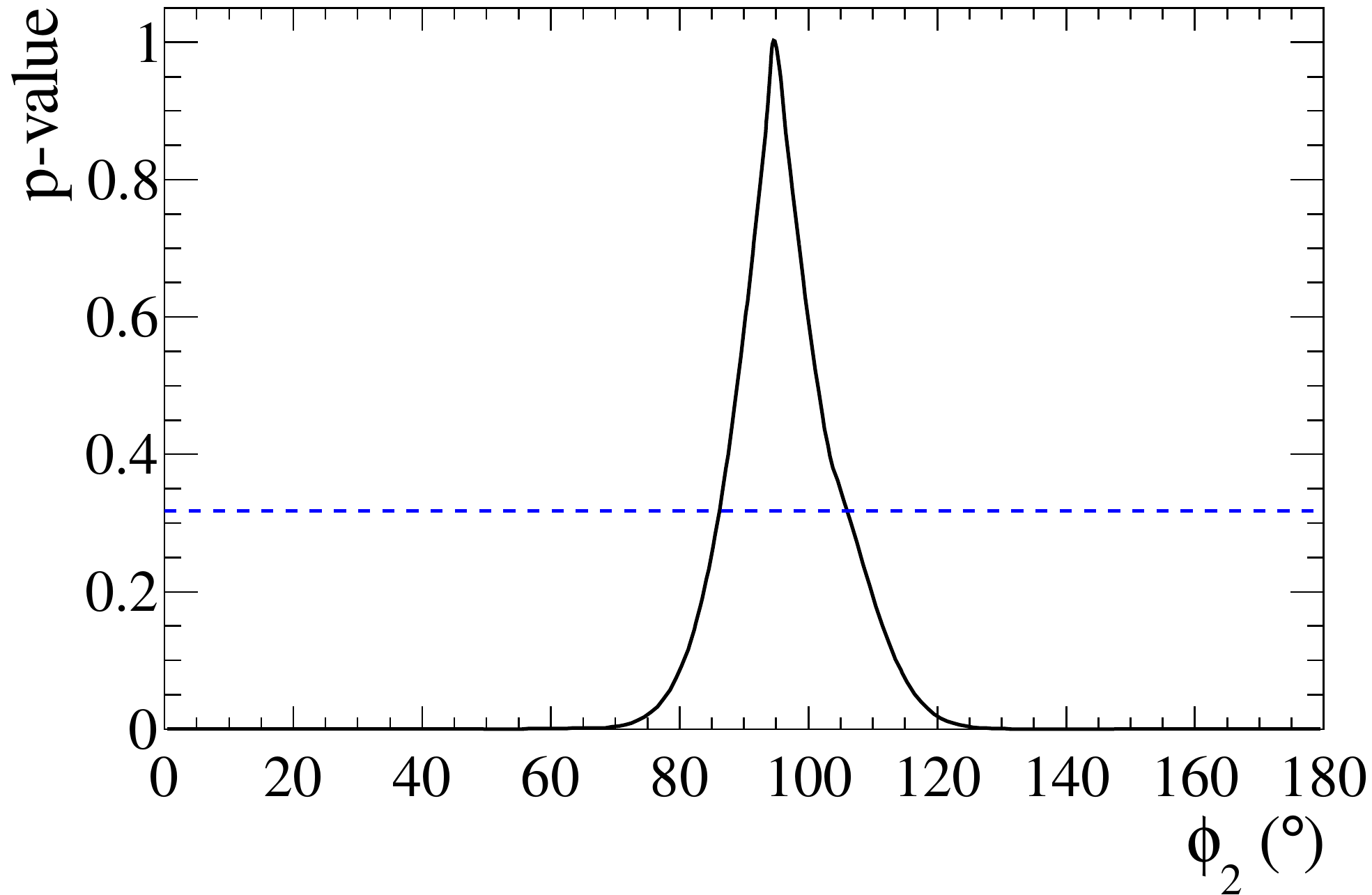}
  \put(-215,105){(e)}
  \put(-28,105){(f)}

  \includegraphics[height=120pt,width=!]{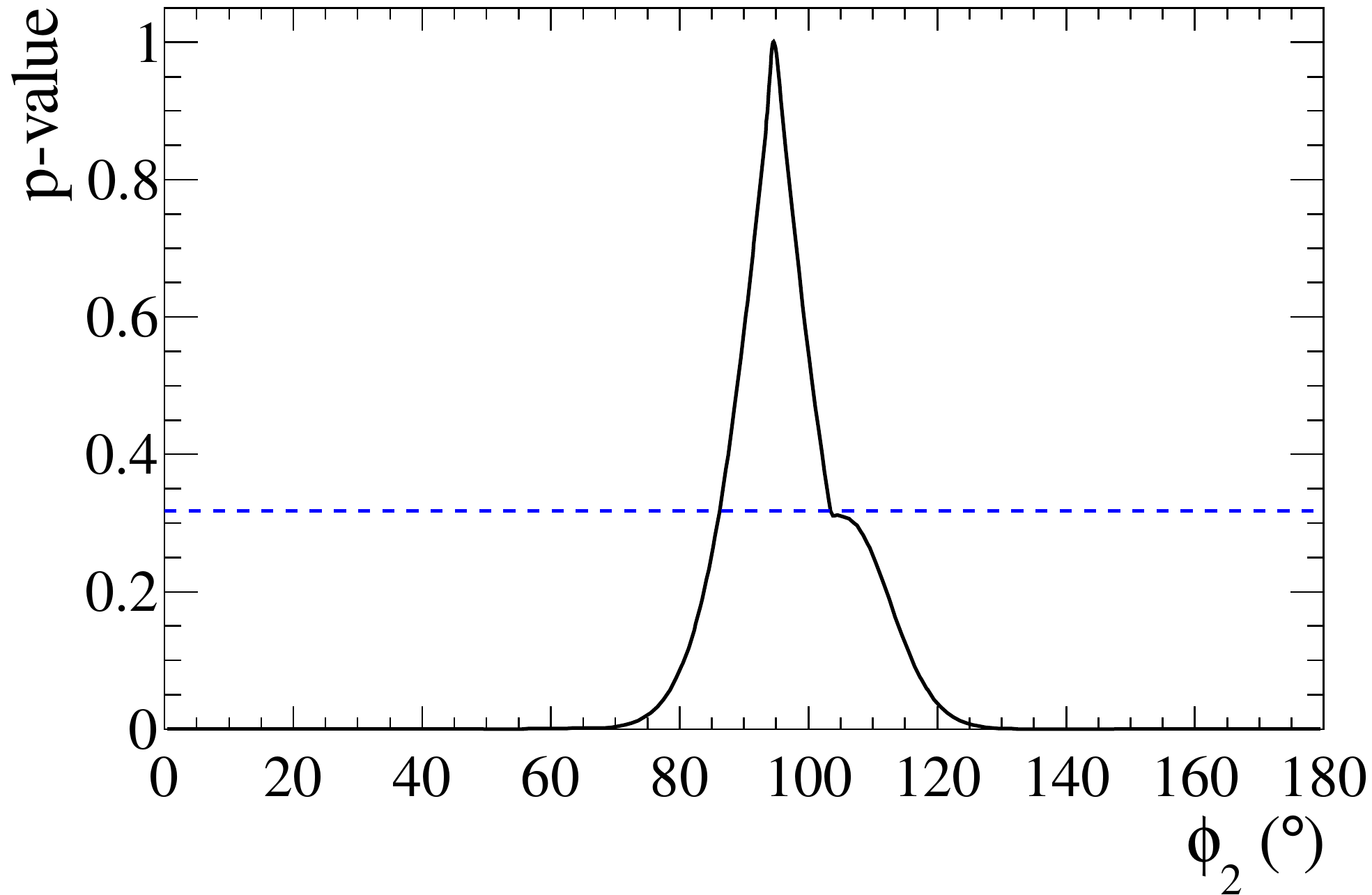}
  \includegraphics[height=120pt,width=!]{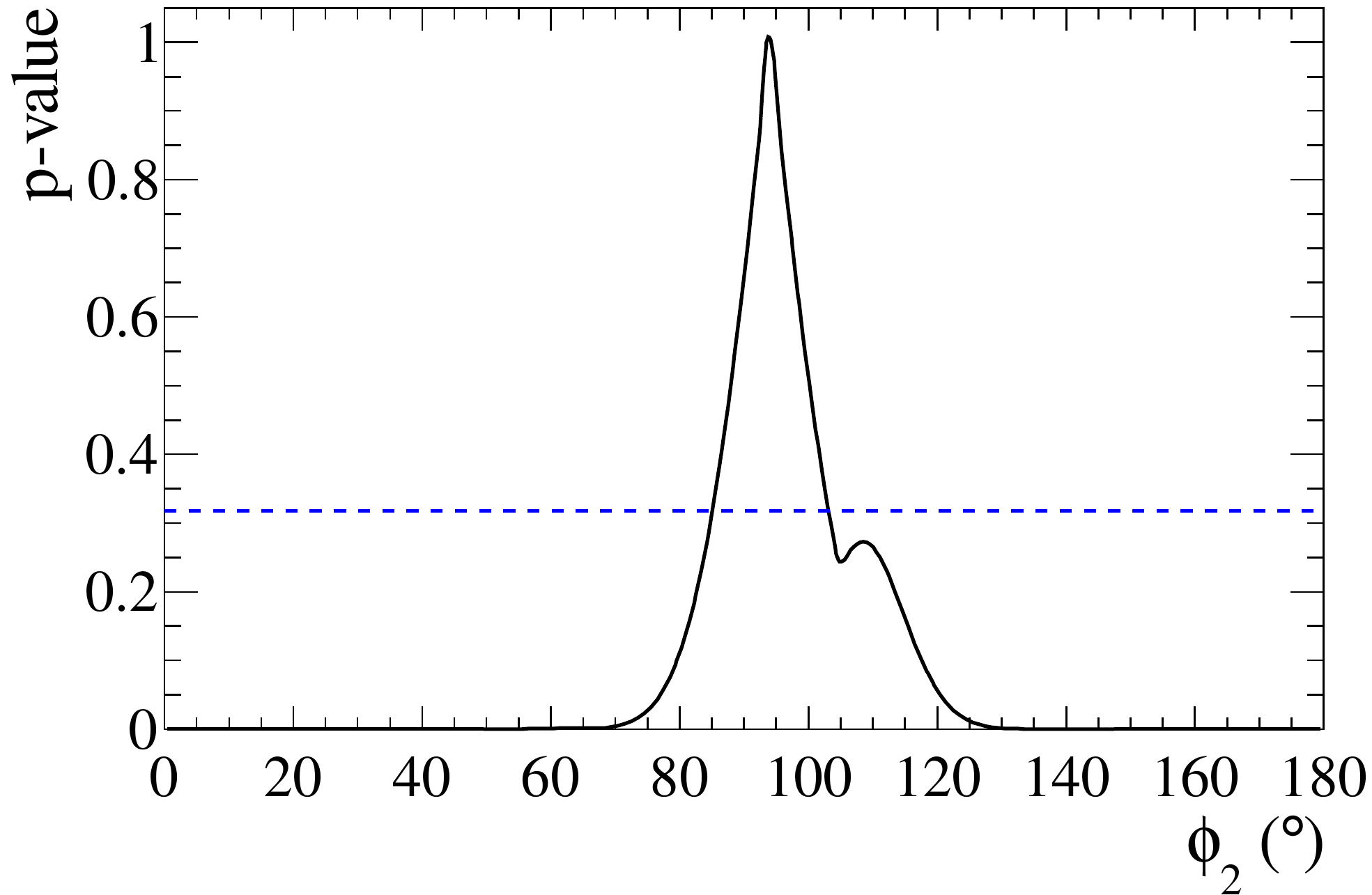}
  \put(-215,105){(g)}
  \put(-28,105){(h)}

  \caption{\label{fig:su3_min:mpv:phase} $p$-value scans of \phitwo\ where the horizontal dashed line shows the $1\sigma$ bound. For the BaBar most probable branching fraction of $\Bp \to \Koneaz \pip$, these scans show the effects of the experimentally determined phase difference between itself and $\Bp \to \Kz \aonep$ when set to (a) $0^\circ$, (b) $45^\circ$, (c) $90^\circ$, (d) $135^\circ$, (e) $180^\circ$, (f) $225^\circ$, (g) $270^\circ$ and (h) $315^\circ$.}
\end{figure}

\begin{figure}[tbp]
  \centering
  \includegraphics[height=120pt,width=!]{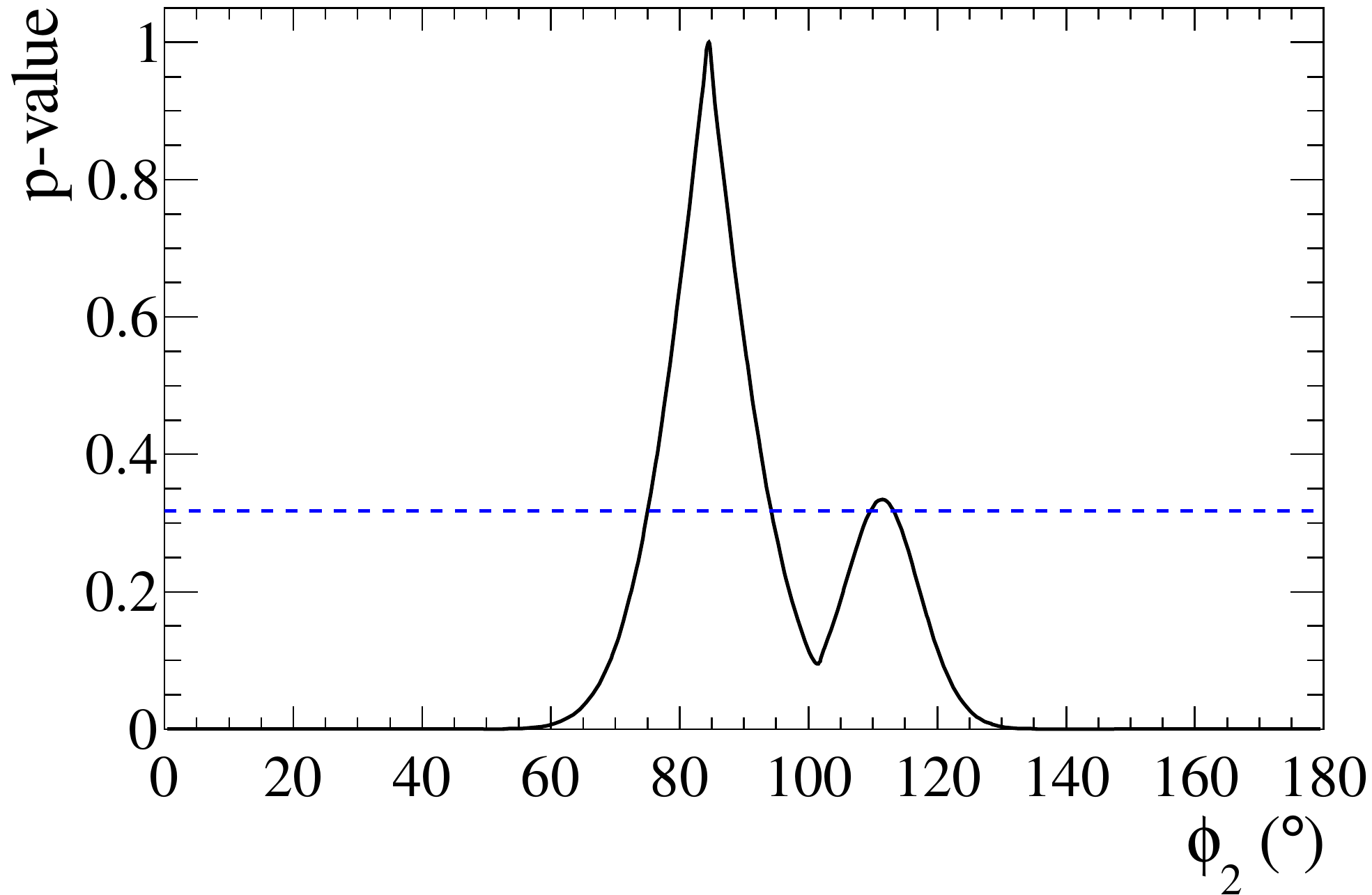}
  \includegraphics[height=120pt,width=!]{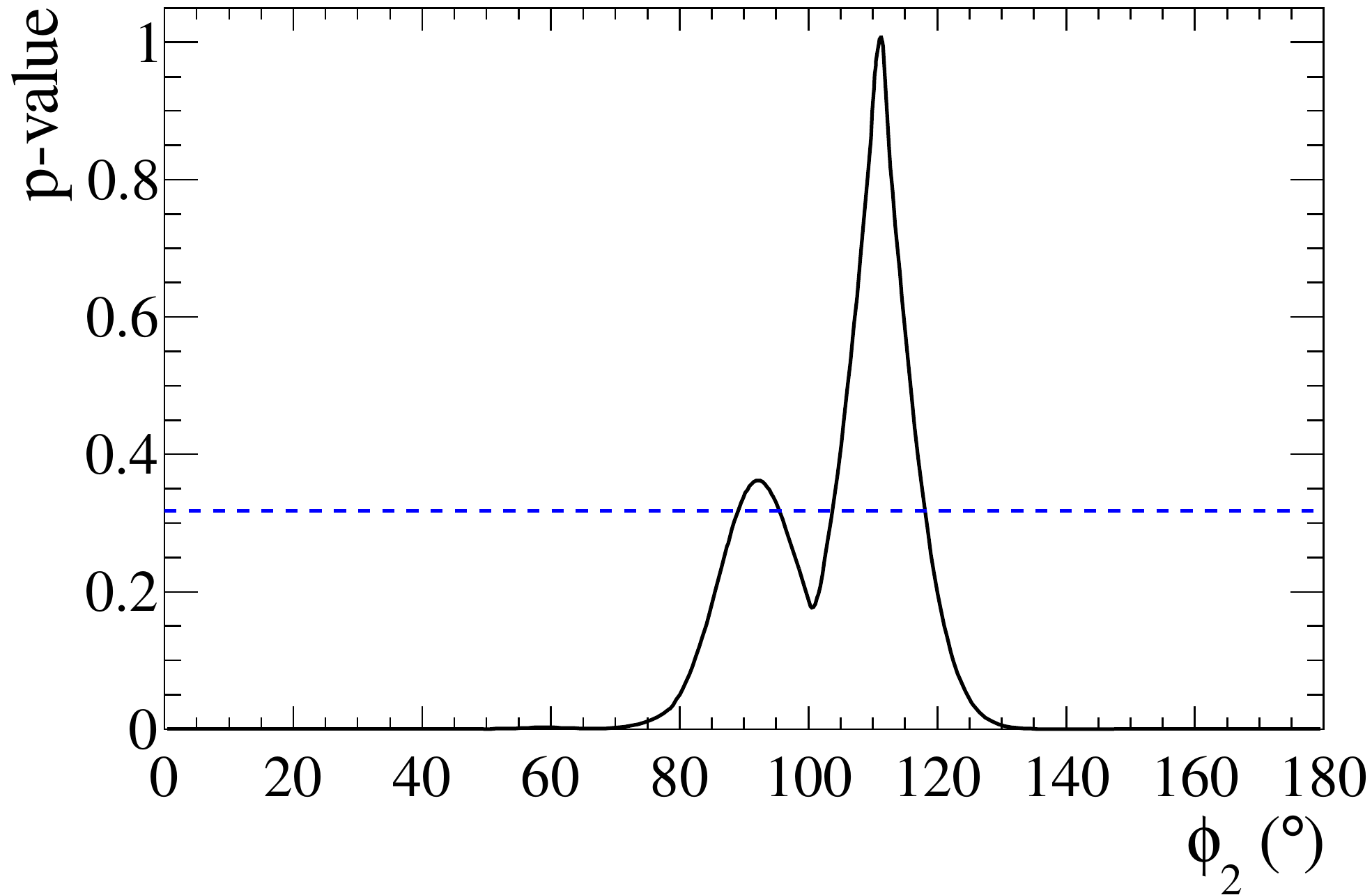}
  \put(-215,105){(a)}
  \put(-28,105){(b)}

  \includegraphics[height=120pt,width=!]{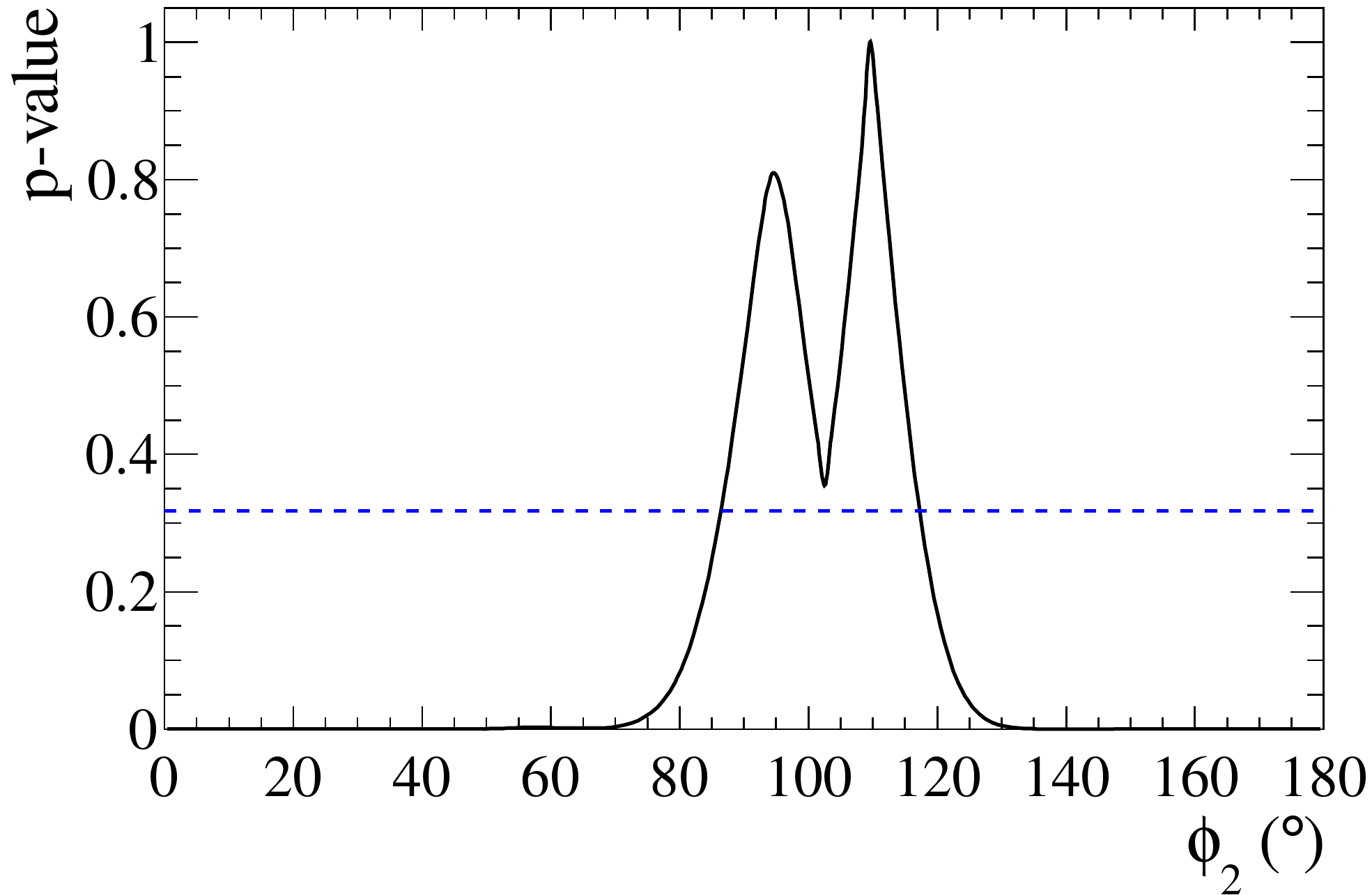}
  \includegraphics[height=120pt,width=!]{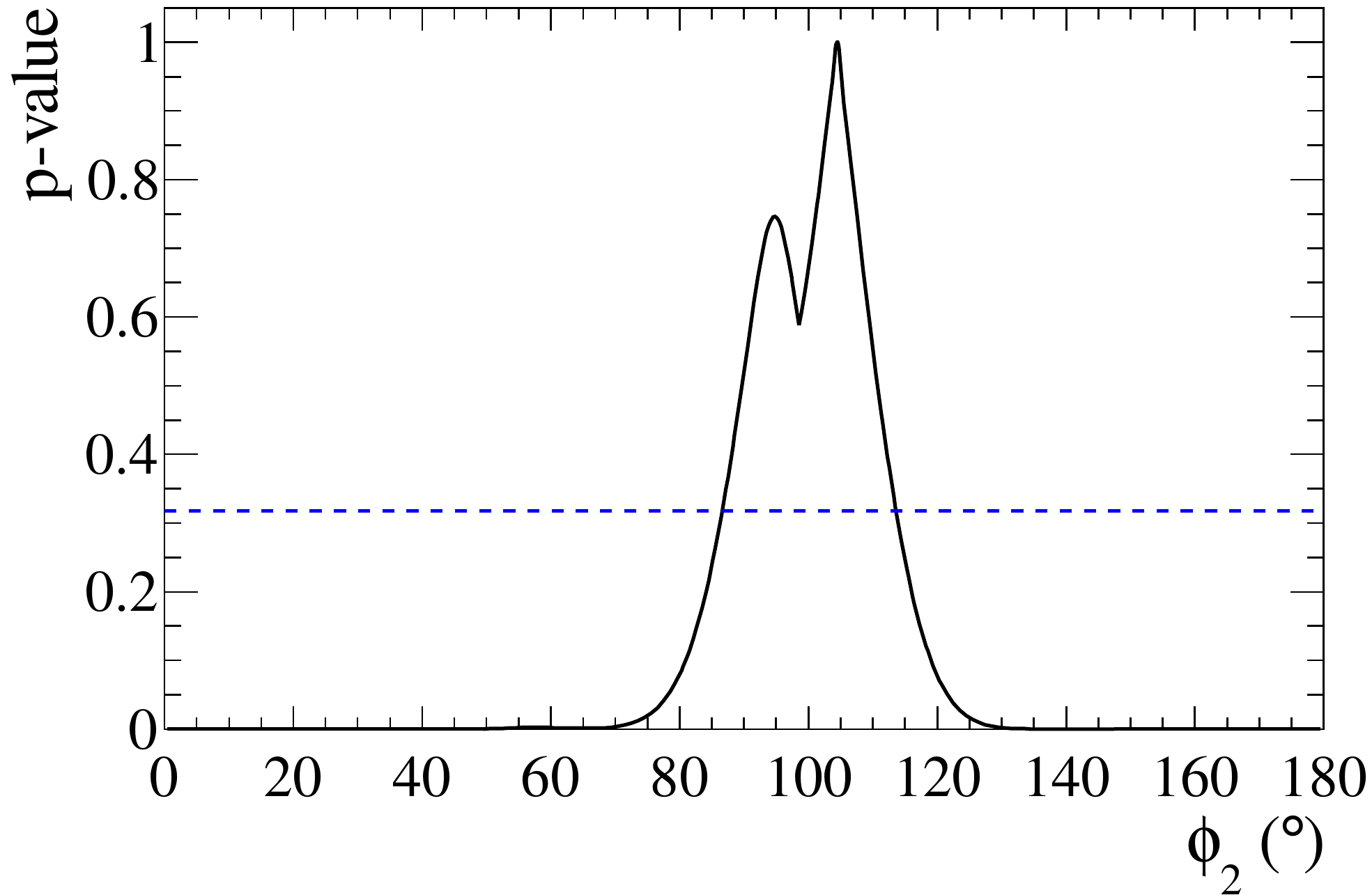}
  \put(-215,105){(c)}
  \put(-28,105){(d)}

  \includegraphics[height=120pt,width=!]{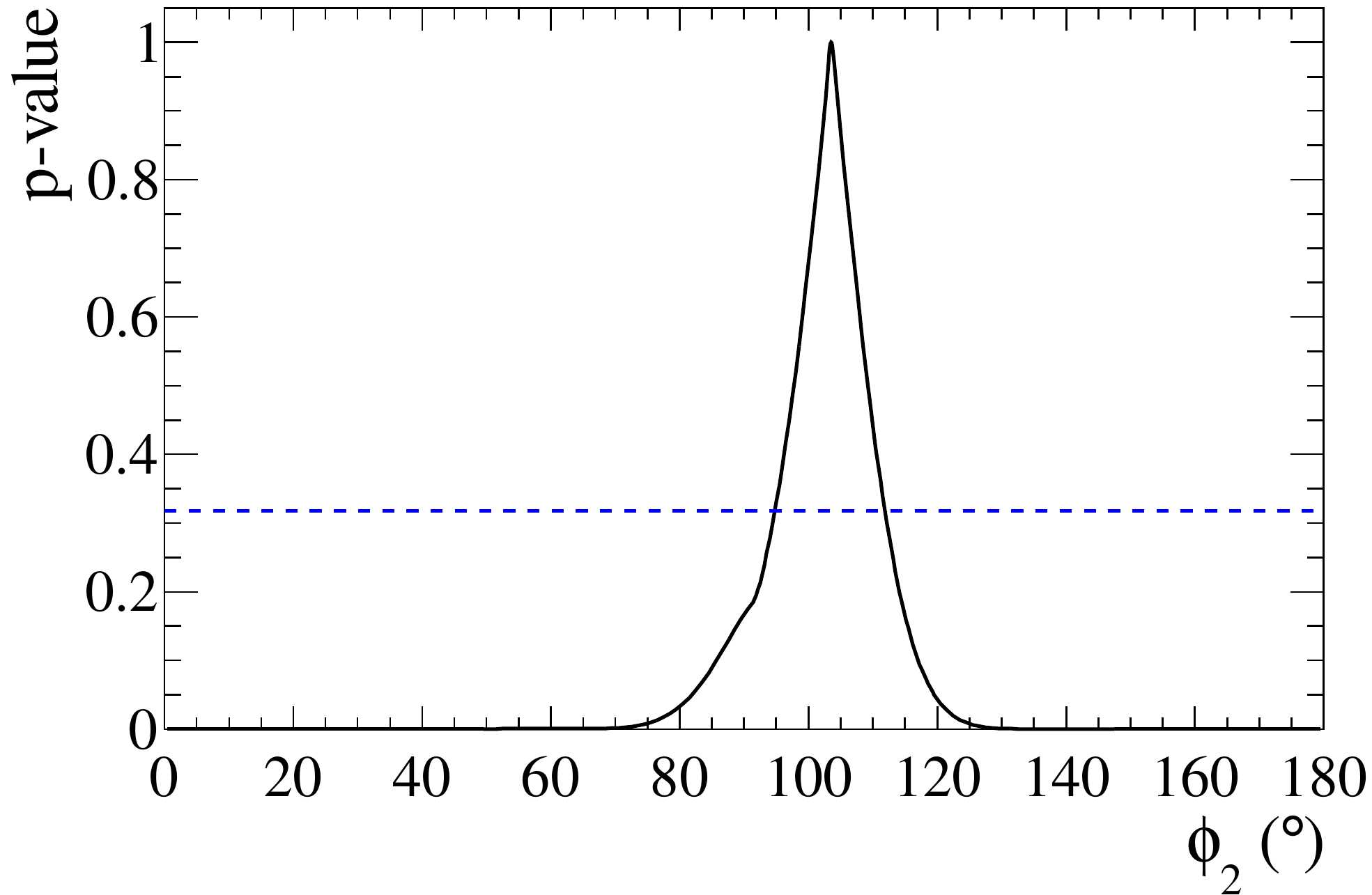}
  \includegraphics[height=120pt,width=!]{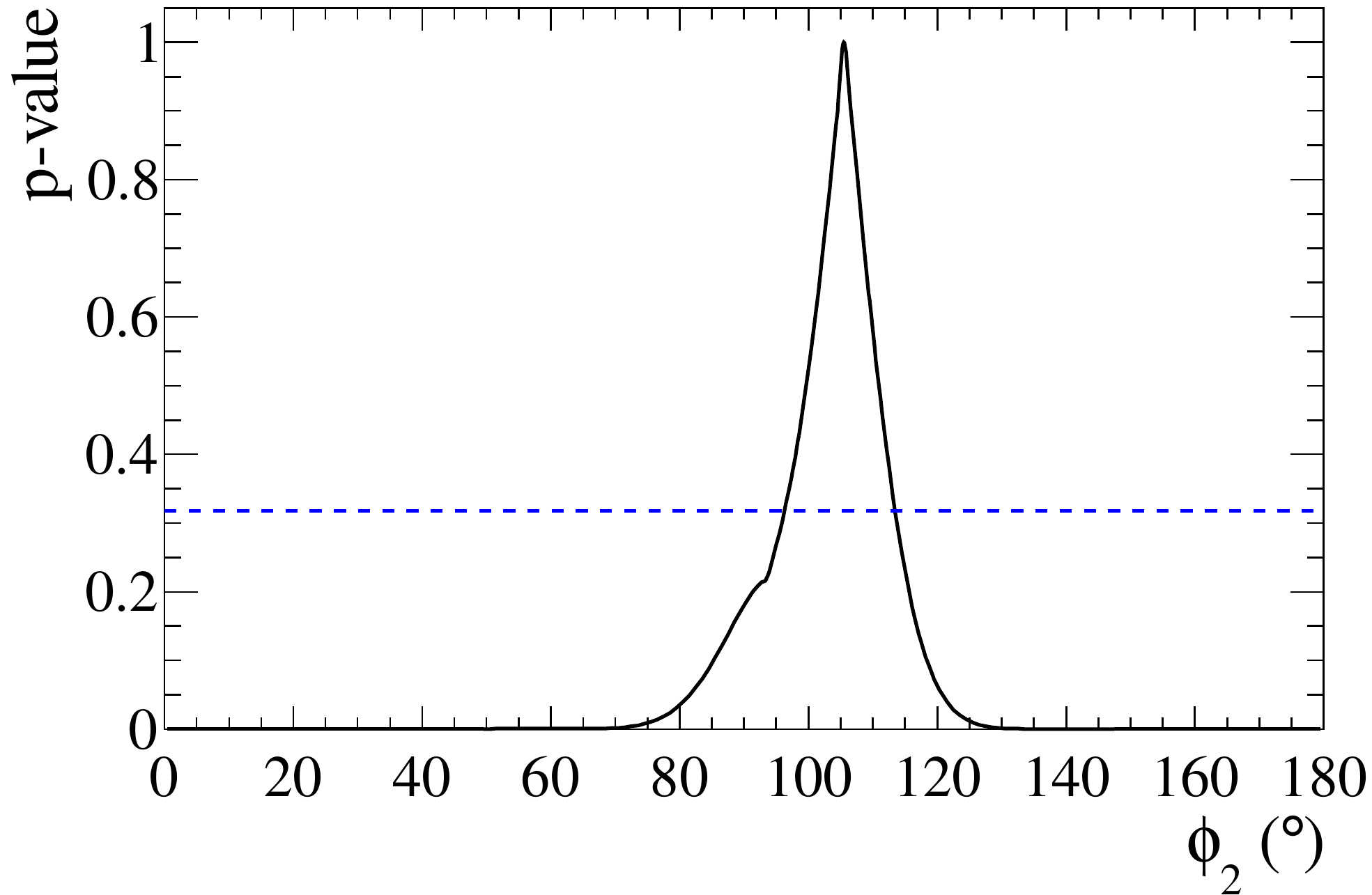}
  \put(-215,105){(e)}
  \put(-28,105){(f)}

  \includegraphics[height=120pt,width=!]{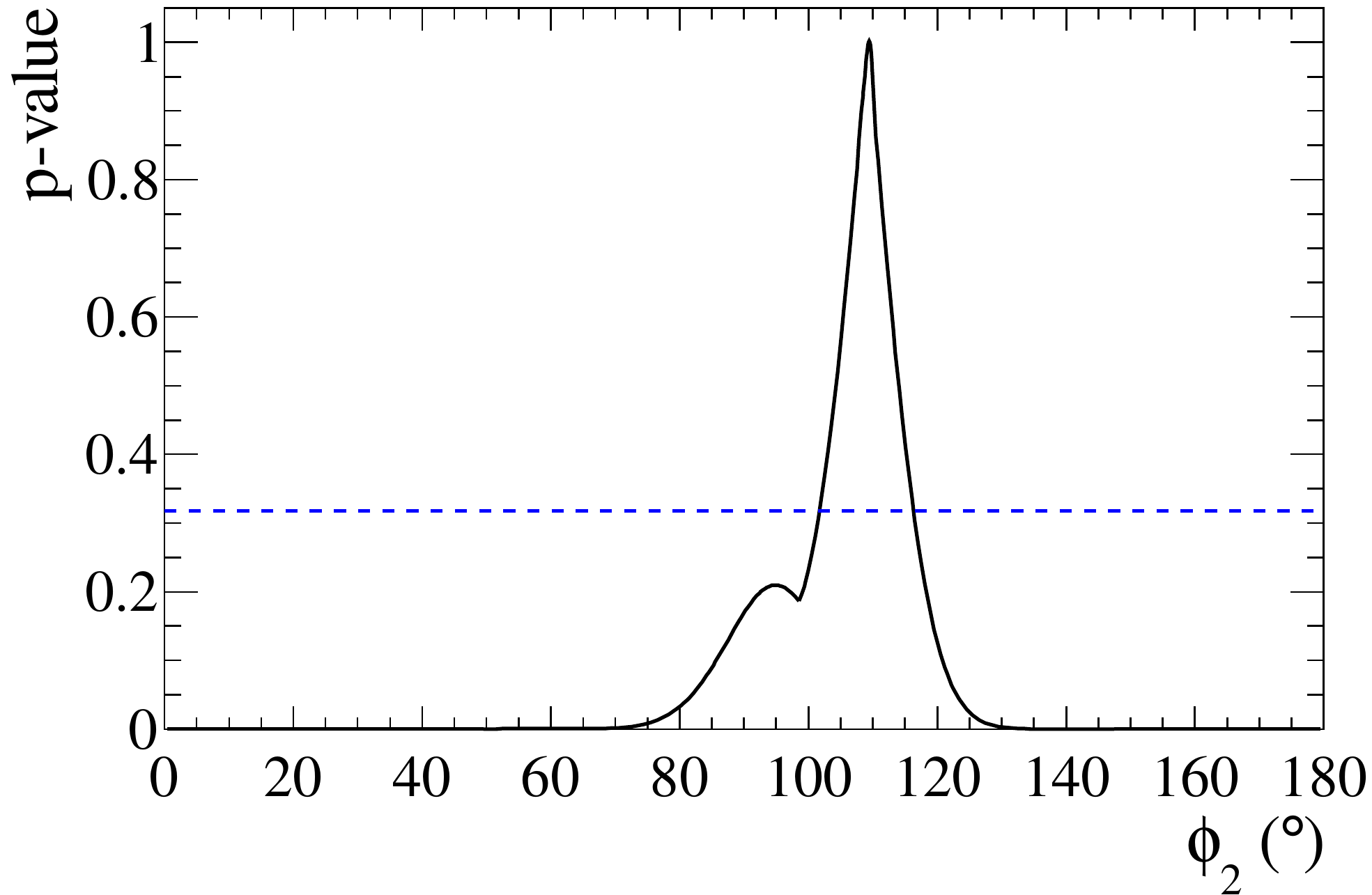}
  \includegraphics[height=120pt,width=!]{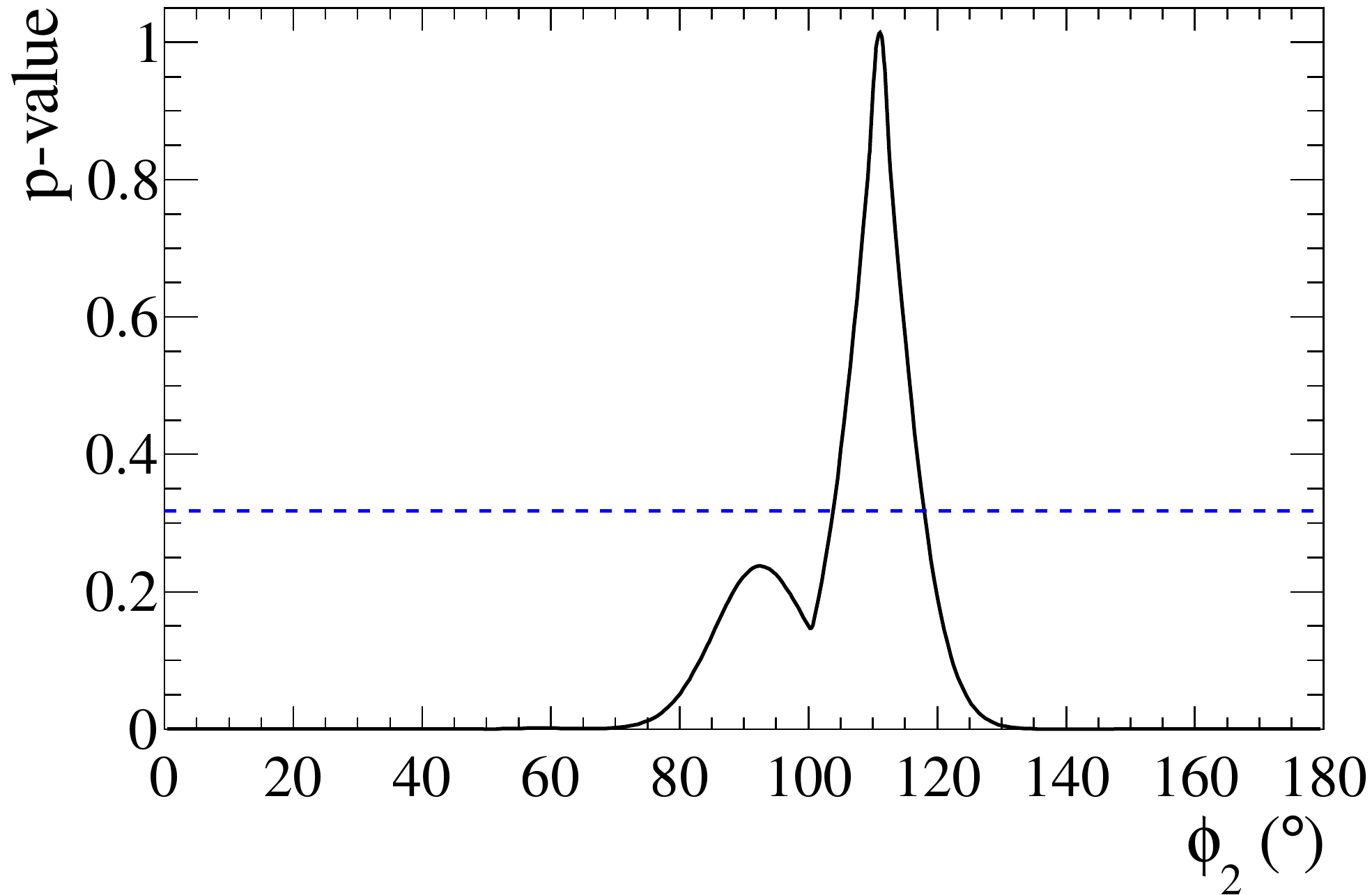}
  \put(-215,105){(g)}
  \put(-28,105){(h)}

  \caption{\label{fig:su3_min:mean:phase} $p$-value scans of \phitwo\ where the horizontal dashed line shows the $1\sigma$ bound. For the BaBar mean branching fraction of $\Bp \to \Koneaz \pip$, these scans show the effects of the experimentally determined phase difference between itself and $\Bp \to \Kz \aonep$ when set to (a) $0^\circ$, (b) $45^\circ$, (c) $90^\circ$, (d) $135^\circ$, (e) $180^\circ$, (f) $225^\circ$, (g) $270^\circ$ and (h) $315^\circ$.}
\end{figure}

Next, I test the impact of the phase difference between $\Bp \to \Kz \aonep$ and $\Bp \to \Koneaz \pip$ for the most probable (figure~\ref{fig:su3_min:mpv:phase}) and mean values (figure~\ref{fig:su3_min:mean:phase}) of the $\Bp \to \Koneaz \pip$ branching fraction. The experimentally determined phase difference is set in steps of $45^\circ$ over the entire range with a serviceable uncertainty of $10^\circ$ for each \phitwo\ scan. A notable improvement can be seen with respect to the scenario in which the phase difference is unknown in the constraint (figure~\ref{fig:su3_min:mpv}), particularly when the phase difference is around $180^\circ$. Despite being over-constrained, the best $\chi^2$ for each phase configuration is less than unity indicating good statistical stability of the model.

There is an interplay between the strong phase differences, $\arg(A(\aonem \pip)/A(\aonep \pim))$ and $\arg(A(\Kz \aonep)/A(\Koneaz \pip))$. If the phase difference between $\Bz \to \aonem \pip$ and $\Bz \to \aonep \pim$ is set to its second solution, the phase difference between $\Bp \to \Kz \aonep$ and $\Bp \to \Koneaz \pip$ would have to shift by the same amount in order to have the same overall impact. For example, the second solution for $\arg(A(\aonem \pip)/A(\aonep \pim))$ combined with no phase difference in $\arg(A(\Kz \aonep)/A(\Koneaz \pip))$ would produce a \phitwo\ scan looking more like figures~\ref{fig:su3_min:mpv:phase}e~and~\ref{fig:su3_min:mean:phase}e.

\subsection{Wider SU(3) analysis}

Despite the system already being over-constrained, it may be desirable to include the neutral $\Bz \to \Kp \pim \pip \pim$ decays, though the axial vector contributions contain both tree and penguin processes. Presumably, this would be of greater interest if the minimal SU(3) analysis failed to give a unique solution for \phitwo. Within SU(3) flavour symmetry, they are related to $\Bz \to \aone \pimp$ through
\begin{eqnarray}
  A(\Bz \to \Koneap \pim) &=& \frac{f_{K_1}}{f_{a_1}} \biggl(e^{+i\phithree} \bar \lambda T^+ -\frac{1}{\bar \lambda} P^+ \biggr), \nonumber \\
  A(\Bz \to \Kp \aonem) &=& \frac{f_{K}}{f_{\pi}} \biggl(e^{+i\phithree} \bar \lambda T^- -\frac{1}{\bar \lambda} P^- \biggr),
\end{eqnarray}
again assuming SU(3) factorisation.

For completeness, the $CP$-conjugate amplitudes are simply
\begin{eqnarray}
  \bar A(\Bz \to \Koneap \pim) &=& \frac{f_{K_1}}{f_{a_1}} \biggl(e^{-i\phithree} \bar \lambda T^+ -\frac{1}{\bar \lambda} P^+ \biggr), \nonumber \\
  \bar A(\Bz \to \Kp \aonem) &=& \frac{f_{K}}{f_{\pi}} \biggl(e^{-i\phithree} \bar \lambda T^- -\frac{1}{\bar \lambda} P^- \biggr).
\end{eqnarray}

In this system, $CP$ violation in the decay is possible and provides additional constraints for both $\Bz \to \Koneap \pim$ and $\Bz \to \Kp \aonem$ through
\begin{equation}
  \Acp = \frac{|\bar A|^2 - |A|^2}{|\bar A|^2 + |A|^2}.
\end{equation}

\begin{figure}[tbp]
  \centering
  \includegraphics[height=120pt,width=!]{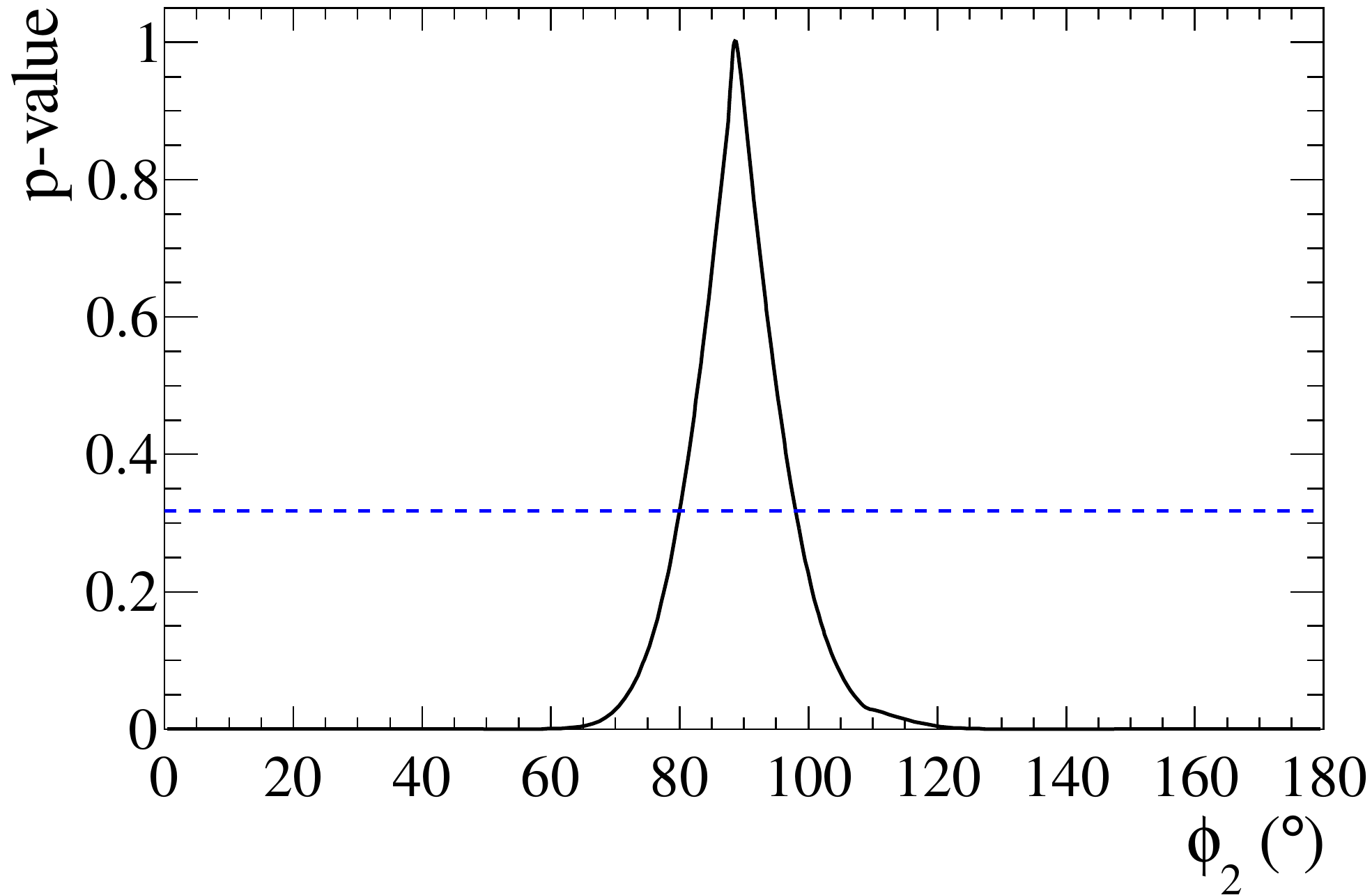}
  \includegraphics[height=120pt,width=!]{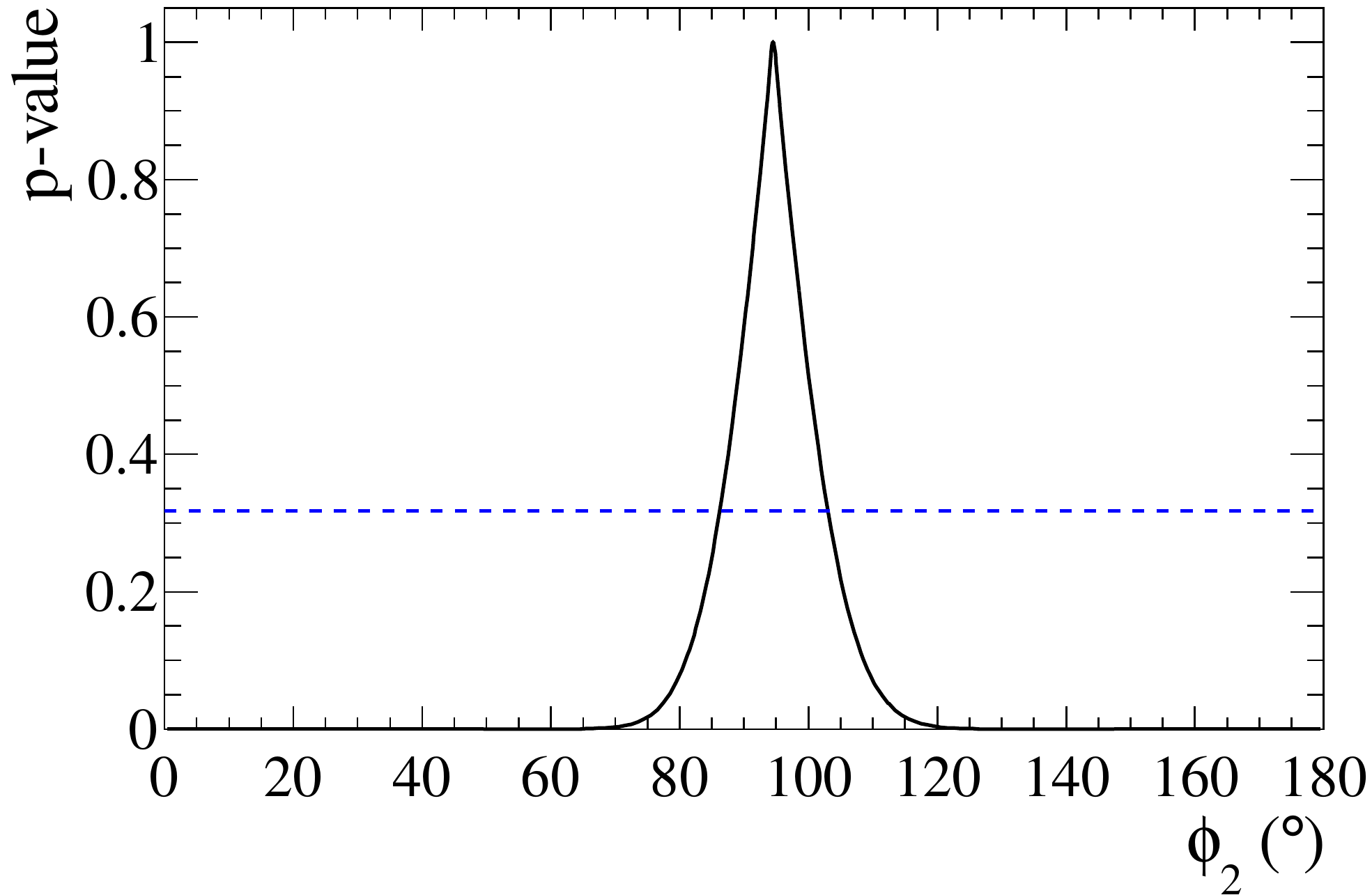}
  \put(-215,105){(a)}
  \put(-28,105){(b)}

  \includegraphics[height=120pt,width=!]{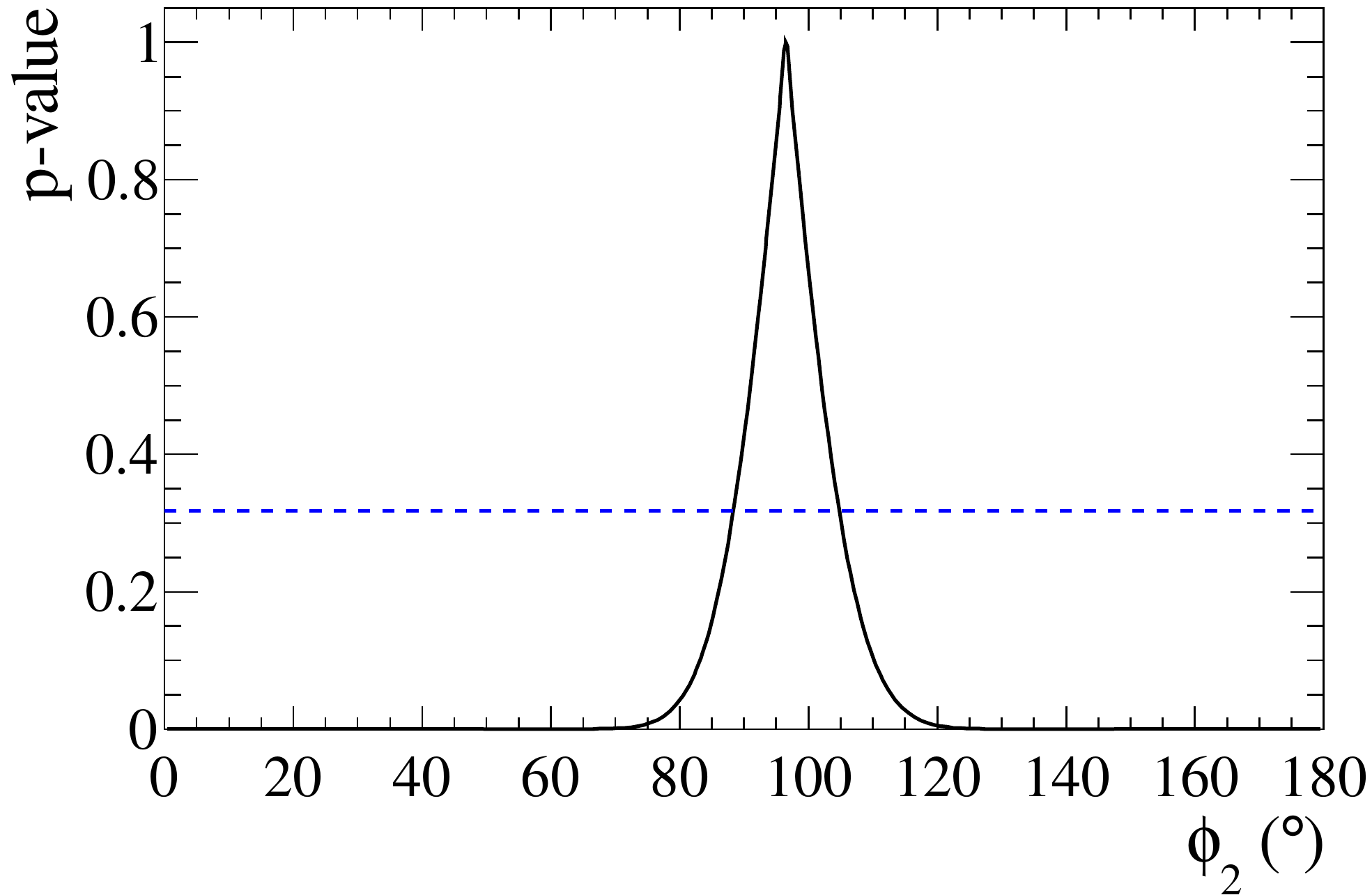}
  \includegraphics[height=120pt,width=!]{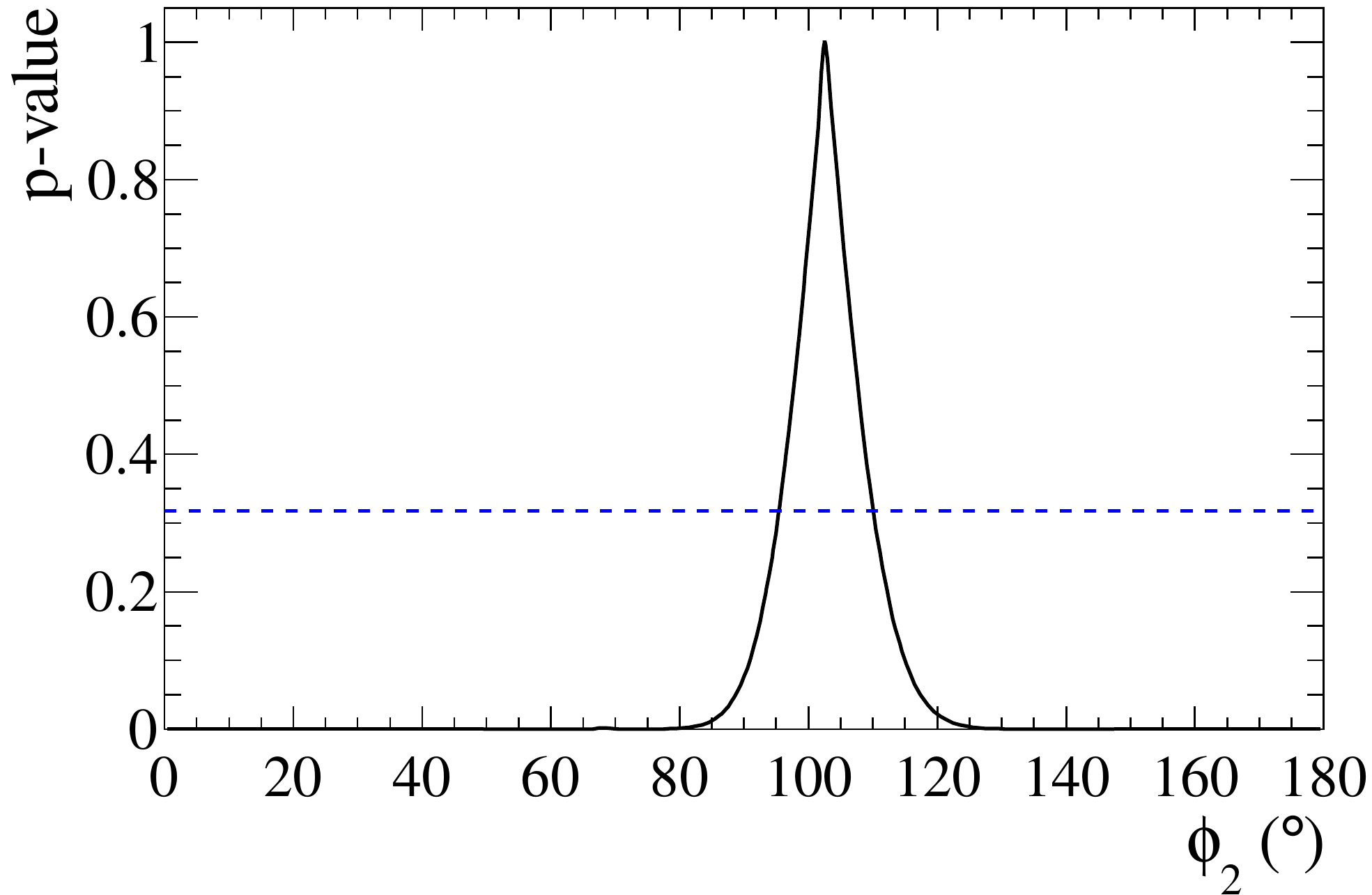}
  \put(-215,105){(c)}
  \put(-28,105){(d)}

  \includegraphics[height=120pt,width=!]{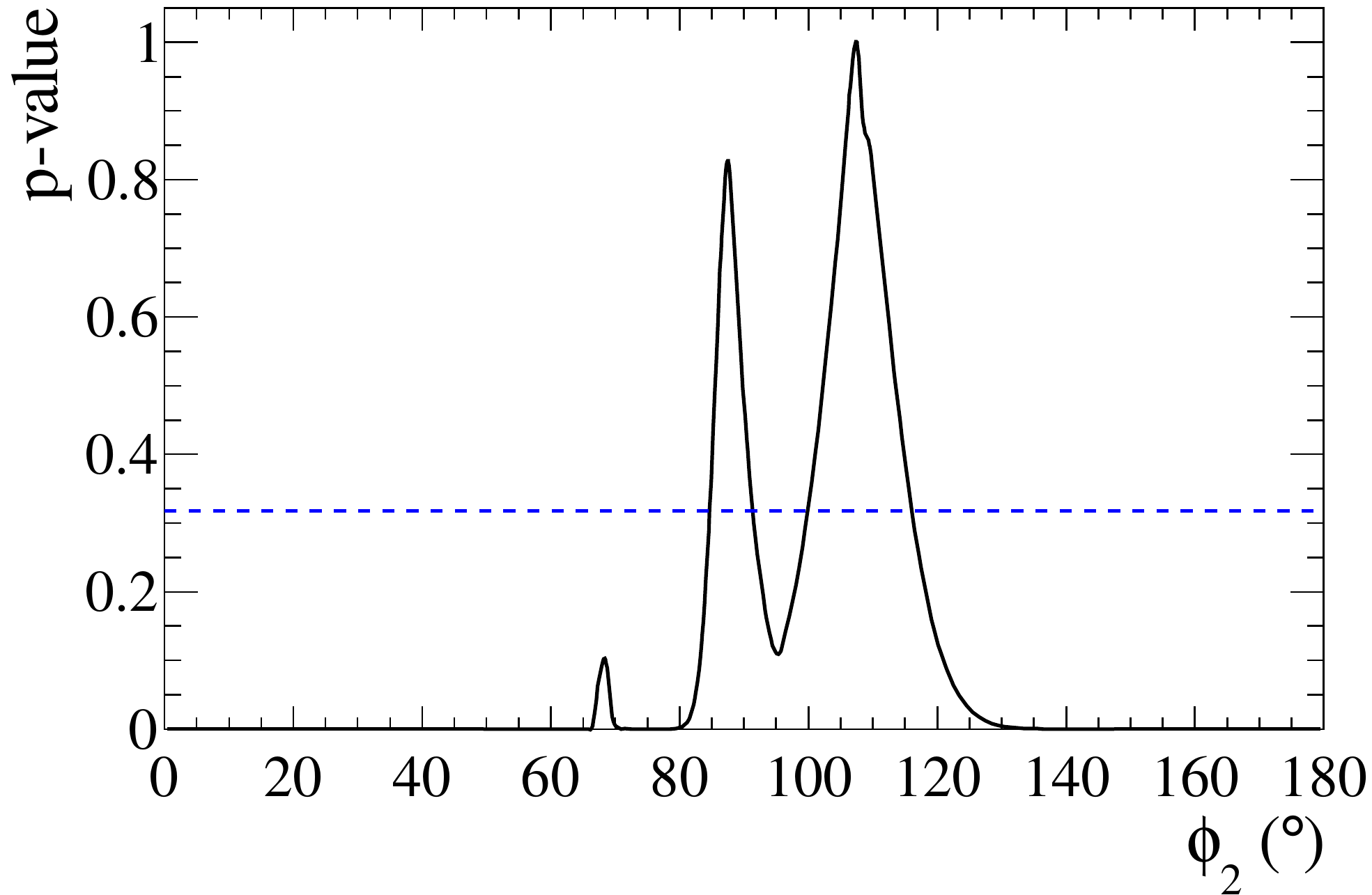}
  \includegraphics[height=120pt,width=!]{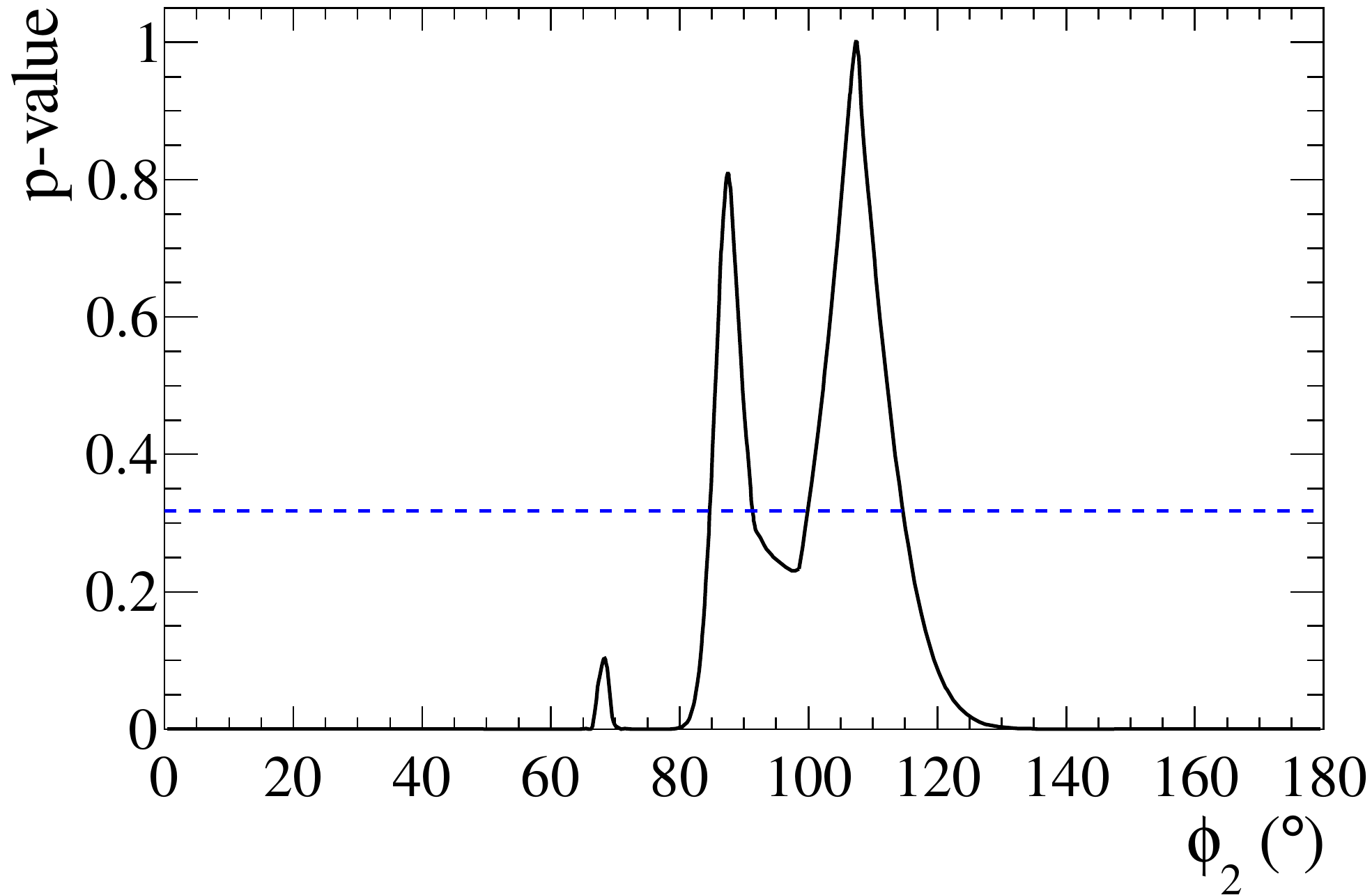}
  \put(-215,105){(e)}
  \put(-28,105){(f)}

  \includegraphics[height=120pt,width=!]{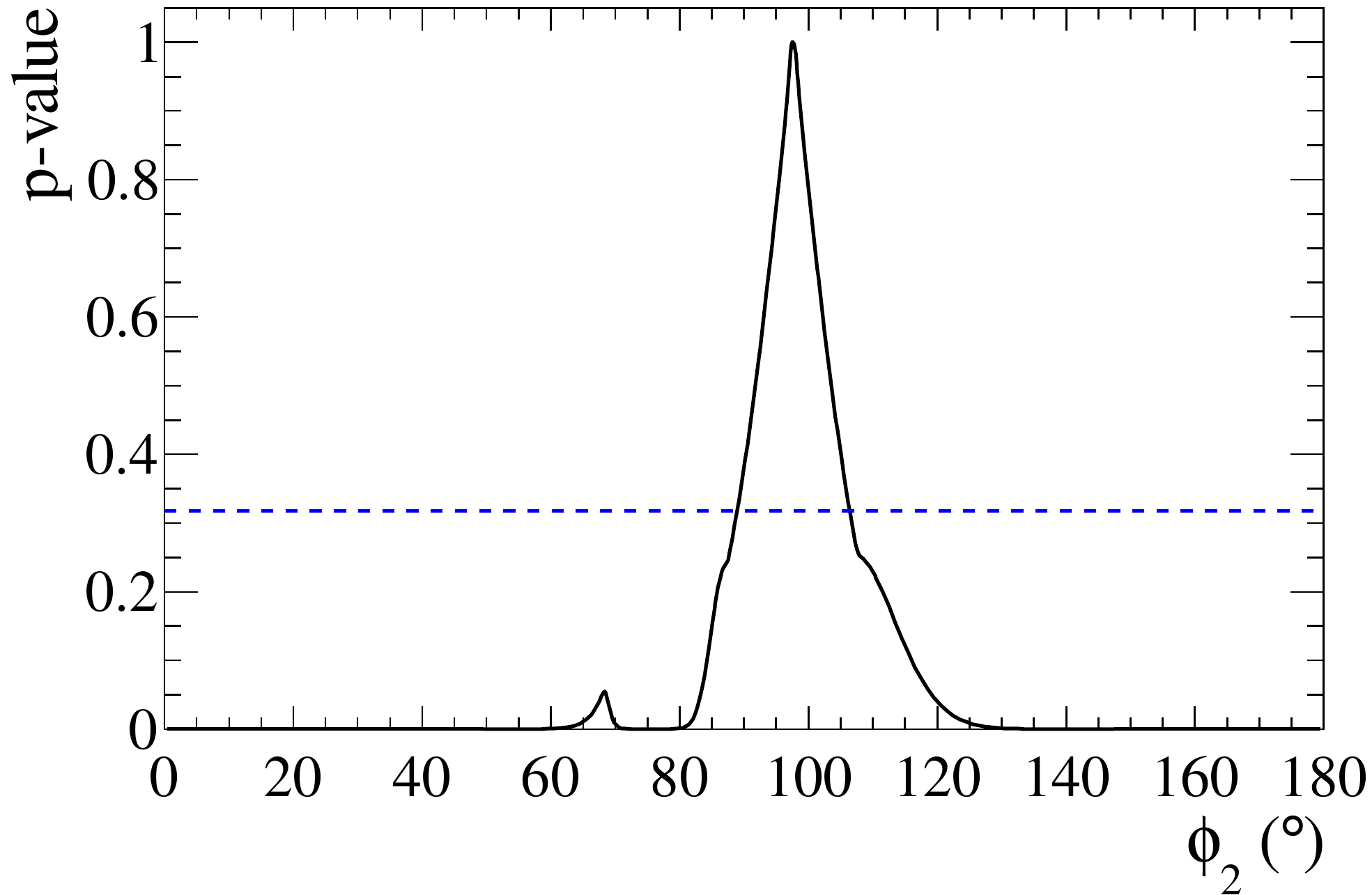}
  \includegraphics[height=120pt,width=!]{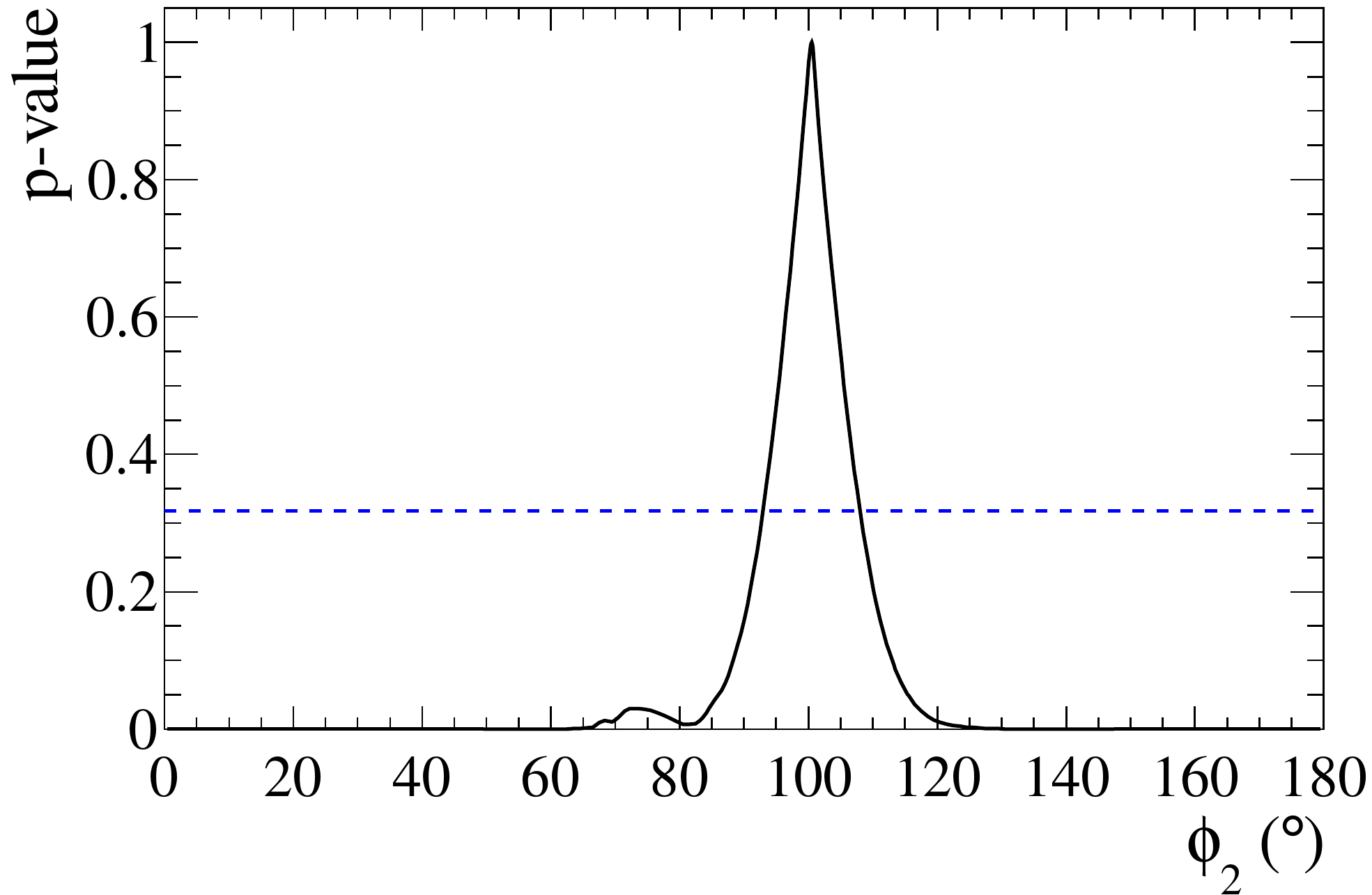}
  \put(-215,105){(g)}
  \put(-28,105){(h)}

  \caption{\label{fig:su3_full} $p$-value scans of \phitwo\ where the horizontal dashed line shows the $1\sigma$ bound. For the BaBar most probable branching fraction of $\Bz \to \Koneap \pim$, these scans show the effects of the experimentally determined phase difference between itself and $\Bz \to \Kp \aonem$ when set to (a) $0^\circ$, (b) $45^\circ$, (c) $90^\circ$, (d) $135^\circ$, (e) $180^\circ$, (f) $225^\circ$, (g) $270^\circ$ and (h) $315^\circ$.}
\end{figure}

I now perform a \phitwo\ scan with the additional parameters given in table~\ref{tab:su3_full} contributing to the $\chi^2$. In this test, the phase difference between $\Bz \to \Kp \aonem$ and $\Bz \to \Koneap \pim$ is varied for the worst-case scenario of the phase difference between $\Bp \to \Kz \aonep$ and $\Bp \to \Koneaz \pip$ which is when it is set to $0^\circ$. Similarly to the minimal SU(3) tests, the phase difference is increased in steps of $45^\circ$ over the entire range with an uncertainty of $10^\circ$. For brevity, only the most probable values for the $B \to \Konea \pi$ branching fractions are considered in the remainder of this section with the ${\cal B}(\Koneap \pim)$ branching fraction set to $14 \times 10^{-6}$ and its \Acp\ excluded from the $\chi^2$. The scans can be found in figure~\ref{fig:su3_full}.

\begin{table}[tbp]
  \centering
  \begin{tabular}{|c|c|c|}
    \hline
    Parameter & Value & Reference\\ \hline
    ${\cal B}(\Koneap \pim)$ & $(14.0 \vee 16.0 \pm 9.5) \times 10^{-6}$ & \cite{phi2_a1pi3}\\
    ${\cal B}(\Kp \aonem)$ & $(16.3 \pm 3.7) \times 10^{-6}$ & \cite{a1k}\\
    $\arg(A(\Kp \aonem)/A(\Koneap \pim))$ & $[-180, +180]^\circ$ & ---\\
    $\Acp(\Koneap \pim)$ & $[-1, 1]$ & ---\\
    $\Acp(\Kp \aonem)$ & $-0.16 \pm 0.12$ & \cite{a1k}\\
    \hline
  \end{tabular}
  \caption{Additional parameters for the wider SU(3) \phitwo\ constraint, where an unknown central value is indicated by a range.}
  \label{tab:su3_full}
\end{table}

The situation immediately improves when the phase difference between $\Bz \to \Kp \aonem$ and $\Bz \to \Koneap \pim$ is $0^\circ$, compared to that shown in figure~\ref{fig:su3_min:mpv:phase}a, however the opposite behaviour with the constraint becoming progressively worse while approaching $180^\circ$ is seen. It should be noted that none of the phase configurations give a reasonable goodness-of-fit with the best $\chi^2$ ranging from 6 units at $0^\circ$ to 16 units at $180^\circ$ indicating that all scenarios considered here are most likely unphysical. As this system is largely over-constrained with 15 observables for only 8 free parameters of the SU(3) flavour model, such analyses could also be used to predict lesser known physical quantities.

\subsection{Non-factorisable SU(3)-breaking effects}
\label{sec:su3:nf}

Non-factorisable refers to sources of SU(3)-breaking effects not already accounted for in the ratios of CKM elements and decay constants. These can include the presence of amplitudes from other Feynman diagrams, theoretical uncertainties or sources that are unknown in origin. These effects can be parametrised with a real factor relating $\Delta S = 1$ to $\Delta S = 0$ amplitudes $F_{\rm SU(3)}$, which is unity in the limit of no non-factorisable SU(3)-breaking sources. Restricting the discussion to the minimal SU(3) analysis, the penguin amplitudes become
\begin{eqnarray}
  A(\Bp \to \Koneaz \pip) &=& -\frac{F_{\rm SU(3)}^+}{\bar \lambda} \frac{f_{K_1}}{f_{a_1}} P^+, \nonumber \\
  A(\Bp \to \Kz \aonep) &=& -\frac{F_{\rm SU(3)}^-}{\bar \lambda} \frac{f_{K}}{f_{\pi}} P^-.
\end{eqnarray}

\begin{figure}[tbp]
  \centering
  \includegraphics[height=120pt,width=!]{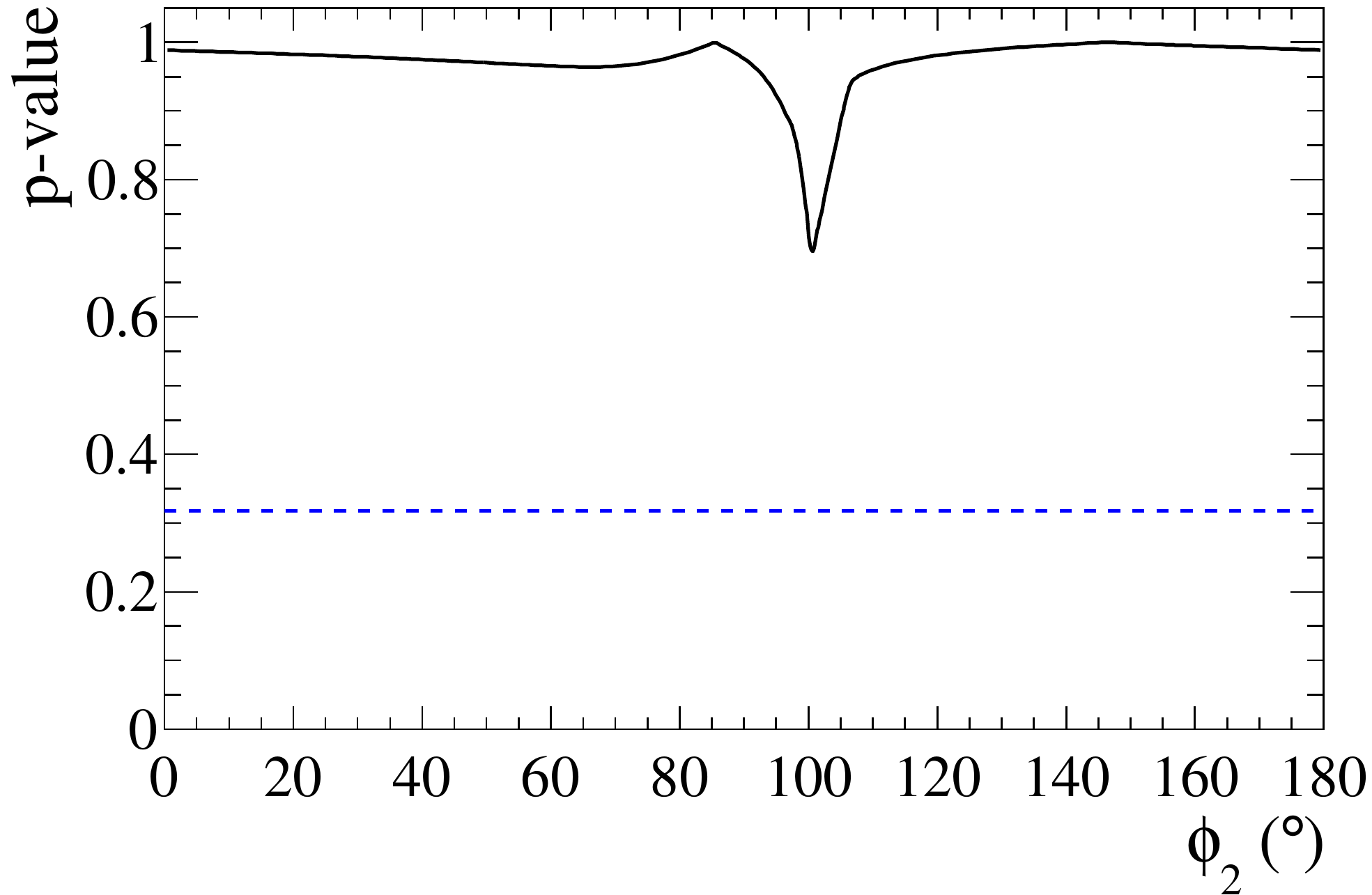}
  \includegraphics[height=120pt,width=!]{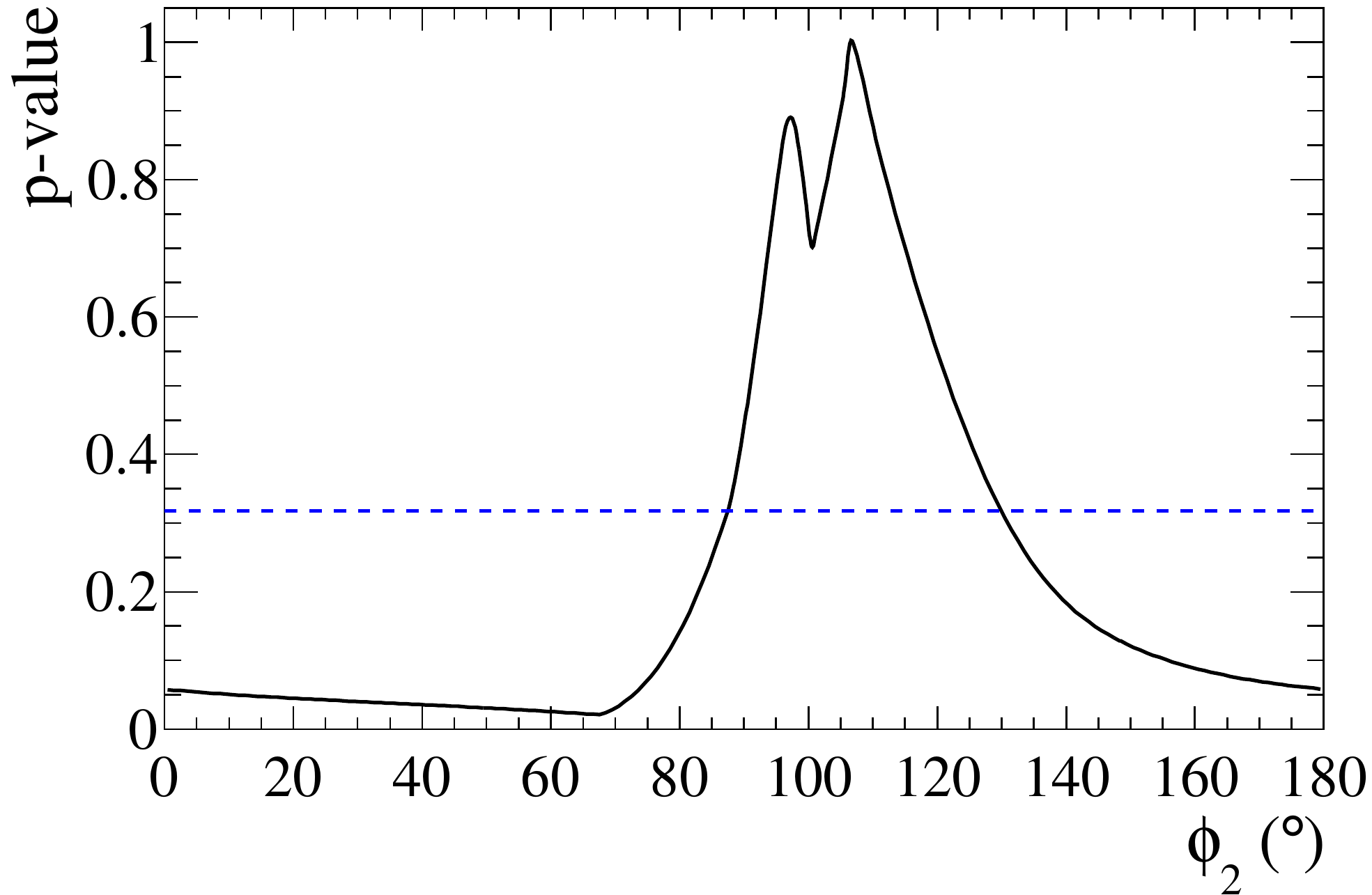}
  \put(-215,105){(a)}
  \put(-28,105){(b)}

  \includegraphics[height=120pt,width=!]{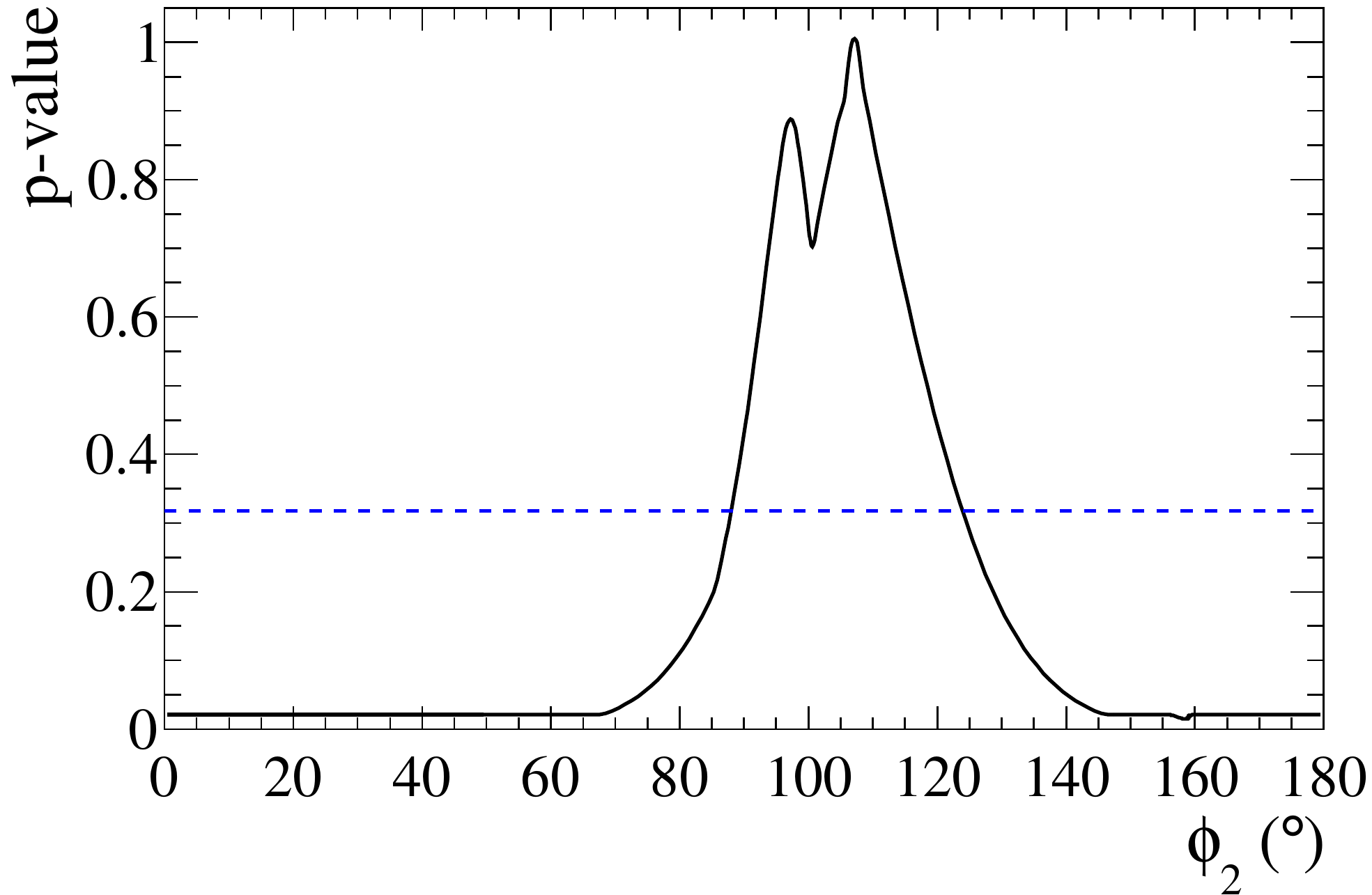}
  \includegraphics[height=120pt,width=!]{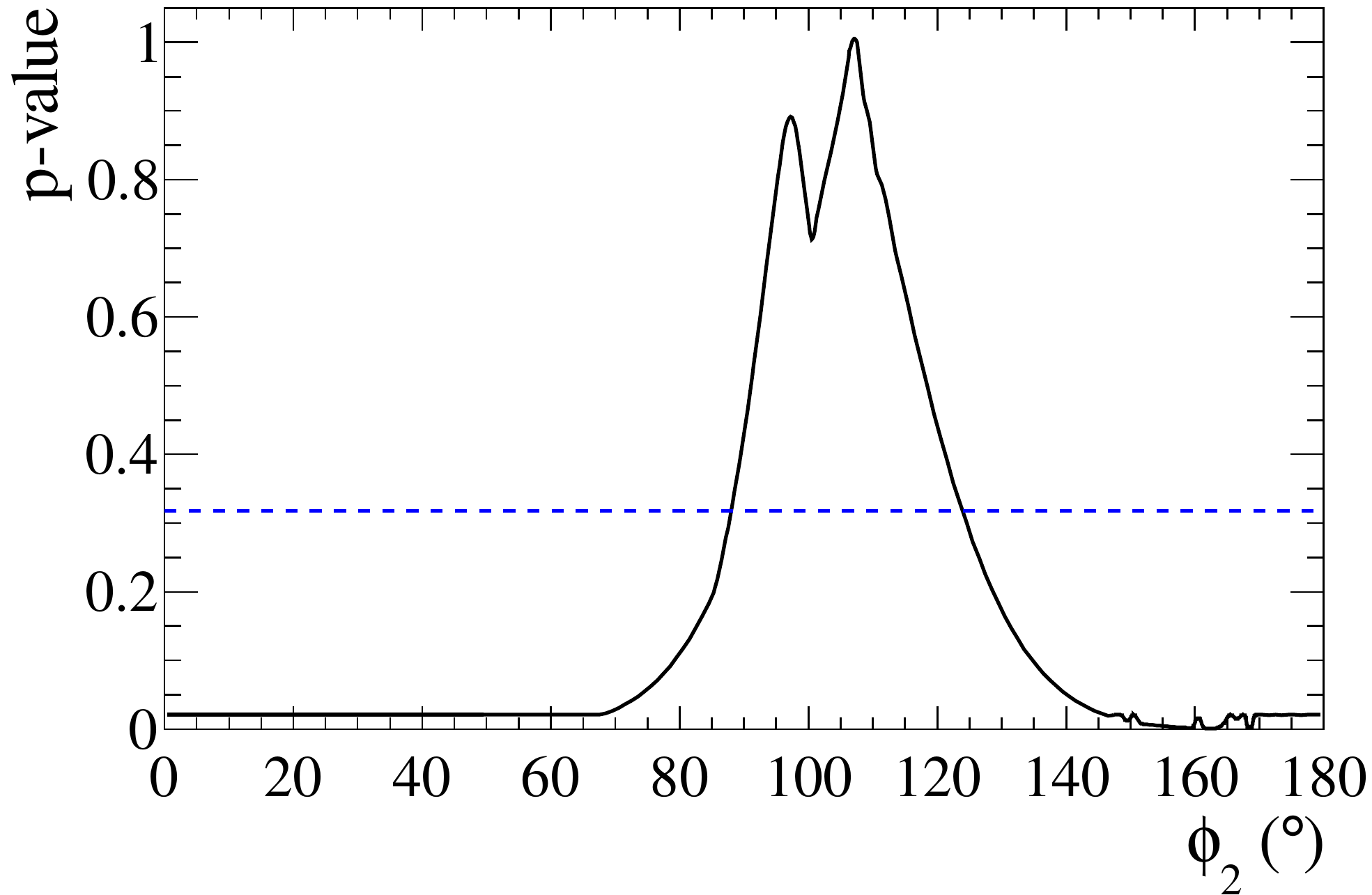}
  \put(-215,105){(c)}
  \put(-28,105){(d)}

  \includegraphics[height=120pt,width=!]{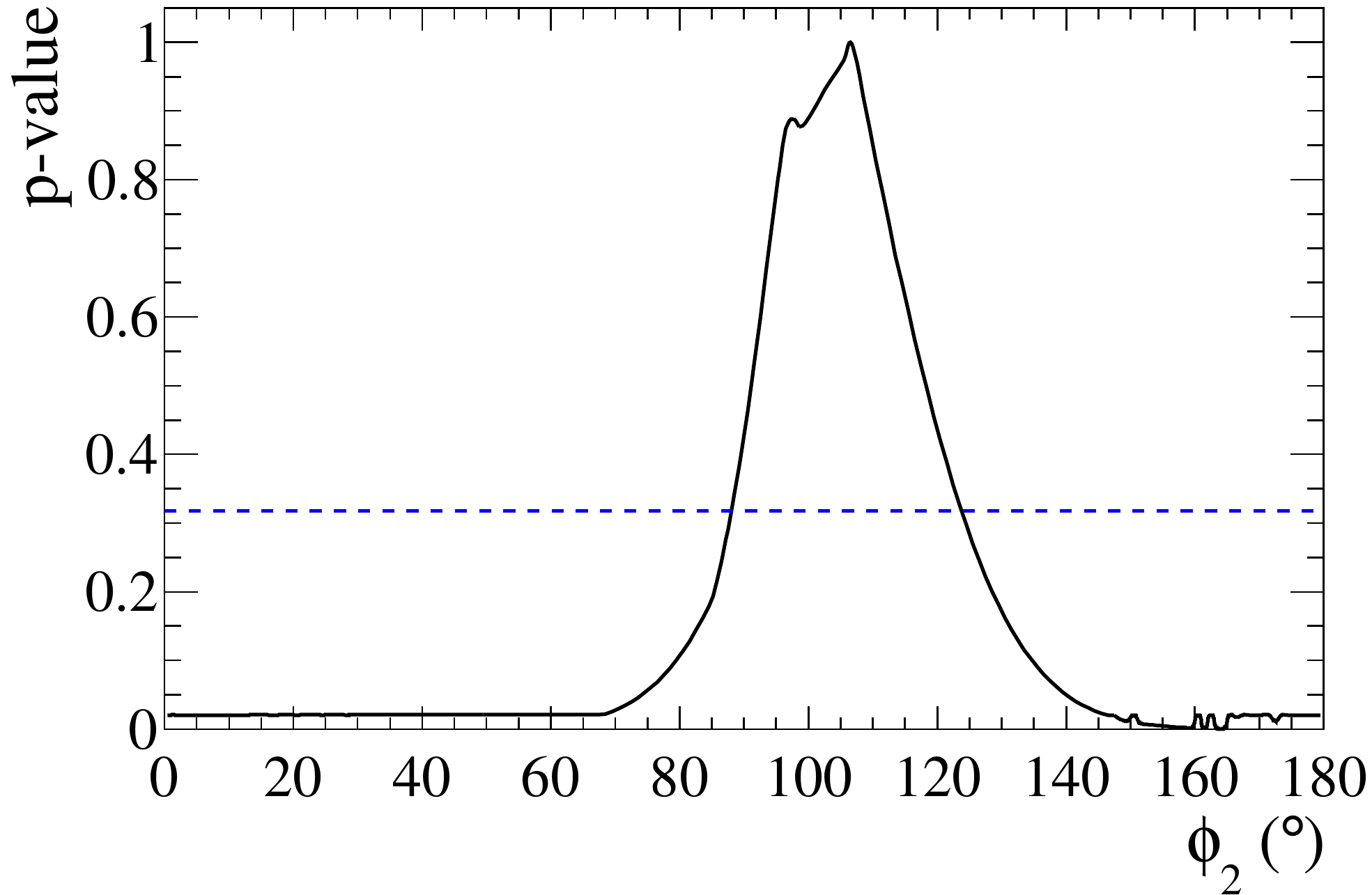}
  \includegraphics[height=120pt,width=!]{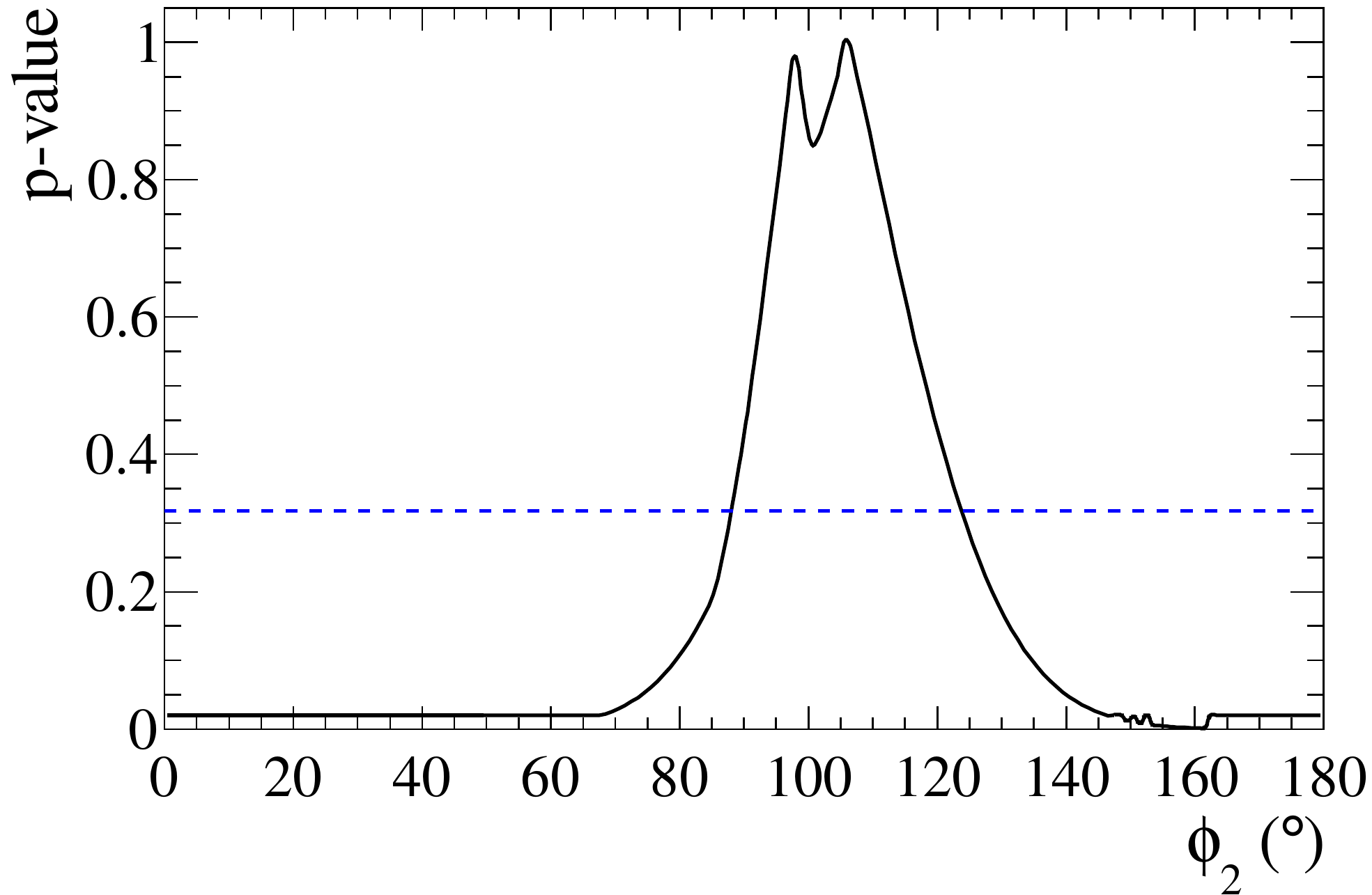}
  \put(-215,105){(e)}
  \put(-28,105){(f)}

  \includegraphics[height=120pt,width=!]{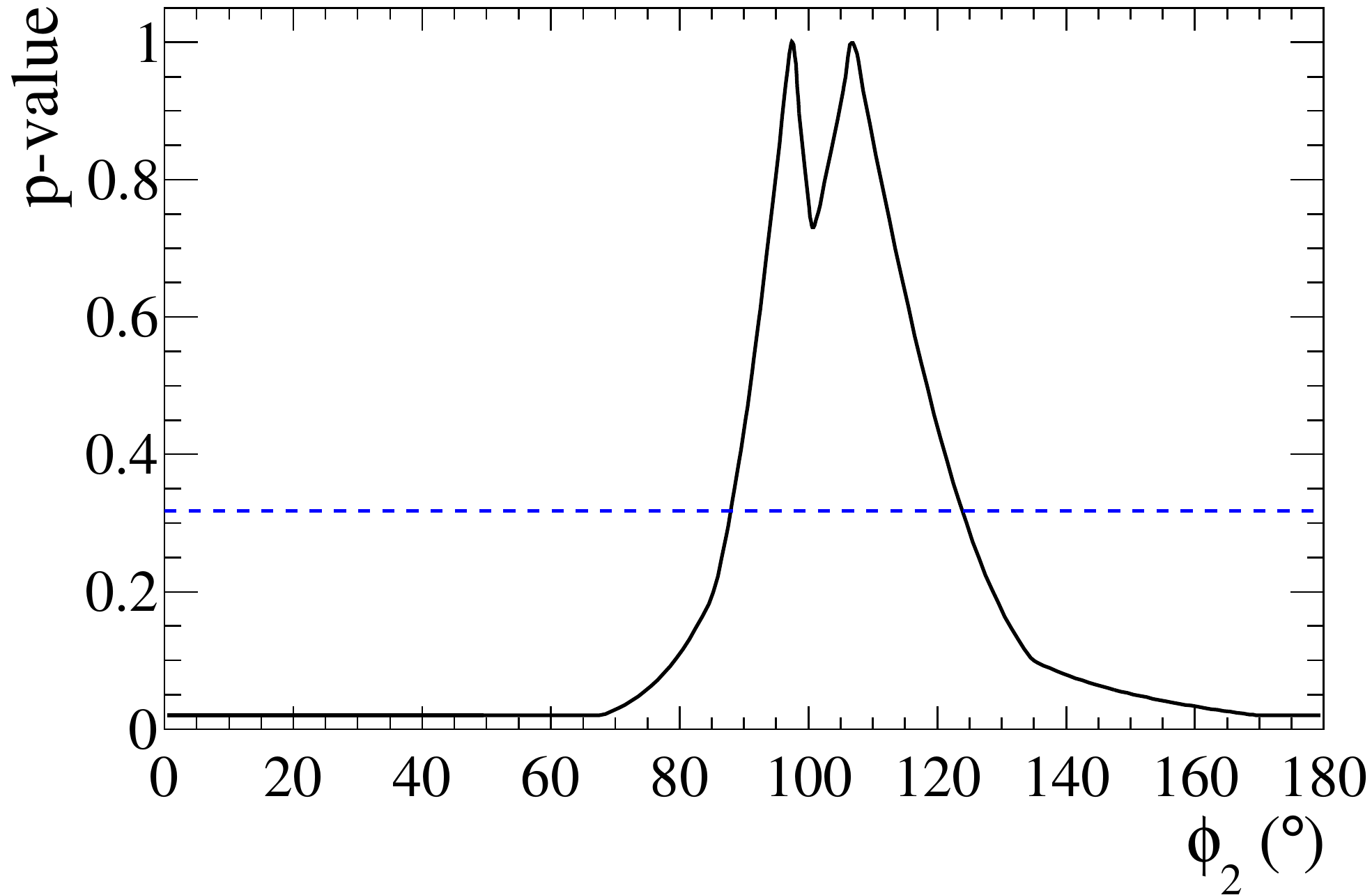}
  \includegraphics[height=120pt,width=!]{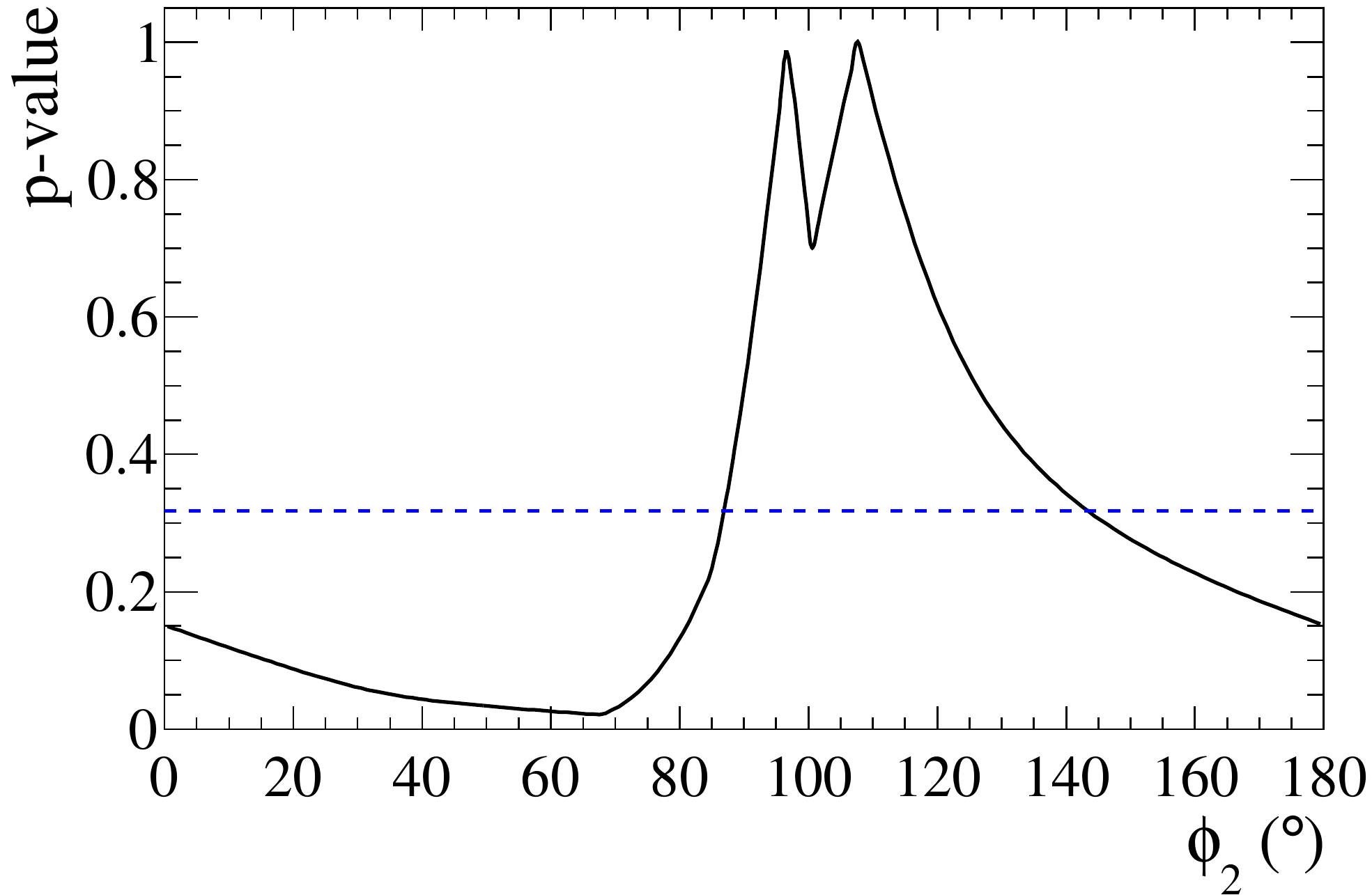}
  \put(-215,105){(g)}
  \put(-28,105){(h)}

  \caption{\label{fig:su3:phi2} $p$-value scans of \phitwo\ with free non-factorisable SU(3)-breaking parameters where the horizontal dashed line shows the $1\sigma$ bound. For the BaBar most probable branching fraction of $\Bp \to \Koneaz \pip$, these scans show the effects of the experimentally determined phase difference between itself and $\Bp \to \Kz \aonep$ when set to (a) $0^\circ$, (b) $45^\circ$, (c) $90^\circ$, (d) $135^\circ$, (e) $180^\circ$, (f) $225^\circ$, (g) $270^\circ$ and (h) $315^\circ$.}
\end{figure}

\begin{figure}[tbp]
  \centering
  \includegraphics[height=120pt,width=!]{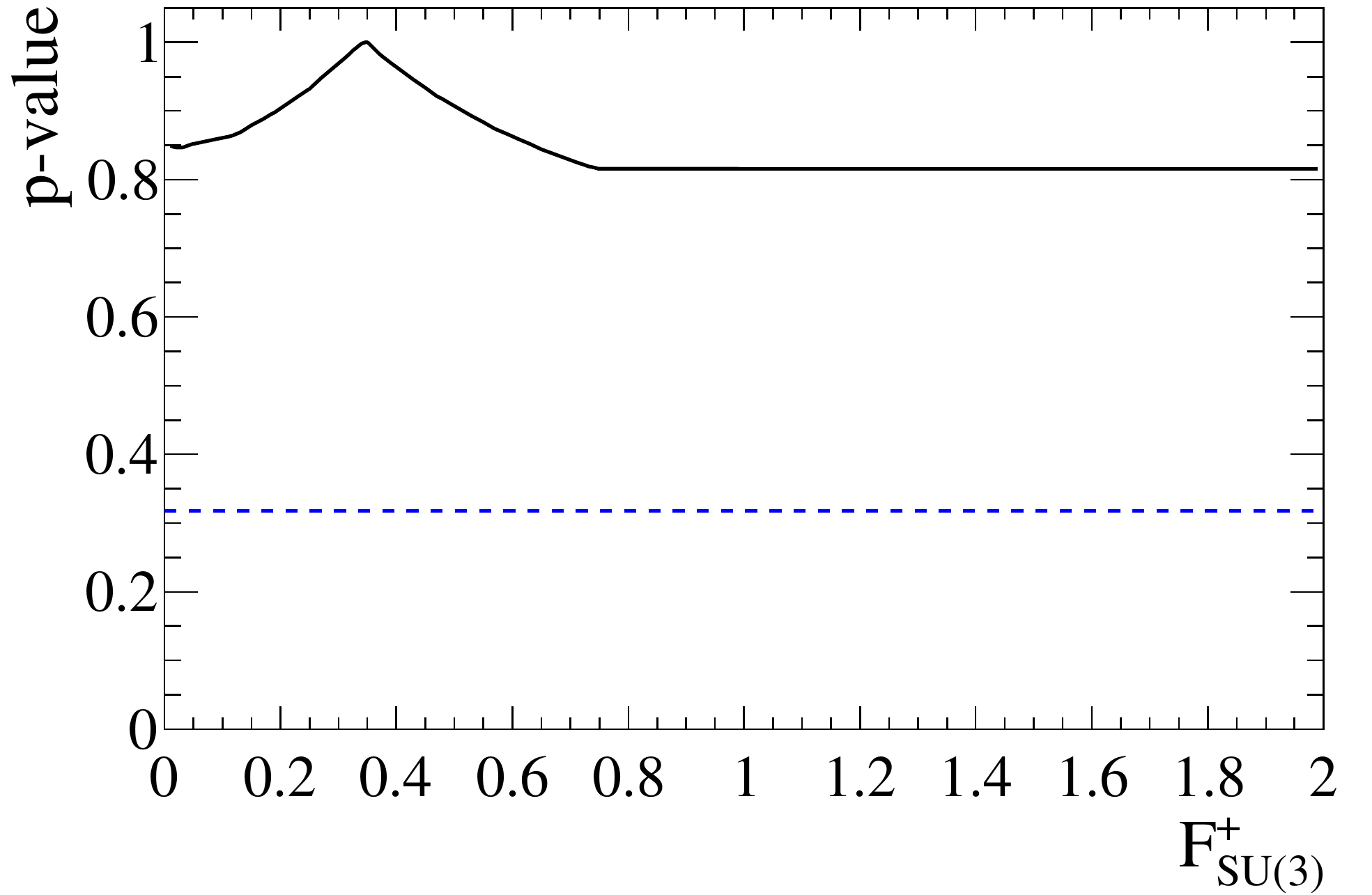}
  \includegraphics[height=120pt,width=!]{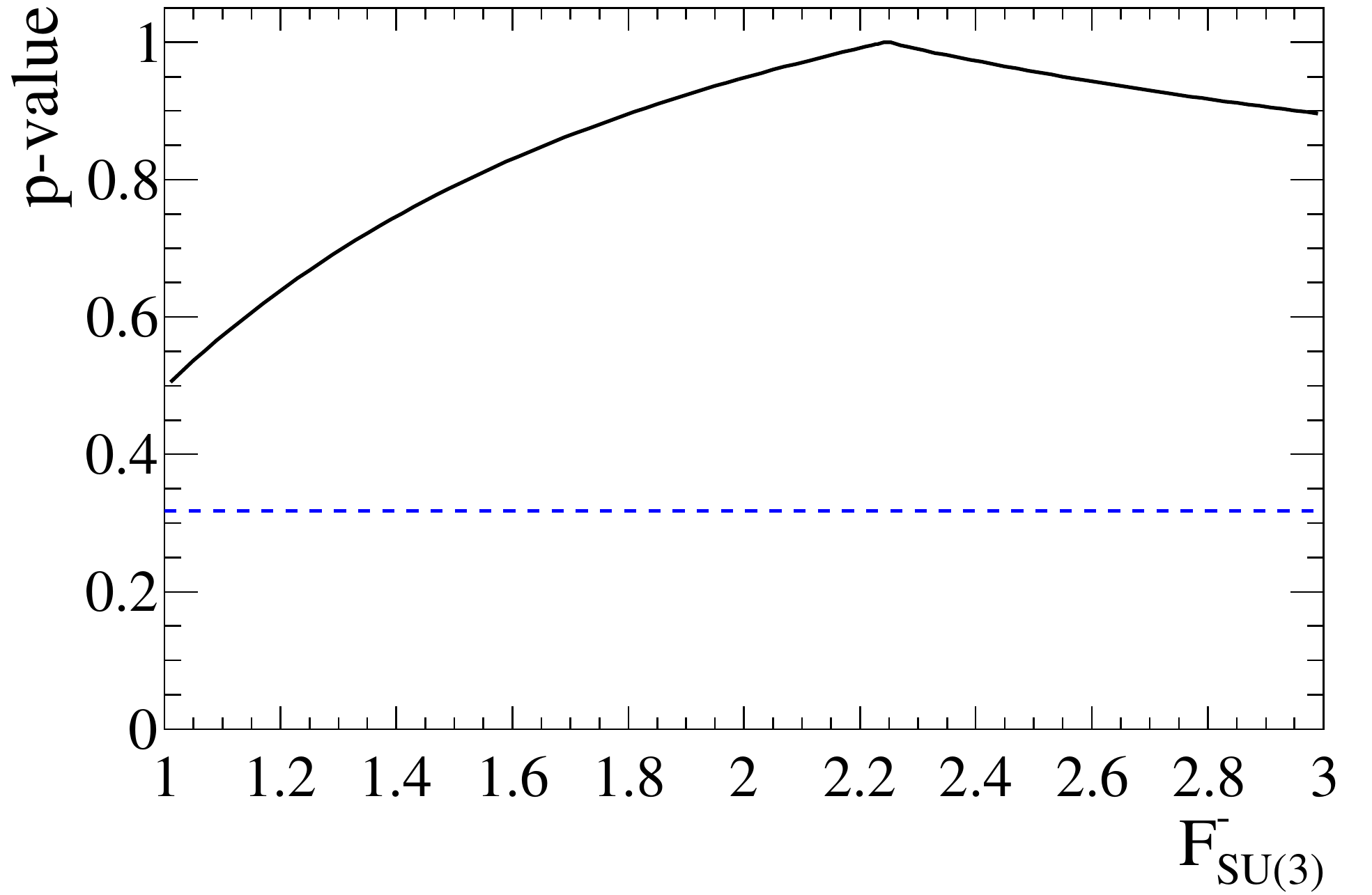}
  \put(-215,30){(a)}
  \put(-28,30){(b)}

\caption{\label{fig:su3:f} $p$-value scans of (a) $F_{\rm SU(3)}^+$ and (b) $F_{\rm SU(3)}^-$, where the horizontal dashed line shows the $1\sigma$ bound. For the BaBar most probable branching fraction of $\Bp \to \Koneaz \pip$, these scans show the effects of the experimentally determined phase difference between itself and $\Bp \to \Kz \aonep$ when set to $45^\circ$.}
\end{figure}

As the system without SU(3)-breaking parameters is already over-constrained, there may be a possibility to release these in the \phitwo\ constraint as shown in figure~\ref{fig:su3:phi2}. Clearly, the ability to constrain non-factorisable SU(3) breaking within the analysis will come at a cost of precision on \phitwo, largely depending on the phase difference between $\Bp \to \Kz \aonep$ and $\Bp \to \Koneaz \pip$. For a phase difference of $45^\circ$, scans for the non-factorisable SU(3) breaking factors are shown in figure~\ref{fig:su3:f}, so chosen as they just happen to best indicate the possible emerging sensitivity given current experimental uncertainties. If this turns out to be a viable approach in future analyses, the advantage is that \phitwo\ uncertainties will continue to scale with further increases in data sample sizes. Otherwise, it is always possible to fall back on the usual method of nominally fixing the non-factorisable SU(3)-breaking factors to unity and varying their values to quantify their impact on the \phitwo\ constraint as a systematic uncertainty that is irreducible from the experimental perspective without theoretical intervention.

\section{Amplitude model}
\label{sec:model}

Now that the possibility to constrain a single solution for \phitwo\ in $B \to \aone \pimp$ is established under an ideal set of circumstances, it is important to estimate the feasibility of measuring the phase difference between $\Bp \to \Koneaz \pip$ and $\Bp \to \Kz \aonep$ given the current experimental knowledge of the $\Bp \to \Kz \pip \pim \pip$ final state. A primary concern regarding this proposed method is whether the hadronic uncertainties, particularly those arising from the axial vectors, can be controlled at a level that will still permit a meaningful measurement of the phase difference. To demonstrate, I generate a set of pseudo-experiments varying the hadronic model within its current experimental uncertainties in the isobar approach. I then set the yields to those expected at various milestones expected during the timelines of the Belle II and LHCb experiments and record the width of the phase difference distribution obtained from a maximum likelihood fit to each pseudo-experiment, which is a measure of the total expected uncertainty including systematic sources pertaining to the physical amplitude model. Experimental systematic effects such as detection efficiency and background contributions to the phase space will be neglected in this study.

\subsection{Hadronic form factors}
I consider a rudimentary model with five contributions to the 4-body phase space coming only from the channels we know to exist, $\Bp \to \Kone(1270)^0 \pip$, $\Kone(1400)^0 \pip$, $\Kz \aonep$, $\Kstp \rhoz$ and $\Kstp \fz(980)$. The amplitude for each intermediate state is parametrised as
\begin{equation}
  A_i(\Phi_4) = B_{L_B}(\Phi_4) \cdot [B_{L_{R_1}}(\Phi_4) T_{R_1} (\Phi_4)] \cdot [B_{L_{R_2}}(\Phi_4) T_{R_2} (\Phi_4)] \cdot S_i(\Phi_4),
\end{equation}
where $B_{L_B}$ represents the production Blatt-Weisskopf barrier factor~\cite{bwbf} depending on the orbital angular momentum between the products of the \Bp\ decays, $L_B$. Two resonances will appear in each isobar, denoted by $R_1$ and $R_2$, for which respective decay barrier factors are also assigned. The Breit-Wigner propagators are represented by $T$, while the overall spin amplitude is given by $S$. Each isobar is Bose-symmetrised so that the total amplitude is symmetric under the exchange of like-sign pions.

The Blatt-Weisskopf penetration factors account for the finite size of the decaying resonances by assuming a square-well interaction potential with radius $r$. They depend on the breakup momentum between the decay products $q$, and the orbital angular momentum between them $L$. Their explicit expressions used in this analysis are
\begin{eqnarray}
  B_0(q) &=& 1, \nonumber \\
  B_1(q) &=& \frac{1}{\sqrt{1+(qr)^2}}, \nonumber\\
  B_2(q) &=& \frac{1}{\sqrt{9+3(qr)^2+(qr)^4}}.
\end{eqnarray}

Spin amplitudes are constructed with the covariant tensor formalism based on the Rarita-Schwinger conditions~\cite{spin}. The spin $S$, of some state with 4-momentum $p$, and spin projection $s_z$, is represented by a rank-$S$ polarisation tensor that is symmetric, traceless and orthogonal to $p$. These conditions reduce the number of independent elements to $2S+1$ in accordance with the number of degrees of freedom available to a spin-$S$ state. The sum over these polarisation indices of the inner product of polarisation tensors form the fundamental basis on which all spin amplitudes are built. Called the spin projection operator $P$, it projecst an arbitrary tensor onto the subspace spanned by the spin projections of the spin-$S$ state.

Another particularly useful object is the relative orbital angular momentum spin tensor $L$, which for some process $R \to P_1 P_2$, is the relative momenta of the decay products $q_R \equiv p_1 - p_2$ projected to align with the spin of $R$,
\begin{equation}
  L_{\mu_1 \mu_2 ... \mu_L}(p_R, q_R) = P_{\mu_1 \mu_2 ... \mu_L \nu_1 \nu_2 ... \nu_L} (p_R) q_R^{\nu_1} q_R^{\nu_2} ... q_R^{\nu_L},
\end{equation}
where the number of indices representing the tensor rank is equal to the value of $L$. Finally, to ensure that the spin amplitude behaves correctly under parity transformation, it is sometimes necessary to include the Levi-Cevita totally antisymmetric tensor $\epsilon_{abcd}p_R^d$. Each stage of a decay is represented by a Lorentz scalar obtained by contracting an orbital tensor between the decay products with a spin wavefunction of equal rank representing the final state.

Five topologies are necessary for this analysis, including two for the axial cascade decays, $A \to VP$ and $A \to SP$. While a relative orbital angular momentum $D$-wave between the vector and pseudoscalar is possible, these have yet to be definitively seen, so only the $S$-wave configuration is considered at this time. The only possibility between the scalar and pseudoscalar is a $P$-wave, thus the spin densities for the entire decay chains are given by
\begin{eqnarray}
  && A \to VP: \hspace{10pt} S \propto L_a(p_{\Bp}, q_{\Bp}) P^{ab}(p_{A}) L_b(p_{V}, q_{V}), \nonumber\\
  && A \to SP: \hspace{10pt} S \propto L_a(p_{\Bp}, q_{\Bp}) L^a(p_{A}, q_{A}).
\end{eqnarray}
For $\Bp \to \Kstp \rhoz$ decays, $S$-, $P$- and $D$-waves are permitted between the vector resonances, with total spin densities,
\begin{eqnarray}
  &&S\text{-wave}: \hspace{10pt} S \propto L_a(p_{\Kstp}, q_{\Kstp})L^a(p_{\rhoz}, q_{\rhoz}), \nonumber\\
  &&P\text{-wave}: \hspace{10pt} S \propto \epsilon_{abcd} L^d(p_{\Bp}, q_{\Bp}) L^c(p_{\Kstp}, q_{\Kstp}) L^b(p_{\rhoz}, q_{\rhoz}) p^a_{\Bp}, \nonumber\\
  &&D\text{-wave}: \hspace{10pt} S \propto L_{ab}(p_{\Bp}, q_{\Bp}) L^b(p_{\Kstp}, q_{\Kstp}) L^a(p_{\rhoz}, q_{\rhoz}).
\end{eqnarray}
However, as the $S$- and $D$-waves are found to be indistinguishable due to the size of the \Bp\ phase space, these are considered together using only the $S$-wave term. The final topology comes from $\Bp \to \Kstp \fz$ decays, with a spin density given by,
\begin{equation}
  S \propto L_a(p_{\Bp}, q_{\Bp}) L^a(p_{\Kstp}, q_{\Kstp}).
\end{equation}

In general, resonance lineshapes are described by Breit-Wigner propagators as a function of the energy-squared $s$,
\begin{equation}
  T(s) = \frac{1}{M^2(s) - s - im_0\Gamma(s)},
\end{equation}
where $M^2(s)$ is the energy-dependent mass and $\Gamma(s)$ is the total width which is normalised such that it represents the nominal width $\Gamma_0$, at the pole mass $m_0$.

For the \rhoz\ resonance, the Gounaris-Sakurai parametrisation is used to provide an analytic expression for $M^2(s)$ and $\Gamma(s)$~\cite{gs}. Otherwise, I ignore potential dispersive effects in the axial vectors, setting $M^2(s)$ to the pole-mass squared. The $\fz$ is modelled with the Flatt\'{e} lineshape~\cite{flatte}, which takes into account the opening of the inelastic $KK$ channel coupled to the elastic $\pi\pi$ contribution.

\subsubsection{Energy-dependent width of axial vector resonances}

In amplitude analysis, the total energy-dependent width of a resonance is typically approximated by the partial width corresponding to the decay channel being studied. While this can be somewhat justified for resonances decaying predominantly into 2-body final states, neglecting the contributions from all partial widths in cascade decays can lead to large variations in determinations of their pole parameters, rendering comparison difficult. In order to achieve the best possible description of their lineshapes, I calculate distributions for the total energy-dependent widths of the $\Kone(1270)$, $\Kone(1400)$ and $a_1(1260)$ resonances.

The total energy-dependent width of a resonance is essentially a measure of all of its possible decay channels. Assuming that some form for the energy-dependent volume of phase space available $\rho_i(s)$, of decay process $i$, is known, then the total width can be composed from the sum of partial widths as
\begin{equation}
  \Gamma(s) = \sum_i \Gamma_i(s) = \Gamma_0\sum_i g_i \rho_i(s),
\end{equation}
with the couplings $g_i$, comprising each partial width to be determined. With the normalisation convention at the pole mass $\Gamma(m^2_0) = \Gamma_0$, this implies $\sum_i g_i = 1$. This total width can then act in the Breit-Wigner denominator to produce a Flatt\'{e}-like lineshape. In the absence of theoretical predictions for $g_i$, they can be reverse-engineered from the known branching fractions of the resonance,
\begin{equation}
  {\cal B}^{\rm pred}_i \propto \int_{s_{\rm min}}^{\infty} \frac{m_0 \Gamma_i(s)}{|m_0^2 - s - im_0 \Gamma(s)|^2} ds = \int_{s_{\rm min}}^{\infty} \frac{m_0 g_i \rho_i(s)}{|m_0^2 - s - im_0 \Gamma(s)|^2} ds.
\end{equation}
For practical purposes, the numerical integral calculation is terminated at 10 pole widths above the pole mass where the Breit-Wigner shape is negligibly above zero. Imposing the normalisation condition $\sum_i {\cal B}_i = 1$, the couplings are found by minimising a $\chi^2$
\begin{equation}
  \chi^2 = \sum_i \biggl[\frac{{\cal B}^{\rm exp}_i - {\cal B}^{\rm pred}_i (g_i)}{\Delta {\cal B}^{\rm exp}_i}\biggr]^2,
\end{equation}
where ${\cal B}^{\rm exp}_i$ and $\Delta {\cal B}^{\rm exp}_i$ represent the central values and experimental uncertainties, respectively.

The energy-dependent phase space of a spin-1 decay is given by
\begin{equation}
  \rho_n(s) = \frac{1}{2\sqrt{s}} \int \sum_{\lambda=0,\pm 1}|A_{\lambda}(\Phi_n)|^2 d\Phi_n (s),
  \label{eq:phasespace}
\end{equation}
where $A$ is the transition amplitude of the cascade, itself being comprised of barrier factors, a spin density and lineshape, with a coherent sum taken over the open polarisation indices of the initial state. The differential $d\Phi_n$, is the phase space density for $n$-body decays,
\begin{equation}
  d\Phi_n(s, p_1,...,p_n) = \delta^4(P-\sum_{i=1}^np_i)\prod_{i=1}^n \frac{d^3p_i}{(2\pi)^32E_i}.
\end{equation}
Essentially this means that the decay structure in the phase space of the resonance must be known across its mass range. 

\begin{figure}[tbp]
  \centering
  \includegraphics[height=130pt,width=!]{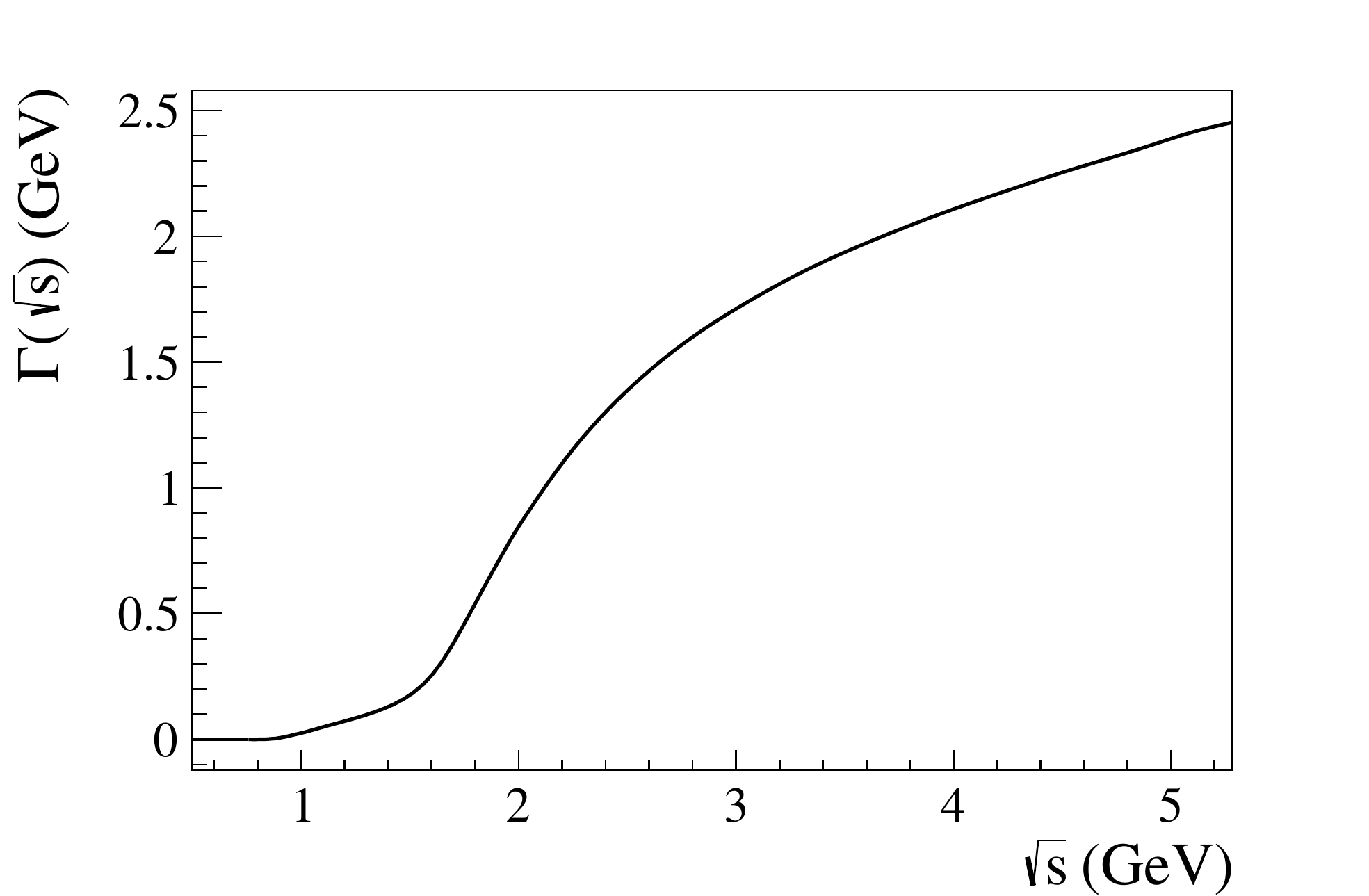}
  \includegraphics[height=130pt,width=!]{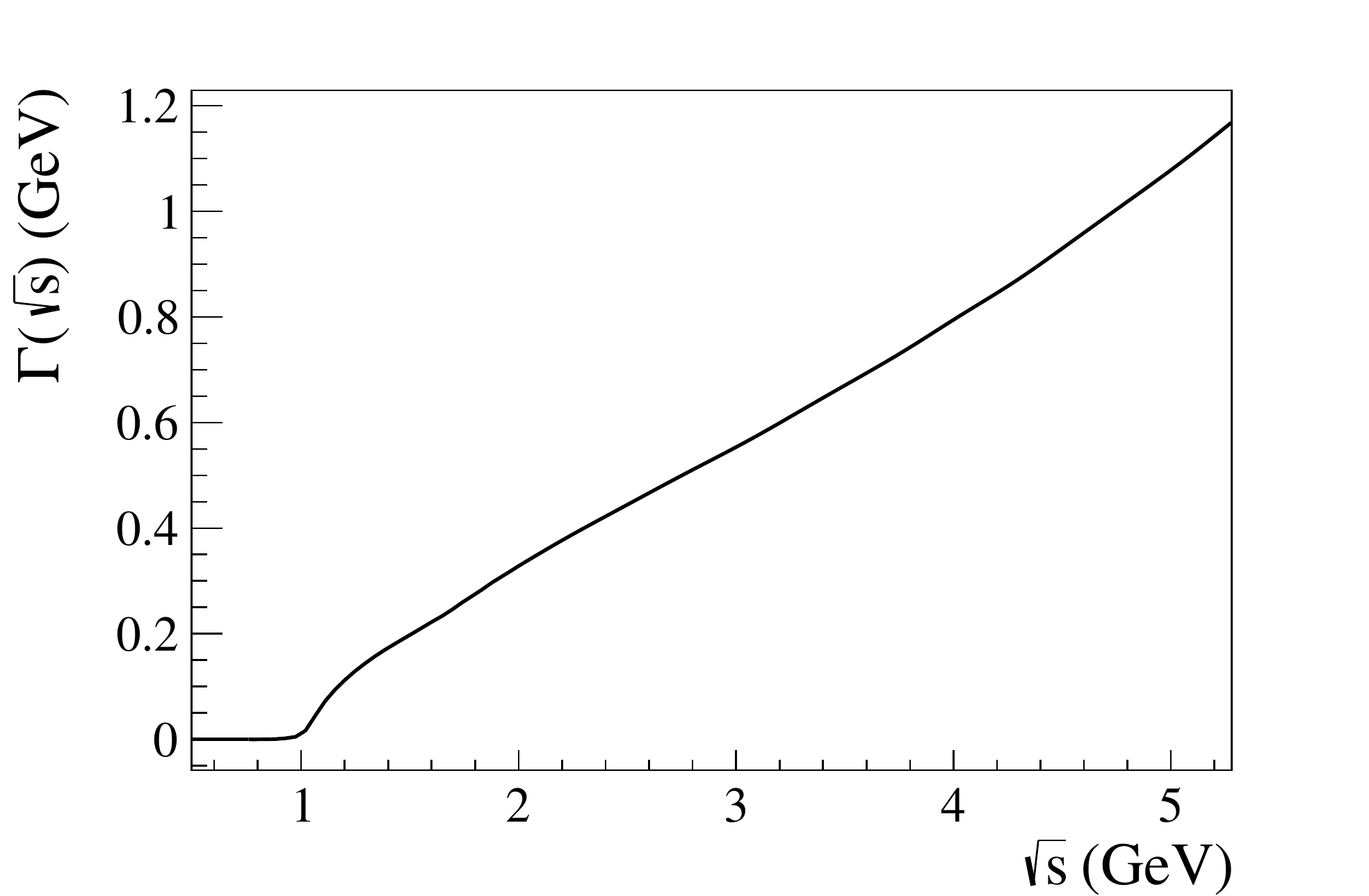}
  \put(-365,105){(a)}
  \put(-163,105){(b)}

  \caption{\label{fig:k1} Total energy-dependent widths of the (a) $\Kone(1270)^0$ and (b) $\Kone(1400)^0$.}
\end{figure}

I now determine the couplings for each decay of the $\Kone(1270)$ and $\Kone(1400)$ for which a branching fraction is given in the PDG~\cite{PDG}. In principle, some of the decay channels could interfere, however as their complex couplings are unknown, they are assumed to be incoherent and as such, $\rho_n(s)$ is calculated for each decay channel individually. A $\chi^2$ minimisation is performed to determine the coupling in each partial width for $\Kone(1270)$ and $\Kone(1400)$, where the sum of couplings are constrained to unity. Good solutions are found with the fit results collected in table~\ref{tab:k1width} and the total widths shown in figure~\ref{fig:k1}.

\begin{table}[tbp]
  \centering
  \begin{tabular}{|c|c|c|}
    \hline
    Decay Channel & $\Kone(1270)^0$ Couplings& $\Kone(1400)^0$ Couplings\\ \hline
    $K \rho$ & $0.473$ (constrained) & $0.029$ (constrained)\\
    $K \omega$ & $0.124 \pm 0.022$ & $0.010 \pm 0.010$\\
    $K^*(892) \pi$ & $0.184 \pm 0.047$ & $0.956 \pm 0.028$\\
    $K f_0(1370)$ & $0.016 \pm 0.010$ & $0.005 \pm 0.005$\\
    $K^*_0(1430) \pi$ & $0.203 \pm 0.030$ & ---\\\hline
    $\chi^2$ & $1.4 \times 10^{-4}$ & $1.2 \times 10^{-4}$\\ \hline
  \end{tabular}
  \caption{Fit results for the partial width couplings of the decays of the $\Kone(1270)^0$ and $\Kone(1400)^0$. The constrained parameters are fixed in the fit to be unity less the sum of the remaining couplings.}
  \label{tab:k1width}
\end{table}

For consistency, the treatment of the total energy-dependent of the $a_1(1260)^+$ is kept identical to that from its previous use in the study of the $\Bz \to \rhoz \rhoz$ system~\cite{rhorho_dalseno}, where only the $S$-wave decay to $\rho^0 \pip$ is  considered. However, future analyses will also have to consider the overlapping $\aonep \to \sigma \pip$ channel which contributes at a level of $\sim 25\%$ the $\aonep \to \rho^0 \pip$ rate~\cite{CLEO_4pi}.

\subsubsection{Treatment of $\Kone(1270)$--$\Kone(1400)$ mixing}
\label{sec:model:k1}

This work calls for a direct measurement of the phase difference between $\Bp \to \Koneaz \pip$ and $\Bp \to \Kz \aonep$, with the \Konea\ being the strange ${}^3P_1$ partner of the $a_1$. However, the \Konea\ is not a mass eigenstate, but is mixed together with the \Koneb\ from the ${}^1P_1$ axial-vector nonet, due to the strange and non-strange light-quark mass difference. The mixing of these two flavour eigenstates manifests as the physical mass eigenstates, the $\Kone(1270)$ and $\Kone(1400)$, through
\begin{equation}
  \left(
    \begin{array}{c}
      | \Kone(1270) \rangle\\
      | \Kone(1400) \rangle 
    \end{array}
  \right)
  =
  \left(
    \begin{array}{cc}
      \sin \theta_{K_1} & \phantom{-}\cos \theta_{K_1}\\
      \cos \theta_{K_1} & -\sin \theta_{K_1} 
    \end{array}
  \right)
  \left(
    \begin{array}{c}
      | \Konea \rangle\\
      | \Koneb \rangle 
    \end{array}
  \right),
\end{equation}
where $\theta_{K_1}$ is the mixing angle. Standard practice is to independently determine the complex couplings of the $\Kone(1270)$ and $\Kone(1400)$ resonances directly in an amplitude analysis, deriving physical information on the \Konea\ and \Koneb\ states from the fit results. Such treatment will not be as useful here as the measurement of phases relative to the \Konea\ is required.

For the purposes of this study, this equation can be inverted to rather express the \Konea\ and \Koneb\ in terms of a relation between the physical mass eigenstates, giving direct access to the phase of the \Konea\ in an amplitude analysis,
\begin{eqnarray}
    A(\Bp \to \Koneaz \pip) &=& \sin \theta_{K_1} A(\Bp \to \Kone(1270)^0 \pip) + \cos \theta_{K_1} A(\Bp \to \Kone(1400)^0 \pip) \nonumber \\
    A(\Bp \to \Konebz \pip) &=& \cos \theta_{K_1} A(\Bp \to \Kone(1270)^0 \pip) - \sin \theta_{K_1} A(\Bp \to \Kone(1400)^0 \pip). \nonumber \\
\end{eqnarray}
This construct poses an additional challenge as it implies that the $A(\Bp \to \Kone(1270)^0 \pip)$ and $A(\Bp \to \Kone(1400)^0 \pip)$ amplitudes are individually normalised across the \Bp\ phase space. Because the $\Kone(1270)^0$ and $\Kone(1400)^0$ decays have resonant substructure of their own whose couplings are also free parameters of the model, the normalisation of the total amplitude cannot be calculated in each function minimisation loop until those of the separate $\Bp \to \Kone(1270)^0 \pip$ and $\Bp \to \Kone(1400)^0 \pip$ contributions are. Thus, this analysis should be viewed in technical terms as two nested 3-body amplitude analyses of the $\Kone(1270)^0$ and $\Kone(1400)^0$ resonances within a 4-body \Bp\ amplitude analysis with all the additional computing overhead this entails.

The mixing angle itself is largely unknown at this time. Phenomenological analyses tend to indicate that its magnitude is $|\theta_{K_1}| \approx 33^\circ \vee 57^\circ$~\cite{k1mixing1,k1mixing2,k1mixing3,k1mixing4}, for a 4-fold degeneracy. I will also examine the impact of the mixing angle on measurements of the phase difference between $\Bp \to \Koneaz \pip$ and $\Bp \to \Kz \aonep$.

\subsection{Pseudo-experiment generation method}

In order to get some early indication of the impact of hadronic uncertainties on the phase difference measurement between the $\Bp \to \Koneaz \pip$ and $\Bp \to \Kz \aonep$, I adopt a procedure of varying these within current experimental uncertainties for each pseudo-experiment in an ensemble test. %The width of the phase difference residual can then be interpreted as the quadratic sum of the expected statistical error with the sources of model uncertainty considered in this analysis. 
The Monte Carlo (MC) is based on the decay rate in phase space,
\begin{equation}
    {\cal P}(\Phi_4, q) = \frac{1 - q}{2} |A(\Phi_4)|^2 + \frac{1 + q}{2} |\bar A(\bar \Phi_4)|^2,
\end{equation}
where $q = +1 (-1)$ for \Bp\ (\Bm). The total amplitude $A$, can be written in the typical isobar approach as the coherent sum over the number of intermediate states in the model with amplitude $A_i$, as a function of 4-body phase space $\Phi_4$,
\begin{equation}
  \label{eq:a}
  A \equiv \sum_i a_iA_i(\Phi_4),
\end{equation}
where $a_i$ is a strong complex coupling determined directly from the data. For simplicity, a reduced set of $K_1$ decay channels are considered as isobars in the 3-body amplitude analyses as opposed to those used to earlier determine their widths, with a cutoff for inclusion set at $3\sigma$ significance. At the 4-body level of the amplitude analysis, contributions that have achieved observation status are additionally included.

Incorporating a complex $CP$ violation parameter $\lambda_i$, for each weak contribution in the phase space, the total $\bar A$ can be written as
\begin{equation}
  \label{eq:abar}
  \bar A \equiv \sum_i a_i \lambda_i \bar A_i(\bar \Phi_4) = \sum_i a_i \lambda_i A_i(\Phi_4),
\end{equation}
where the phase space of the $CP$-conjugated process $\bar \Phi_4$, is set by convention to be equivalent to $\Phi_4$ in a flavour-specific final state, thus leaving $A_i$ to contain only strong dynamics blind to flavour. Naturally, no $CP$ violating parameters are built into the embedded 3-body Dalitz analyses as they decay strongly.

The MC generation requires a 2-stage process. The first stage sets the Breit-Wigner pole parameters of each resonance and the phase of each isobar relative to $\Bp \to \Koneaz \pip$, which is fixed along the real axis.  For processes with a tree diagram possible, a $CP$ violating phase is uniformly distributed and the magnitude of $\lambda_i$ calculated from
\begin{equation}
    |\lambda_i| = \sqrt{\frac{1+\Acp^i}{1-\Acp^i}},
\end{equation}
where $\Acp^i$ is a $CP$ violation in decay parameter known from experiment. Table~\ref{tab:stage1} records the specific list of isobars considered in this analysis along with all generated parameters needed for stage 1.

\begin{table}[tbp]
  \centering
  \begin{tabular}{|c|c|c|}
    \hline
    Parameter & Range & Reference\\ \hline
    $|a_{\Bp \to \Koneaz \pip}|$ & 1 & ---\\
    $\arg(a_{\Bp \to \Koneaz \pip})$ & $0^\circ$ & ---\\
    $\arg(a_{\Bp \to \Konebz \pip})$ & $[-180, +180]^\circ$ & ---\\

    $|a_{\Kone(1270)^0 \to \Kz \rhoz}|$ & 1 & ---\\
    $\arg(a_{\Kone(1270)^0 \to \Kz \rhoz})$ & $0^\circ$ & ---\\
    $\arg(a_{\Kone(1270)^0 \to \Kstp \pim})$ & $[-180, +180]^\circ$ & ---\\
    $\arg(a_{\Kone(1270)^0 \to K^*_0(1430)^+ \pim})$ & $[-180, +180]^\circ$ & ---\\

    $|a_{\Kone(1400)^0 \to \Kstp \pim}|$ & 1 & ---\\
    $\arg(a_{\Kone(1400)^0 \to \Kstp \pim})$ & $0^\circ$ & ---\\

    $\arg(a_{\Bp \to \Kz a_1(1260)^+})$ & $[-180, +180]^\circ$ & ---\\
    $\arg(a_{\Bp \to [\Kstp \rhoz]_{S+D}})$ & $[-180, +180]^\circ$ & ---\\
    $\arg(a_{\Bp \to [\Kstp \rhoz]_{P}})$ & $[-180, +180]^\circ$ & ---\\
    $\arg(a_{\Bp \to \Kstp \fz})$ & $[-180, +180]^\circ$ & ---\\ \hline

    %$\Acp(\Bp \to \Kz a_1(1260)^+)$ & $+0.12 \pm 0.11$ & \cite{a1k_babar}\\
    %$\arg(\lambda_{\Bp \to \Kz a_1(1260)^+})$ & $[-90, +90]^\circ$ & ---\\
    $\Acp(\Bp \to \Kstp \rhoz)$ & $+0.31 \pm 0.13$ & \cite{kstrhof0}\\
    $\arg(\lambda_{\Bp \to \Kstp \rhoz})$ & $[-90, +90]^\circ$ & ---\\
    $\Acp(\Bp \to \Kstp \fz)$ & $-0.15 \pm 0.12$ & \cite{kstrhof0}\\
    $\arg(\lambda_{\Bp \to \Kstp \fz})$ & $[-90, +90]^\circ$ & ---\\ \hline

    $r$ & $[2, 6]$ $c/$GeV & ---\\

    $m_0(\rhoz)$ & $0.7690 \pm 0.0009$ GeV$/c^2$ & \cite{PDG}\\
    $\Gamma_0(\rhoz)$ & $0.1509 \pm 0.0017$ GeV & \cite{PDG}\\
    $m_0(\fz)$ & $0.965 \pm 0.010$ GeV$/c^2$ & \cite{f0}\\
    $g_{\pi}(\fz)$ & $0.165 \pm 0.018$ & \cite{f0}\\
    $g_K/g_{\pi}(\fz)$ & $4.21 \pm 0.33$ & \cite{f0}\\
    $m_0(\fz(1370))$ & $1.4751 \pm 0.0063$ GeV$/c^2$ & \cite{LHCbf0}\\
    $\Gamma_0(\fz(1370))$ & $0.113 \pm 0.011$ GeV & \cite{LHCbf0}\\
    $m_0(\Kstp)$ & $0.89176 \pm 0.00025$ GeV$/c^2$ & \cite{PDG}\\
    $\Gamma_0(\Kstp)$ & $0.0503 \pm 0.0008$ GeV & \cite{PDG}\\    $m_0(K^*_0(1430)^+)$ & $1.425 \pm 0.050$ GeV$/c^2$ & \cite{PDG}\\
    $\Gamma_0(K^*_0(1430)^+)$ & $0.270 \pm 0.080$ GeV & \cite{PDG}\\

    $m_0(a_1(1260)^+)$ & $1.225 \pm 0.022$ GeV$/c^2$ & \cite{CLEO_4pi}\\
    $\Gamma_0(a_1(1260)^+)$ & $0.430 \pm 0.039$ GeV & \cite{CLEO_4pi}\\
    $m_0(\Kone(1270)^0)$ & $1.272 \pm 0.007$ GeV$/c^2$ & \cite{PDG}\\
    $\Gamma_0(\Kone(1270)^0)$ & $0.090 \pm 0.020$ GeV & \cite{PDG}\\
    $m_0(\Kone(1400)^0)$ & $1.403 \pm 0.007$ GeV$/c^2$ & \cite{PDG}\\
    $\Gamma_0(\Kone(1400)^0)$ & $0.174 \pm 0.013$ GeV & \cite{PDG}\\
    $\theta_{K_1}$ & $35^\circ$ & \cite{k1mixing4} \\ \hline
  \end{tabular}
  \caption{Stage 1 parameters. Uncertainties indicate the parameter was Gaussian distributed, square brackets indicate uniform generation within the range while a single value is a constant of generation.}
  \label{tab:stage1}
\end{table}

A second stage is then required to reverse engineer the magnitudes of the complex couplings missing from table~\ref{tab:stage1}, for which the unknown phase is uniformly distributed in stage 1. This is realised through a $\chi^2$ fit relating the generated branching fractions for each isobar scaled to unity, to the fit fractions of each isobar calculated for the generated model in the 4-body phase space,
\begin{equation}
  {\cal F}^{\rm pred}_i = \frac{\int (|A_i|^2 + |\bar A_i|^2) d\Phi_4}{\int \sum_i(|A_i|^2 + |\bar A_i|^2) d\Phi_4}.
\end{equation}

The assignment of generated branching fractions for each isobar is mostly straightforward, applying product branching fractions down the decay chain and isospin decomposition in strong decays where necessary. Noteworthy are the $\Bp \to \Kone \pip$ decay channels, for which the only experimental information on their branching fractions comes from BaBar~\cite{phi2_a1pi3}. For this study, their values are generated according to an asymmetric-width Gaussian constructed from the reported mean values and their distance to their respective 68\% C.L. boundaries. In $\Bp \to \Kstp \rhoz$, the fraction of longitudinal polarisation $f_L$, is also available from BaBar~\cite{kstrhof0}, representing the minimum fraction of the $P$-even $S+D$ component. The parameters required to generate the isobar model branching fractions are listed in table~\ref{tab:stage2}.

\begin{table}[tbp]
  \centering
  \begin{tabular}{|c|c|c|}
    \hline
    Parameter & Range & Reference\\ \hline
    ${\cal B}(\Bp \to \Kone(1270)^0 \pip)$ & $(17.0^{+4.0}_{-17.0})\times10^{-6}$ & \cite{phi2_a1pi3}\\
    ${\cal B}(\Kone(1270)^0 \to \Kz \rhoz)$ & $\frac{1}{3}(42 \pm 6)\%$ & \cite{PDG}\\
    ${\cal B}(\rhoz \to \pip \pim)$ & 1 & ---\\
    ${\cal B}(\Kone(1270)^0 \to \Kstp \pim)$ & $\frac{2}{3}(16 \pm 5)\%$ & \cite{PDG}\\
    ${\cal B}(\Kstp \to \Kz \pip)$ & $\frac{2}{3}$ & ---\\
    ${\cal B}(\Kone(1270)^0 \to K^*_0(1430)^+ \pim)$ & $\frac{2}{3}(28 \pm 4)\%$ & \cite{PDG}\\
    ${\cal B}(K^*_0(1430)^+ \to \Kz \pip)$ & $\frac{2}{3}(93 \pm 10)\%$ & \cite{PDG}\\\hline

    ${\cal B}(\Bp \to \Kone(1400)^0 \pip)$ & $(20.0^{+5.0}_{-20.0})\times10^{-6}$ & \cite{phi2_a1pi3}\\
    ${\cal B}(\Kone(1400)^0 \to \Kstp \pim)$ & $\frac{2}{3}(94 \pm 6)\%$ & \cite{PDG}\\\hline

    ${\cal B}(\Bp \to \Kz a_1(1260)^+)$ & $(34.6 \pm 6.7)\times10^{-6}$ & \cite{a1k_babar}\\
    ${\cal B}(a_1(1260)^+ \to \rhoz \pip)$ & $\frac{1}{2}$ & ---\\\hline

    ${\cal B}(\Bp \to \Kstp \rhoz)$ & $(4.6 \pm 1.1)\times10^{-6}$ & \cite{kstrhof0}\\
    ${f_L}(\Bp \to \Kstp \rhoz)$ & $0.78 \pm 0.12$ & \cite{kstrhof0}\\
    ${\cal B}(\Kstp \rhoz \to [\Kstp \rhoz]_P)$ & $[0.0, 1.0]\times[1-f_L(\Bp \to \Kstp \rhoz)]$ & ---\\
    ${\cal B}(\Kstp \rhoz \to [\Kstp \rhoz]_{S+D})$ & 1-${\cal B}(\Kstp \rhoz \to [\Kstp \rhoz]_P)$ & ---\\\hline

    ${\cal B}(\Bp \to \Kstp \fz) {\cal B}(\fz \to \pip \pim)$ & $(4.2 \pm 0.7)\times10^{-6}$ & \cite{kstrhof0}\\
    \hline
  \end{tabular}
  \caption{Stage 2 parameters. Uncertainties indicate the parameter was Gaussian distributed while square brackets indicate uniform generation within the range.}
  \label{tab:stage2}
\end{table}

\subsection{Expected yields}

Estimates of the size of future event samples can be obtained by extrapolating from existing results. Belle and LHCb have not presented results on any of the five \Bp\ decay channels listed in table~\ref{tab:stage2}, so I rely on information purely from BaBar to guess the Belle~II yield. Based on the reported $\Bp \to \Kone(1270)^0 \pip + \Kone(1400)^0 \pip$ branching fraction and efficiency in ref.~\cite{phi2_a1pi3}, BaBar's signal yield should correspond to roughly 70 events in $\Ks \pip \pim \pip$. For $\Bp \to \Kz a_1(1260)^+$, $\Kstp \rhoz$ and $\Kstp \fz$, the yields are 241~\cite{a1k_babar}, 85~\cite{kstrhof0} and 69~\cite{kstrhof0} events, respectively. Considering that Belle collected almost twice the amount of data as BaBar and the wider analysis region required to study all these channels together in an amplitude analysis, I set the Belle yield to approximately 1000 events in the \Ks \pip \pim \pip\ final state.

An estimate of the Run~1 yield at LHCb requires a little more gymnastics with published quantities. A recent analysis of the neutral $\Bz \to (\Kp \pim)(\pip \pim)$ final state found 11066 events~\cite{k3pi_lhcb}. Now, accounting for the different branching fractions of $\Bz \to \Ks \pip \pim$ and $\Bp \to \Kp \pip \pim$, the \Ks\ to \Kp\ detection ratio in that 3-body system is less than 5\%~\cite{kspipi_lhcb,kpipi_lhcb}. Assuming this will translate to their 4-body system, LHCb should have around 300 $\Bp \to (\Ks \pip)(\pip \pim)$ events in an analysis region which cuts away the axial vectors to negligible levels. Expanding to the entire phase space doubles that yield as mentioned in the LHCb analysis of the $\Bz \to (\pip \pim) (\pip \pim)$ system~\cite{phi2_rhorho7}. Assuming uniform detection efficiency, the total yield also including the axial vectors should then be roughly 1200 events in Run~1.

I generate an ensemble of pseudo-experiments based on individual amplitude models with the \verb|GENBOD| phase space function~\cite{phsp} and \verb|qft++| to provide the spin densities~\cite{qft}. Tests are performed with crude estimates of the expected total $\Bp \to \Kp \pip \pim \pip$ yields for 10~ab$^{-1}$ and the full 50~ab$^{-1}$ of data expected with Belle~II~\cite{belle2}, as well as the amount of data recorded by LHCb at the end of Run~2 and expected to be recorded in Run~3~\cite{run3}. These are recorded in table~\ref{tab:yields}. The Belle~II projections are based on a naive scaling of the Belle expected yield calculated earlier according to the desired integrated luminosity. The LHCb Run~2 projection assumes a similar yield in Run~1 and the 2015+2016 data set accounting for the change in $b$ production cross-section with $\sqrt{s}$, with the 2017+2018 sample doubling that yet again. Run~3 is roughly estimated to procure an order of magnitude more data over that expected by naively scaling expected integrated luminosities, due to the planned removal of the hardware (L0) trigger and migration to a fully software-based system~\cite{trigger}.

\begin{table}[tbp]
  \centering
  \begin{tabular}{|c|c|c|c|}
    \hline
    Experiment & Milestone & Year & Projected Yield\\ \hline
    LHCb & \phantom{0}9 fb$^{-1}$ (Run 2) & 2018 & \phantom{0}5000\\
    Belle II & 10 ab$^{-1}$ & 2021 & 10000\\
    LHCb & 23 fb$^{-1}$ (Run 3) & 2023 & 50000\\
    Belle II & 50 ab$^{-1}$ & 2024 & 50000\\
    \hline
  \end{tabular}
  \caption{Expected yields of $\Bp \to \Ks \pip \pim \pip$ at various stages of data taking at Belle~II and LHCb.}
  \label{tab:yields}
\end{table}

\section{Results}
\label{sec:results}

For each pseudo-experiment in the ensemble, I record the fit residual for the relative magnitude and phase difference between $\Bp \to \Koneaz \pip$ and $\Bp \to \Kz \aonep$. The spread of these distributions is a measure of the statistical and specifically considered systematic effects at play in this study, namely those arising from the uncertainties on the hadronic parameters listed at the bottom of in table~\ref{tab:stage1} that are varied in generation but fixed to their nominal values in the fit. The fit residuals are also recorded for the ensemble where these model parameters are instead fixed to their generated values in order to isolate the statistical portion of the total error. In this analysis, what is referred to as the statistical error also includes deviations propagated from the current experimental knowledge of the considered branching fractions listed in table~\ref{tab:stage2} in addition to those expected purely from the expected total yields. These are recorded in table~\ref{tab:results} and demonstrate that the hadronic uncertainties from current experimental knowledge dominate the overall error. The hadronic uncertainty itself can be inferred from the quadratic subtraction of the statistical error from the total error, which for the 50000 signal events sample amounts to a 13.9\% uncertainty on the relative magnitude between $\Bp \to \Koneaz \pip$ and $\Bp \to \Kz \aonep$ and a $9.3^\circ$ uncertainty on their phase difference.
\begin{table}[tbp]
  \centering
  \begin{tabular}{|c|c|c|}
    \hline
    Projected Yield & $\delta |a_{\Bp \to \Kz a_1(1260)^+}|$ (\%) & $\delta \arg(a_{\Bp \to \Kz a_1(1260)^+})$ $({}^\circ)$\\ \hline
    %5000 & 1.4 & 8.2\\ %7.1
    %10000 & 0.9 & 5.6\\ %5.1
    %50000 & 0.4 & 3.9\\ %2.8
    \phantom{0}5000 & 17.7 (7.2) [12.3] & 17.3 (13.3) [13.1]\\
    10000 & 15.5 (4.6) \phantom{0}[8.6] & 14.4 (10.5) \phantom{0}[9.4]\\
    50000 & 14.1 (2.7) \phantom{0}[4.2] & 12.0 \phantom{0}(7.6) \phantom{0}[7.3]\\
    \hline
  \end{tabular}
  \caption{Expected uncertainties including systematic effects from the considered hadronic sources for the relative magnitude and phase difference between $\Bp \to \Koneaz \pip$ and $\Bp \to \Kz \aonep$ for different projected yields with fixed resonance pole parameters. The values inside the round brackets indicate the statistical uncertainties only while the square brackets indicate the total uncertainty when the axial pole parameters are released in the fit.}
  \label{tab:results}
\end{table}

This situation where the total error barely decreases with data sample size is clearly untenable in the long term for a precision \phitwo\ measurement, so obviously steps must be taken to absorb the model uncertainty into the statistical error. The pole parameters of the axial vectors are relatively poorly known at this time, with the uncertainty on the $K_1(1270)$ width alone being over 20\%. As these states are expected to dominate the phase space, it is therefore imperative that their masses and widths are released in the fit. I repeat the ensemble test with this premise as shown in table~\ref{tab:results} and find that the magnitude of the $\Bp \to \Kz \aonep$ relative to $\Bp \to \Koneaz \pip$ is now only affected by hadronic uncertainties at the same level as the statistical error for the 50000 signal event sample, while the phase difference is largely immune. The slightly smaller uncertainty on the phase difference when compared to the pure statistical error is due to a small fit bias which prefers a larger $\Bp \to \Kz \aonep$ magnitude relative to the $\Bp \to \Koneaz \pip$ when the axial pole parameters are released. Fortunately, this discrepancy becomes less pronounced as the data sample size increases, as can be seen by the reduction in the corresponding uncertainty of the magnitude parameter relative to its statistical uncertainty.

Given the potentially devastating impact of hadronic uncertainties in axial vector states on the precision of the measured amplitude, there are implications for other critical analyses that intend to involve the $K\pi\pi$ mass spectrum. For example, an amplitude analysis of $B \to K\pi\pi\gamma$ decays can be used to test the SM by probing the polarisation of the emitted photon~\cite{kpipigamma,kpipigamma2}, and an improved understanding of the hadronic uncertainties associated to the $K\pi\pi$ system will benefit also such studies.

The phase space of $\Bp \to \Ks \pip \pim \pip$ does not provide enough degrees of freedom for an amplitude analysis to determine the $K_1$ mixing angle as it is essentially a multiplicative factor to the free amplitude coefficients assigned to each decay channel. As this parameter must therefore be fixed in the fit, I study any potential impact its central value may have on the ability to determine the amplitude parameters of the $\Bp \to \Kz \aonep$ relative to $\Bp \to \Koneaz \pip$. This is achieved by scanning $\theta_{K_1}$ in the range $[-90,+90]^\circ$, and performing each fit to a pseudo-experiment picked from the ensemble. Firstly, the $-2\log {\cal L}$ remains constant across the range of $\theta_{K_1}$, confirming that no discrimination power is possible for this parameter.

I also track the central values of $|a_{\Bp \to \Kz a_1(1260)^+}|$ and $\arg(a_{\Bp \to \Kz a_1(1260)^+})$ as a function of $\theta_{K_1}$. For the most part, the phase difference fluctuation is negligible at a level below $0.1^\circ$, which is expected as the mixing parameter is real. At some point in the negative region of $\theta_{K_1}$ however, its value becomes large enough such that the reference $\Bp \to \Koneaz \pip$ amplitude acquires a negative sign at which the phase difference flips by $180^\circ$ to compensate. Conversely, the relative magnitude of $\Bp \to \Kz a_1(1260)^+$ varies wildly with the mixing angle, however this cannot impact its branching fraction measurement which is needed for the \phitwo\ constraint as fit fractions themselves are absolute. Of course the branching fraction of $\Bp \to \Koneaz \pip$, another input to the \phitwo\ measurement, is greatly affected by the mixing angle. By inspection of the amplitude parametrisation, it can be seen that some values of the mixing angle will cause the branching fraction to go to zero, while it would be maximised for other values. Essentially, this forces experiment to fix a value for $\theta_{K_1}$ in the interim without being able to assign an associated uncertainty as occurred in ref.~\cite{phi2_a1pi3}. If a consensus from the theoretical community can be found on the value for the $K_1$ mixing angle along with an uncertainty profile, this would greatly assist experimentalists in providing measurements of physical parameters that depend on it.

\section{Conclusion}
\label{sec:conc}

I present an extension to the time-dependent amplitude analysis of $\Bz \to \pip \pim \pip \pim$ decays first proposed in ref.~\cite{rhorho_dalseno}, that has the capacity to resolve the \phitwo\ solution degeneracy in the $\Bz \to \aone\pimp$ system within SU(3) flavour symmetry. This can ultimately be achieved through an amplitude analysis of the $\Bp \to \Ks \pip \pim \pip$ final state, where the primary goal is to measure the phase difference between the pure penguin channels, $\Bp \to \Koneaz \pip$ and $\Bp \to \Kz \aonep$. Complementary input from $\Bz \to \Kp \pim \pip \pim$ could also be explored by experiment in future analyses if the charged mode fails to provide a unique solution for \phitwo. The formalism by which to parametrise the $K_1$ sector in order to accomplish this is outlined and no dependency on the $K_1$ mixing angle is foreseen for the phase difference. By contrast, the $\Bp \to \Koneaz \pip$ branching fraction dependency will benefit from an external measurement of the mixing angle such that its impact on the \phitwo\ constraint can be quantified.

Accounting for the main hadronic systematic uncertainties, this phase difference is expected to be measured to the level of around $7^\circ$ with Run~3 data from LHCb and the final Belle~II sample and will be statistics limited as long as the axial pole parameters are released in the fit. Given a favourable central value, this parameter can lead to the removal of all degenerate \phitwo\ solutions in $\Bz \to \aone\pimp$ through combination of the proposed SU(3)-related amplitude analyses. Furthermore, the number of physical observables available in these two systems may even be sufficient so as to allow experiment to directly constrain non-factorisable SU(3)-breaking effects. While Belle II already has a comprehensive \phitwo\ programme, this proposal may be particularly attractive to LHCb as the only other possibility for a \phitwo\ constraint conducted exclusively with their own data is the experimentally-challenging time-dependent amplitude analysis of $\Bz \to \pip \pim \piz$.

\acknowledgments{
  I am grateful to P.~Vanhoefer for our discussions arising from his work on $\Bz \to \rhoz \rhoz$ while I studied $\Bz \to \aone \pimp$ at Belle, which inspired this idea. Thanks also goes out to J.~Rademacker for commenting on the paper draft, and whose pioneering methodology developed for 4-body Dalitz Plot analyses provided me with the knowledge necessary to evaluate the feasibility of this method. I am indebted to M.~Charles who checked the projections on behalf of the LHCb physics coordination and to H.-Y.~Cheng for useful exchanges on the $K_1$ mixing angle. Finally, very special thanks to T.~Gershon, whose careful reading improved this work immensely. This work was supported by XuntaGal (Spain).
}


\begin{thebibliography}{99}

\bibitem{Cabibbo}
  N.~Cabibbo, \emph{Unitary Symmetry and Leptonic Decays}, \emph{Phys.~Rev.~Lett.} {\bf 10} (1963) 531.

\bibitem{KM}
  M.~Kobayashi and T.~Maskawa, \emph{$CP$ Violation in the Renormalizable Theory of Weak Interaction}, \emph{Prog.~Theor.~Phys.} {\bf 49} (1973) 652.

\bibitem{phi2_pipi1}
  BaBar collaboration, J.P.~Lees et al., \emph{Measurement of $CP$ Asymmetries and Branching Fractions in Charmless Two-Body $B$-Meson Decays to Pions and Kaons}, \emph{Phys.~Rev.} {\bf D 87} (2013) 052009 [arXiv:1206.3525].

\bibitem{phi2_pipi2}
  Belle collaboration, J.~Dalseno et al., \emph{Measurement of the $CP$-violation parameters in $B^0 \to \pi^+ \pi^-$ decays}, \emph{Phys.~Rev.} {\bf D 88} (2013) 092003 [arXiv:1302.0551].

\bibitem{phi2_pipi3}
  LHCb collaboration, R.~Aaij et al., \emph{Measurement of $CP$ asymmetries in two-body $B^0_{(s)}$-meson decays to charged pions and kaons}, \emph{Phys.~Rev.} {\bf D 98} (2018) 032004 [arXiv:1805.06759 (2018)].

\bibitem{phi2_pipi4}
  BaBar collaboration, B.~Aubert et al., \emph{Study of $B^0 \to \pi^0 \pi^0$, $B^\pm \to \pi^\pm \pi^0$, and $B^\pm \to K^\pm \pi^0$ Decays and Isospin Analysis of $B \to \pi \pi$ Decays}, \emph{Phys.~Rev.} {\bf D 76} (2007) 091102 [arXiv:0707.2798].

\bibitem{phi2_pipi5}
  Belle collaboration, Y.-T.~Duh et al., \emph{Measurements of branching fractions and direct $CP$ asymmetries for $B \to K\pi$, $B \to \pi \pi$ and $B \to KK$ decays}, \emph{Phys.~Rev.} {\bf D 87} (2013) 031103 [arXiv:1210.1348].

\bibitem{phi2_pipi6}
  Belle collaboration, T.~Julius et al., \emph{Measurement of the branching fraction and $CP$ asymmetry in $B^0 \to \pi^0 \pi^0$ decays and an improved constraint on $\phi_2$}, \emph{Phys.~Rev.} {\bf D 96} (2017) 032007 [arXiv:1705.02083].

\bibitem{phi2_rhopi1}
  BaBar collaboration, J.P.~Lees et al., \emph{Measurement of $CP$-violating asymmetries in $B^0 \to (\rho\pi)^0$ decays using a time-dependent Dalitz plot analysis}, \emph{Phys.~Rev.} {\bf D 88} (2013) 012003 [arXiv:1304.3503].

\bibitem{phi2_rhopi2}
  Belle collaboration, A.~Kusaka et al., \emph{Measurement of $CP$ Asymmetry in a Time-Dependent Dalitz Analysis of $B^0 \to (\rho\pi)^0$ and a Constraint on the CKM Angle $\phi_2$}, \emph{Phys.~Rev.~Lett.} {\bf 98} (2007) 221602 [hep-ex/0701015].

\bibitem{phi2_rhorho1}
  BaBar collaboration, B.~Aubert et al., \emph{A Study of $B^0 \to \rho^+ \rho^-$ Decays and Constraints on the CKM Angle $\alpha$}, \emph{Phys.~Rev.} {\bf D 76} (2007) 052007 [arXiv:0705.2157].

\bibitem{phi2_rhorho2}
  Belle collaboration, P.~Vanhoefer et al., \emph{Study of $B^0 \to \rho^+ \rho^-$ decays and implications for the CKM angle $\phi_2$}, \emph{Phys.~Rev.} {\bf D 93} (2016) 032010 [\emph{Erratum ibid.} {\bf D 94} (2016) 099903] [arXiv:1510.01245].

\bibitem{phi2_rhorho3}
  BaBar collaboration, B.~Aubert et al., \emph{Improved Measurement of $B^+ \to \rho^+ \rho^0$ and Determination of the Quark-Mixing Phase Angle $\alpha$}, \emph{Phys.~Rev.~Lett.} {\bf 102} (2009) 141802 [arXiv:0901.3522].

\bibitem{phi2_rhorho4}
  Belle collaboration, J.~Zhang et al., \emph{Observation of $B^\mp \to \rho^\mp \rho^0$ Decays}, \emph{Phys.~Rev.~Lett.} {\bf 91} (2003) 221801 [hep-ex/0306007].

\bibitem{phi2_rhorho5}
  BaBar collaboration, B.~Aubert et al., \emph{Measurement of the Branching Fraction, Polarization and $CP$ Asymmetries in $B^0 \to \rho^0 \rho^0$ Decay and Implications for the CKM Angle $\alpha$}, \emph{Phys.~Rev.} {\bf D 78} (2008) 071104 [arXiv:0807.4977].

\bibitem{phi2_rhorho6}
  Belle collaboration, P.~Vanhoefer et al., \emph{Study of $B^0 \to \rho^0 \rho^0$ decays, implications for the CKM angle $\phi_2$ and search for other $B^0$ decay modes with a four-pion final state}, \emph{Phys.~Rev.} {\bf D 89} (2014) 072008 [\emph{Erratum ibid.} {\bf D 89} (2014) 119903] [arXiv:1212.4015].

\bibitem{phi2_rhorho7}
  LHCb collaboration, R.~Aaij et al., \emph{Observation of the $B^0 \to \rho^0 \rho^0$ decay from an amplitude analysis of $B^0 \to (\pi^+ \pi^-)(\pi^+ \pi^-)$ decays}, \emph{Phys.~Lett.} {\bf B 747} (2015) 468 [arXiv:1503.07770].

\bibitem{phi2_a1pi1}
  BaBar collaboration, B.~Aubert et al., \emph{Measurements of $CP$-Violating Asymmetries in $B^0 \to a_1^\pm(1260) \pi^\mp$ decays}, \emph{Phys.~Rev.~Lett.} {\bf 98} (2007) 181803 [hep-ex/0612050].

\bibitem{phi2_a1pi2}
  Belle collaboration, J.~Dalseno et al., \emph{Measurement of Branching Fraction and First Evidence of $CP$-violation in $B^0 \to a_1^\pm(1260) \pi^\mp$ Decays}, \emph{Phys.~Rev.} {\bf D 86} (2012) 092012 [arXiv:1205.5957].

\bibitem{phi2_a1pi3}
  BaBar collaboration, B.~Aubert et al., \emph{Measurement of branching fractions of $B$ decays to $K_1(1270) \pi$ and $K_1(1400) \pi$ and determination of the CKM angle $\alpha$ from $B^0 \to a_1(1260)^\pm \pi^\mp$}, \emph{Phys.~Rev.} {\bf D 81} (2010) 052009 [arXiv:0909.2171].

\bibitem{phi2_gronau}
  M.~Gronau and J.L~Rosner, \emph{Improving the measurement of the CKM phase $\phi_2 = \alpha$ in $B \to \pi\pi$ and $B \to \rho\rho$ decays}, \emph{Phys.~Lett.} {\bf B 763} (2016) 228 [arXiv:1608.06224].

\bibitem{CKMFitter1}
  J.~Charles, O.~Deschamps, S.~Descotes-Genon and V.~Niess, \emph{Isospin analysis of charmless $B$-meson decays}, \emph{Eur.~Phys.~J.} {\bf C 77} (2017) 574 [arXiv:1705.02981].

\bibitem{UTfit}
  UTfit collaboration, M. Bona et al., \emph{The Unitarity Triangle Fit in the Standard Model and Hadronic Parameters from Lattice QCD: A Reappraisal after the Measurements of $\Delta m_s$ and $BR(B \to \tau \nu_\tau)$}, \emph{JHEP} {\bf 10} (2006) 081 [hep-ph/0606167] and online at \verb|http://utfit.org/UTfit/WebHome|.

\bibitem{rhorho_dalseno}
  J. Dalseno, \emph{Resolving the \phitwo\ ($\alpha$) ambiguity in $B \to \rho \rho$}, \emph{JHEP} {\bf 11} (2018) 193 [arXiv:1808.09391].

\bibitem{iso_mixing}
  M.~Gronau and J.~Zupan, \emph{Isospin-breaking effects on $\alpha$ extracted in $B \to \pi \pi$, $\rho \rho$, $\rho \pi$}, \emph{Phys.~Rev.} {\bf D 71} (2005) 074017 [hep-ph/0502139].

\bibitem{rhowidth1}
  A.F.~Falk, Z.~Ligeti, Y.~Nir and H.~Quinn, \emph{Comment on extracting $\alpha$  from $B \to \rho \rho$}, \emph{Phys.~Rev.} {\bf D 69} (2004) 011502(R) [hep-ph/0310242].

\bibitem{rhowidth2}
  M.~Gronau and J.L~Rosner, \emph{Controlling $\rho$ width effects for a precise value of $\alpha$ in $B \to \rho \rho$}, \emph{Phys.~Lett.} {\bf 766} (2017) 345 [arXiv:1612.08524].

\bibitem{pipi_th}
  M.~Gronau and D.~London, \emph{Isospin analysis of $CP$ asymmetries in $B$ decays}, \emph{Phys.~Rev.~Lett.} {\bf 65} (1990) 3381.

\bibitem{rhopi_th1}
  H.J.~Lipkin, Y.~Nir, H.R.~Quinn and A.~Snyder, \emph{Penguin trapping with isospin analysis and $CP$ asymmetries in $B$ decays}, \emph{Phys.~Rev.} {\bf D 44} (1991) 1454.

\bibitem{rhopi_th2}
  A.E.~Snyder and H.R.~Quinn, \emph{Measuring $CP$ asymmetry in $B \to \rho \pi$ decays without ambiguities}, \emph{Phys.~Rev.} {\bf D 48} (1993) 2139.

\bibitem{a1pi_th}
  M.~Gronau and J.~Zupan, \emph{Weak phase $\alpha$ from $B^0 \to a_1^\pm(1260) \pi^\mp$}, \emph{Phys. Rev.} {\bf D 73} (2006) 057502 [hep-ph/0512148].

\bibitem{PDG}
  Particle Data Group, M.~Tanabashi et al., \emph{Review of Particle Physics}, \emph{Phys.~Rev.} {\bf D 98} (2018) 030001.

\bibitem{CKMFitter2}
  CKMfitter Group, J. Charles et al., \emph{CP violation and the CKM matrix: assessing the impact of the asymmetric B factories}, \emph{Eur. Phys. J.} {\bf C 41} (2005) 1 [hep-ph/0406184] and online at \verb|http://ckmfitter.in2p3.fr|.

\bibitem{HFAG}
  Heavy Flavour Averaging Group, Y.~Amhis et al., \emph{Averages of $b$-hadron, $c$-hadron, and $\tau$-lepton properties as of summer 2016}, \emph{Eur. Phys. J.} {\bf C 77} (2017) 895 [arXiv:1612.07233].

\bibitem{fa1}
  J.C.R.~Bloch, Yu.L.~Kalinovsky, C.D.~Roberts and S.M.~Schmidt, \emph{Describing $a_1$ and $b_1$ decays}, \emph{Phys. Rev.} {\bf D 60} (1999) 111502(R) [nucl-th/9906038].

\bibitem{fk1}
  H.-Y.~Cheng and K.-C.~Yang, \emph{Hadronic charmless $B$ decays $B \to A P$}, \emph{Phys. Rev.} {\bf D 76} (2007) 114020 [arXiv:0709.0137].

\bibitem{a1k}
  BaBar collaboration, B.~Aubert et al., \emph{Observation of $B^+ \to a_1^+(1260) K^0$ and $B^0 \to a_1^-(1260) K^+$}, \emph{Phys.~Rev.~Lett.} {\bf 100} (2008) 051803 [arXiv:0709.4165].

\bibitem{fpi1}
  RBC and UKQCD collaborations, T. Blum et al., \emph{Domain wall QCD with physical quark masses}, \emph{Phys.~Rev.} {\bf D 93} (2016) 074505 [arXiv:1411.7017].
    
\bibitem{fpi2}
  RBC and UKQCD collaborations, R. Arthur et al., \emph{Domain wall QCD with near-physical pions}, \emph{Phys.~Rev.} {\bf D 87} (2013) 094514 [arXiv:1208.4412].

\bibitem{fpi3}
  J. Laiho and R. Van de Water, \emph{Pseudoscalar decay constants, light-quark masses,
and $B_K$ from mixed-action lattice QCD}, \emph{PoS LATTICE2011} (2011) 293.

\bibitem{fpi4}
  MILC collaboration, A. Bazavov et al., \emph{Results for light pseudoscalar mesons}, \emph{PoS LATTICE2010} (2010) 074.

\bibitem{fpi5}
  BMW collaboration, S. Durr et al., \emph{Ratio $F_K/F_\pi$ in QCD}, \emph{Phys. Rev.} {\bf D 81} (2010) 054507 [arXiv:1001.4692].

\bibitem{fpi6}
  HPQCD and UKQCD collaborations, E. Follana et al., \emph{High-Precision Determination of the $\pi$, $K$, $D$, and $D_s$ Decay Constants from Lattice QCD}, \emph{Phys. Rev. Lett.} {\bf 100} (2008) 062002 [arXiv:0706.1726].

\bibitem{fK1}
  ETM collaboration, N. Carrasco et al., \emph{Leptonic decay constants $f_K$, $f_D$, and $f_{D_s}$ with $N_f = 2+1+1$ twisted-mass lattice QCD}, \emph{Phys. Rev.} {\bf D 91} (2015) 054507 [arXiv:1411.7908].

\bibitem{fK2}
  Fermilab Lattice and MILC collaborations, A. Bazavov et al., \emph{Charmed and light pseudoscalar meson decay constants from four-flavor lattice QCD with physical light quarks}, \emph{Phys. Rev.} {\bf D 90} (2014) 074509 [arXiv:1407.3772].

\bibitem{fK3}
  HPQCD collaboration, R.J. Dowdall et al., \emph{$V_{us}$ from $\pi$ and $K$ decay constants in full lattice QCD with physical $u$, $d$, $s$, and $c$ quarks}, \emph{Phys. Rev.} {\bf D 88} (2013) 074504 [arXiv:1303.1670].

\bibitem{bwbf}
  F.~Von~Hippel and C.~Quigg, \emph{Centrifugal-barrier effects in resonance partial decay widths, shapes and production amplitudes}, \emph{Phys. Rev.} {\bf D 5} (1972) 624.

\bibitem{spin}
  W.~Rarita and J.~Schwinger, \emph{On a theory of particles with half integral spin}, \emph{Phys.~Rev.} {\bf 60} (1941) 61.

\bibitem{gs}
  G.J.~Gounaris and J.J.~Sakurai, \emph{Finite width corrections to the vector meson dominance prediction for $\rho \to e^+ e^-$}, \emph{Phys. Rev. Lett.} {\bf 21} (1968) 244.

\bibitem{flatte}
  S.M.~Flatt\'{e}, \emph{Coupled-channel analysis of the $\pi\eta$ and $K \bar K$ systems near $K \bar K$ threshold}, \emph{Phys. Lett. B} {\bf 63} (1976) 244.

\bibitem{CLEO_4pi}
  P.~d'Argent and N.~Skidmore et al., \emph{Amplitude Analyses of $D^0 \to \pi^+ \pi^- \pi^+ \pi^-$ and $D^0 \to K^+ K^- \pi^+ \pi^-$ Decays}, \emph{JHEP} {\bf 05} (2017) 143 [arXiv:1703.08505].

\bibitem{k1mixing1}
  M.~Suzuki, \emph{Strange axial-vector mesons}, \emph{Phys. Rev. D} {\bf 47} (1993) 1252.

\bibitem{k1mixing2}
  L.~Burakovsky and T.~Goldman, \emph{Regarding the enigmas of $P$-wave meson spectroscopy}, \emph{Phys. Rev. D} {\bf 57} (1998) 2879 [hep-ph/9703271].

\bibitem{k1mixing3}
  H. Hatanaka and K.-C. Yang, \emph{$B \to K_1 \gamma$ decays in the light-cone QCD sum rules}, \emph{Phys. Rev. D} {\bf 77} (2008) 094023 [\emph{Erratum ibid.} {\bf D 78} (2008) 059902] [arXiv:0804.3198].

\bibitem{k1mixing4}
  H.-Y. Cheng, \emph{Revisiting axial-vector meson mixing}, \emph{Phys. Lett. B} {\bf 707} (2012) 116 [arXiv:1110.2249].
  
\bibitem{kstrhof0}
  BaBar collaboration, P.~del Amo Sanchez et.al., \emph{Measurements of branching fractions, polarizations, and direct $CP$-violation asymmetries in $\Bp \to \rhoz \Kstp$ and $\Bp \to \fz(980) \Kstp$ decays}, \emph{Phys. Rev. D} {\bf 83} (2011) 051101(R) [arXiv:1012.4044].

\bibitem{f0}
  BES collaboration, M.~Ablikim et al., \emph{Resonances in $J/\psi \to \phi \pip \pim$ and $\phi \Kp \Km$}, \emph{Phys. Lett. B} {\bf 607} (2005) 243 [arXiv:hep-ex/0411001].

\bibitem{LHCbf0}
  LHCb collaboration, R.~Aaij et. al., \emph{Analysis of the resonant components in $\bar B^0_s \to J/\psi \pip \pim$}, \emph{Phys. Rev. D} {\bf 86} (2012) 052006 [arXiv:1204.5643].

\bibitem{a1k_babar}
  BaBar collaboration, B.~Aubert et.al., \emph{Observation of $\Bp \to \aonep(1260) \Kz$ and $\Bz \to \aonem(1260) \Kp$}, \emph{Phys. Rev. Lett.} {\bf 100}, (2008) 051803 [arXiv:0709.4165].

\bibitem{k3pi_lhcb}
  LHCb collaboration, R.~Aaij et. al., \emph{Study of the $\Bz \to \rho(770)^0 K^*(892)^0$ decay with an amplitude analysis of $\Bz \to (\pip \pim)(\Kp \pim)$ decays}, \emph{JHEP} {\bf 05} (2019) 026 [arXiv:1812.07008].

\bibitem{kspipi_lhcb}
  LHCb collaboration, R.~Aaij et. al., \emph{Updated branching fraction measurements of $B_{(s)}^0 \to K_S^0 h^+ h^{\prime -}$ decays}, \emph{JHEP} {\bf 11} (2017) 027 [arXiv:1707.01665].

\bibitem{kpipi_lhcb}
  LHCb collaboration, R.~Aaij et. al., \emph{Measurements of $CP$ violation in the three-body phase space of charmless $B^\pm$ decays}, \emph{Phys. Rev. D} {\bf 90} (2014) 112004 [arXiv:1408.5373].

\bibitem{phsp}
  F.~James, \emph{Monte Carlo phase space}, CERN-68-15.

\bibitem{qft}
  M.~Williams, \emph{Numerical Object Oriented Quantum Field Theory Calculations}, \emph{Comput.~Phys.~Commun.} {\bf 180} (2009) 1847 [arXiv:0805.2956] and online at \verb|https://github.com/jdalseno/qft|.

\bibitem{belle2}
  Belle II collaboration, E.~Kou and P.~Urquijo et al., \emph{The Belle II Physics Book}, BELLE2-PUB-PH-2018-001 (2018) [arXiv:1808.10567].

\bibitem{run3}
  LHCb collaboration, \emph{Physics case for an LHCb Upgrade II --- Opportunities in flavour physics and beyond, in the HL-LHC era}, CERN-LHCC-2018-027 (2018) [arXiv:1808.08865].

\bibitem{trigger}
  LHCb collaboration, \emph{LHCb Trigger and Online Upgrade Technical Design Report}, CERN-LHCC-2014-016 (2014).

\bibitem{kpipigamma}
  M. Gronau, Y. Grossman, D. Pirjol and Anders Ryd, \emph{Measuring the Photon Polarization in $B \to K \pi \pi \gamma$}, \emph{Phys. Rev. Lett.} {\bf 88}, (2002) 051802 [hep-ph/0107254].

\bibitem{kpipigamma2}
  V. Bell\'{e}e et al., \emph{Using an amplitude analysis to measure the photon polarisation in $B \to K \pi \pi \gamma$ decays}, arXiv:1902.09201 (2019).

\end{thebibliography}
\end{document}